\documentclass[preprint,12pt]{elsarticle}

\usepackage[english]{babel}
\usepackage[letterpaper,top=2cm,bottom=2cm,left=3cm,right=3cm,marginparwidth=1.75cm]{geometry}

\usepackage{graphicx}
\usepackage[linesnumbered,ruled, noline]{algorithm2e}
\usepackage{float}
\usepackage{tikz}
\usepackage{amssymb}
\usepackage{amsmath}
\usepackage{listings}
\usepackage{tikz}
\usepackage{subcaption}
\usepackage{rotating}

\biboptions{sort&compress}

\usepackage[colorlinks=true, allcolors=blue]{hyperref}
\usepackage{indentfirst}
\usepackage{color}
\usepackage{comment}

\usepackage{amsthm}
\usepackage{graphicx}
\graphicspath{{figs/}} %% Define the relative path of the images.
\usepackage{booktabs}
\usepackage{multirow}
\usepackage[outline]{contour}
\usepackage{bm} 
\usepackage{svg}
\usepackage[title]{appendix}
\makeatletter
\@addtoreset{table}{section}
\makeatother

\makeatletter
\@addtoreset{figure}{section}
\makeatletter

\begin{document}
	
%%%%%%%%%%%%%%%%%%%%%%%%%%%%%%%%%%%%%%%%%%%%%%%%%%%%%%%%%%%%%

\begin{frontmatter}

\title{An explicit multi-time stepping algorithm for
	multi-time scale coupling problems in SPH}

\author{Xiaojing Tang}
\ead{xiaojing.tang@tum.de}
	\author{Dong Wu}
\ead{dong.wu@tum.de}
	\author{Zhengtong Wang}
\ead{zhentong.wang@tum.de}
\author{Oskar Haidn}
\ead{oskar.haidn@tum.de}
\author{Xiangyu Hu\corref{mycorrespondingauthor}}
\cortext[mycorrespondingauthor]{Corresponding author.}
\ead{xiangyu.hu@tum.de}	
\address{TUM School of Engineering and Design,  Technical University of Munich, 85748 Garching, Germany}
	
\begin{abstract}
Simulating  physical problems involving 
multi-time scale coupling  is  
challenging due to the need
of solving these multi-time scale
processes simultaneously. 
In response to this challenge, 
this paper proposed an explicit multi-time step algorithm coupled with a solid dynamic relaxation scheme. 
The explicit scheme simplifies the equation system  
in contrast to the implicit scheme, 
while the multi-time step algorithm allows the
equations of different physical processes to be solved under different time step sizes. 
Furthermore, an implicit viscous damping relaxation  
technique is  applied to significantly reduce   computational 
iterations required to achieve equilibrium 
in the comparatively fast solid 
response process. 
To validate  the accuracy and efficiency of 
the proposed algorithm, 
two distinct scenarios, i.e., a nonlinear hardening 
bar stretching and a fluid diffusion coupled with
Nafion membrane flexure, are simulated. 
The results show good agreement with experimental
data and results from other numerical methods, 
and the simulation  time is reduced firstly  by 
independently addressing different processes 
with the multi-time step algorithm
and secondly 
decreasing solid dynamic relaxation time
through the incorporation of damping techniques. 

\end{abstract}


\end{frontmatter}

\section{introduction} 
Smoothed Particle Hydrodynamics (SPH), 
a typically  mesh-free method, which is originally introduced 
by Lucy  \cite{lucy1977numerical}, Gigold and Monaghan \cite{gingold1977smoothed} for studying astrophysical problems, has been widely applied
to simulate fluid-flows  \cite{monaghan1994simulating,hu2006multi,shao2006simulation, zhang2019weakly}, 
solid mechanics \cite{libersky1991smooth,benz1995simulations,
monaghan2000sph,randles1996smoothed,zhang2021simple},  fluid-structure interaction  \cite{antoci2007numerical,  han2018sph, zhang2021multi} in recent years. 
Comprehensive reviews can be found in Refs. \cite{liu2010smoothed,monaghan2012smoothed, Luo30Particle, 
	Zhang31Review, Gotoh32On}. 
Even with wide applications, 
SPH has some limitations when it comes to
simulating multi-scale coupling problems 
existing in various engineering fields, 
particularly those involving solid dynamic response which 
is a typically fast process
 \cite{brackbill2014multiple}. 
The disparity in the time scales of fast 
and slow processes presents a continuing 
challenge to numerical simulations \cite{knoll2003balanced}.

To solve multi-time scale problems,
either an implicit or explicit 
scheme can be applied. 
The implicit scheme allows for a larger 
time step in the time integration \cite{prior1994applications,gavalas2018mesh}, 
enabling the monolithic scheme to 
solve the equations for all
fast and slow processes simultaneously. 
For instance, Zhao \cite{zhao2013modeling} used an 
implicit Newmark scheme to model 
the flow through a porous elastic solid, 
where solid dynamics and fluid diffusion occur 
at different time scales.
Gaston \cite{gaston2009moose} employed an implicit 
scheme to analyze the fluid, chemistry, and 
structure coupling behavior in a reactor, 
which is a common phenomenon in the engineering field.
However, since the inversion of the stiffness matrix 
used for solving equations is
required  for each time step \cite{gavalas105brief,sun2000comparison}, 
this approach is quite expensive concerning both computation time and memory consumption \cite{rezaiee2010dynamic}. 

The explicit approach is more favorable for 
solving multi-time scale coupling problems 
due to its direct time integration and simple 
numerical formulation \cite{yaghmaie2020multi,
ragusa2009consistent, beuth2012formulation, harewood2007comparison}. 
Some researchers have used this approach
to simulate material stretching and necking, 
where the load is applied during a long time 
period while the  material's dynamic response
is instant and fast \cite{doll2000volumetric, de2022new, rao2019explicit}.
Since the realistic load is applied in a long time scale,
a long physical simulation time is expected.
However, with a quite small stable time
step size allowed in explicit scheme 
for the fast process, 
usually  millions of time steps are 
required to simulate the entire process, 
which is very often not feasible.
To reduce the overall simulation time, 
loading rate is usually increased artificially \cite{de2022new}. 
However, high non-realistic loading rate 
may lead to certain limitations and 
inaccuracies in the simulation results \cite{yilmaz2014load}. 
 
This paper presents a multi-time stepping algorithm 
in SPH, where a large and a small time steps are
chosen according to the slow and fast processes in the simulation, respectively. 
Two, i.e.,  an outer and an inner 
loops are arranged with these two time steps for time integration.
Specifically, the slow process is integrated with 
a large time step in the outer loop,
while the  
fast solid dynamic process 
with a much smaller time step in the inner loop. 
Since the time step 
size of the fast process is small, many iterations 
of the solid stress relaxation may 
occur within one outer loop
and lead to low computational efficiency.
To address this issue, 
a dynamic relaxation method based on implicit 
operator splitting scheme \cite{zhu2022dynamic} is adopted to 
accelerate the 
convergence rate of the fast dynamic process to an elastic equilibrium state. 
To assess the performance and computational efficiency of the proposed algorithm, 
the simulations of tensile tests, including two dimensional and three dimensional cases,  
are firstly carried out; and then the evolution of fluid diffusion in porous media coupling with elastic deformation is simulated. The latter fluid-structure
coupling process occurs in chemical 
reactors, e.g. in the fuel cell of  battery,
where fluid mixture diffuses through 
a Nafion membrane,
affecting the battery performance due to the 
varying fluid concentration and membrane deformation. 
The obtained results  demonstrate that the proposed algorithm performs better both in accuracy and efficiency compared to previous numerical methods.

The reminder of this paper is organized as follows.
Section 2 summaries the theories and governing equations 
for nonlinear hardening plastic solid mechanics
and fluid-structure interaction.
Section 3 describes the corresponding
SPH discretization.
In Section 4, the proposed multi-time stepping algorithm 
coupling with the dynamic relaxation  are detailed.
Section 5 states the physical problems and the results obtained using the proposed 
algorithm are compared with those
from previous methods and experiments.
Finally, Section 6 presents brief concluding remarks.
The source code and data needed 
for this numerical simulation work
can be found in SPHinXsys, an open-source multi-physics SPH library, available at 
\url{https://www.sphinxsys.org}.

\section{Governing equations}
\subsection{Total Lagrangian  solid dynamics}
\begin{figure*}[htbp]
	\centering
	\includegraphics[trim = 0mm 0mm 0mm 0mm, clip,width=0.85\textwidth]{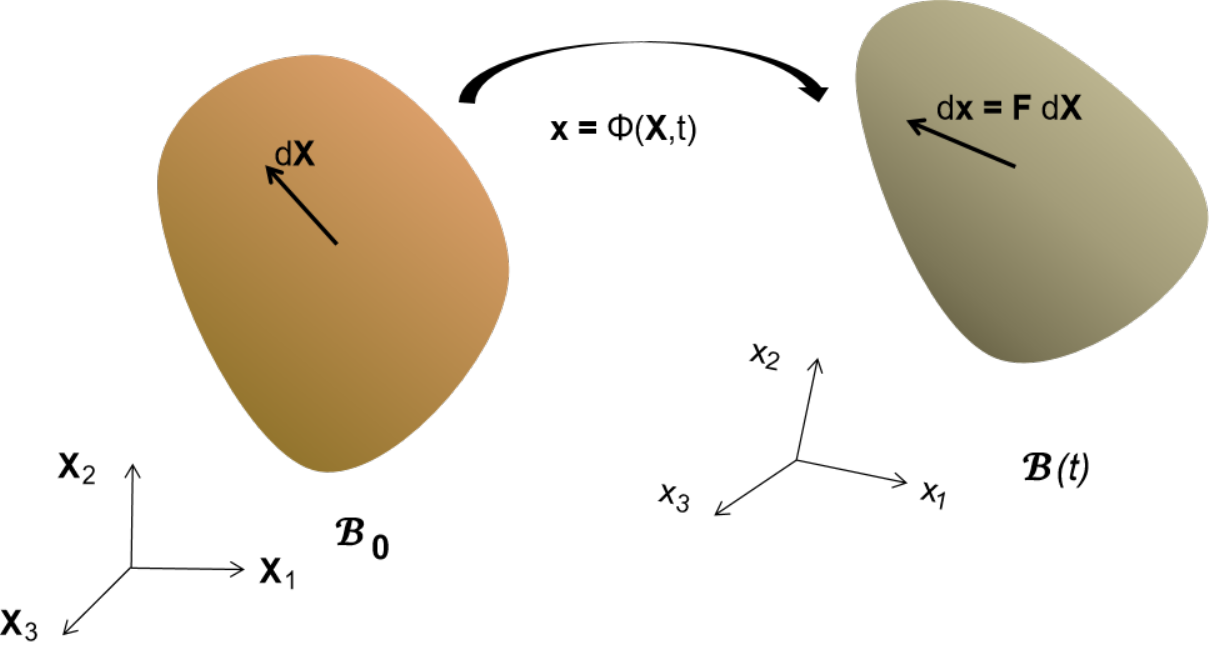}
	\caption{Finite deformation process on a body $ \mathcal{B} $.}
	\label{figure_solid_deformation}
\end{figure*}

In this section, we provide a concise introduction
to solid dynamics
 within the framework of total Lagrange formulation,
along with the relevant notations and and symbols that
will be utilized in the subsequent models.
The analysis focuses on a solid body $\mathcal{B}$, 
which occupies two regions: $\mathcal{R}_0$ and $\mathcal{R}$,
representing the body's configurations at
time $t_0$ ($t=0$) and $t$ respectively.
In the initial configuration $\mathcal{R}_0$,
the position vector of a material point is represented 
by $\mathbf{X}\in \mathcal{R}_0$, 
while in the current configuration, 
it is denoted as $\mathbf{x}\in \mathcal{R}$. 
The motion of the solid body is described 
by the invertible mapping $\phi$, 
which transforms a material point $\mathbf{X}$ 
to its corresponding vector $\mathbf{x}=\phi(\mathbf{X},t)$, as illustrated in Figure. \ref{figure_solid_deformation}.
Based on this definition, 
the Lagrangian velocity of a material point 
is defined as $\mathbf{v}(\mathbf{X},t) = \frac{d\phi(\mathbf{X},t)}{dt}$. 
The deformation gradient $\mathbf{F}$, 
which  characterizes the deviation of 
a material point from its initially 
undeformed position to its deformed position, 
can be computed from the displacement 
vector $\mathbf{u} = \mathbf{x} - \mathbf{X}$
using the following equation: 
\begin{equation} \label{eq:deformationtensor-displacement}
	\mathbf{F} = \frac{d\mathbf{x}}{d\mathbf{X}} = \nabla^{0} {\mathbf{u}}  + \mathbf{I},
\end{equation}
where  $\mathbf{I}$ is the unit matrix, and the superscript 
$\left( {\bullet} \right)^0$ accounts for quantities in 
the initial reference configuration. The corresponding  
Jacobian determinant term $ J = $ det($ \mathbf{F} $) 
indicates the local volume gain  $ J > 1 $ or loss $ J < 1 $.
The governing equations of solid deformation 
within the total Lagrange framework are derived as
\begin{equation}\label{eq:mechanical-mom}
	\begin{cases}
		\rho =  {\rho^0} \frac{1}{J}  \\
		\rho^0 \frac{\text{d} \mathbf{v}}{\text{d} t}  =  \nabla^{0} \cdot \mathbf{P}^T
	\end{cases},
\end{equation}
%. 
where $ \rho $ and $\rho_0$ are the densities 
in the  current configuration $ \mathcal{R} $
 and the 
initial configuration $\mathcal{R}_0$ respectively,	$ \mathbf{v} $ the velocity
and  $\mathbf{P}$ the first 
Piola-Kirchhoff stress tensor.
Different from the Cauchy stress $ \boldsymbol{\sigma} $,
which points to the force measured in 
the deformed configuration,  $\mathbf{P}$ relates to stress within the initial configuration, 
and the two stresses are related by 
\begin{equation}
	\label{stress_trans}
	\mathbf{P}= J \boldsymbol{\sigma}  \mathbf{F}^{-T} = \boldsymbol{\tau}  \mathbf{F}^{-T},
\end{equation} 
where $\boldsymbol{\tau}$ is the Kirchhoff stress tensor, which is obtained from the constitutive relation as given in Appendix \ref{appendixA}.
Also, using a 
multiplicative decomposition technique \cite{simo2006computational, yue2015continuum}, a hardening plastic model is also given in Appendix \ref{appendixA}.

\subsection{Fluid-structure interaction}\label{section_2}

For the fluid diffusion in porous media 
coupling with elastic deformation 
of the porous membrane,  we propose a fluid-structure 
interaction model, where the fluid diffuses in 
the porous solid, leading to an increased fluid pressure 
and solid deformation.
\begin{figure*}[htbp]
	\centering
	\includegraphics[trim = 40mm 40mm 60mm 60mm, clip,width=0.5\textwidth]{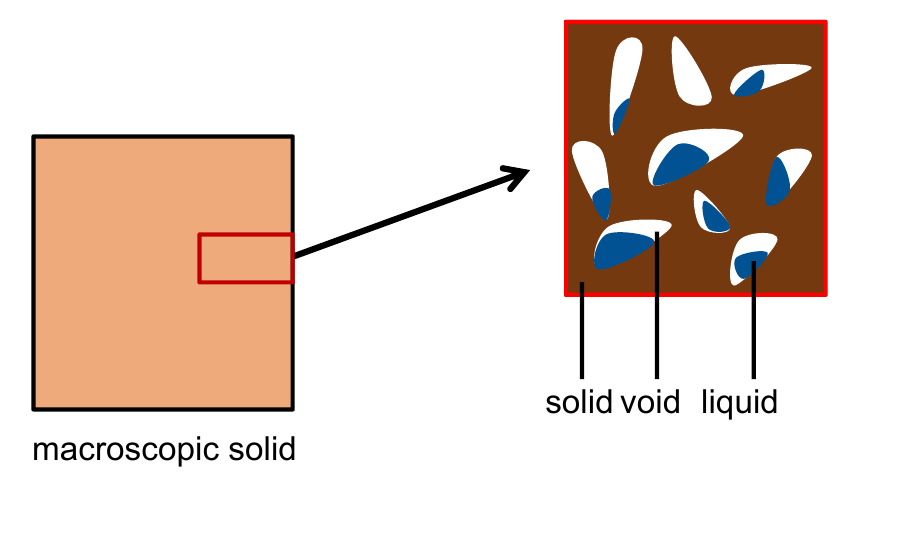}
	\caption{Partially saturated porous medium.}
	\label{continum-sketch}
\end{figure*}
In this model, the heterogeneous body
is considered as a continuous solid medium containing 
uniformly distributed small voids with a homogeneous porosity ${a}$.
When this medium comes 
into contact with a fluid,
fluid flows into these small pores 
and diffuses inside this medium 
due to the presence of the fluid concentration gradient,
resulting in the formation of a mixture 
comprising solid and fluid components, as illustrated in Figure \ref{continum-sketch}.
To simplify this model, 
we adopt the methodology proposed by 
Zhao \cite{zhao2013modeling} to 
to present a mixture momentum equation
while fluid behaviors follow the diffusion law.

\subsubsection{Mass and momentum equations}
With a porosity $ {a}$ and fluid saturation level $ \widetilde{a}$
(see Appendix \ref{appendixB1}), 
the locally effective fluid density $ \rho^l $  
  can be expressed as 
\begin{equation}
	\label{fluid_density}
	\rho^l = \rho^l_0 \widetilde{a},
\end{equation} 
where
$ {\rho}^l_0$ is the 
initial density of the fluid. 
The governing equations for 
the solid body involving the density 
conservation is described as
\begin{equation}\label{eq:density-mom}
\rho^s =  {\rho^s_0} \frac{1}{J},    
\end{equation}
where $ \rho^s$ and  $\rho_0^s$ are the solid density 
defined in current configuration $ \mathcal{R} $ and initial configuration $\mathcal{R}_0$ respectively,   
For a porous solid partially-saturated by fluid,  
the total linear momentum $\mathbf{M}$ in 
the region $\mathcal{R}$ is the sum of fluid momentum 
and solid momentum
\begin{equation}
	\label{P_equation}
	\mathbf{M}= \rho \mathbf{v}=\rho^l \mathbf{v}^{l}+ \rho^s \mathbf{v}^{s},
\end{equation}
where  $ \rho $, $\mathbf{v}$ is the total density and velocity, 
$\mathbf{v}^{l}$ the velocity of fluid,
$\mathbf{v}^{s}$ the velocity 
of dry porous solid. 
Due to the difference between  
  $\mathbf{v}^l$ and  $\mathbf{v}^s$, 
  the fluid flux $\mathbf{q}$ on the element 
boundary $\partial V$ can then be expressed as
\begin{equation}
	\label{defination_q}
	\mathbf{q} = {\rho^l} (\mathbf{v}^{l}-\mathbf{v}^{s}).
\end{equation}
Obviously,  if there is no fluid passing 
through the boundary, $\mathbf{q}=0 $, 
the fluid mass in an element is conserved.
The transfer of fluid mass and momentum 
between micro-scale solid constituents 
happens when fluid flows from regions
with higher fluid saturation to 
 those with lower saturation. 
Therefore, within an element $ dV$ of the mixture, 
the balance of linear momentum implies that 
the time derivative of momentum $\mathbf{M}$ 
is determined by two factors. One is the  stress  
exerting on the element and the other one is 
the fluid flux of linear momentum $\mathbf{v}^{l} \otimes \mathbf{q}$ 
on the boundary $ \partial V $,  
where the symbol $\otimes$ means an outer 
product of two vectors or tensors.  
It follows that 
 the conservation of total linear 
 momentum of the mixture can be expressed as
\begin{equation}
	\frac{D\mathbf{M}}{D t} = \nabla \cdot \boldsymbol{\sigma}- \nabla \cdot\left(\mathbf{v}^{l} \otimes \mathbf{q}\right),
	\label{totalmomentumupdate}
\end{equation}
where $\boldsymbol{\sigma}$ represents 
the cumulative Cauchy stress in the mixture
acting on the solid. 
$ \boldsymbol{\sigma} $ is determined by 
Cauchy stress $\boldsymbol{\sigma}^s$ and 
the pressure stress due to the presence of the fluid phase $\boldsymbol{\sigma}^l$, which is detailed in 
 Appendix \ref{appendixB3}.

\subsubsection{Fick's law} 
In a partially saturated  solid, 
the fluid saturation difference leads to the motion of fluid from higher fluid fraction to lower parts and the flux follows the Fick's law 
\begin{equation}
	\label{grad_q}
	\mathbf{q} = -K \rho ^l \nabla \widetilde{a},
\end{equation}
indicating that  the fluid flux is proportional to the diffusivity $K$, the effective fluid density $\rho^{l} $ as well as  the gradient of the fluid saturation $ \widetilde{a} $.
Consequently,
the time derivative of fluid mass in an element $dV$ is 
due to the fluid flux $\mathbf{q}$ 
on the element 
boundary $\partial V$, written as 
\begin{equation}
\frac{D \rho^{l}}{D t}= - \nabla \cdot \mathbf{q}.
\label{fluid_mass_der}
\end{equation}

\section{SPH implementation} 
In SPH, the continuum is represented by a 
set of Lagrangian particles that carry 
various properties, such as mass, position, 
velocity, and other attributes. 
A variable field is approximated using a 
kernel function that represents the 
influence of neighboring particles 
and the mechanics of the continuum 
are approximated by modeling the 
interactions between these particles. 
In this section, we transform the governing 
equations of two previously discussed models 
into SPH discretization.
 
\subsection{SPH discretization for solid dynamics}
To discretize the solid mechanics, we employ the initial 
undeformed configuration as the reference.  
First,  aiming to restore 1st order consistency,
a correction matrix $\mathbf{B}^0$  
\cite{vignjevic2006sph, randles1996smoothed} 
of particle $ a $ is adopted
as 
\begin{equation} \label{eq:sph-correctmatrix}
	\mathbf{B}^0_a = \left( \sum_b V_b \left( \mathbf{r}^0_b - \mathbf{r}^0_a \right) \otimes \nabla^0_a W_{ab} \right) ^{-1},
\end{equation}
where $V_b$ represents the volume of the 
neighboring particle $b$, 
$\mathbf{r}^0_a$ and $\mathbf{r}^0_b$ denote the positions 
of particles $a$ and $b$ in the reference configuration,
and $\nabla^0_a W_{ab}$ is the gradient of the kernel 
function given by
\begin{equation}\label{strongkernel}
	\nabla^0_a W_{ab} = \frac{\partial W\left( |\mathbf{r}^0_{ab}|, h \right)} {\partial |\mathbf{r}^0_{ab}|} \mathbf{e}^0_{ab},
\end{equation}
where $\mathbf{e}^0_{ab}$ is a unit vector pointing 
from particle $a$ to $b$. 
In total Lagrangian formulation,
the neighborhood of particle $a$ is 
defined in the initial configuration,  
and this set of neighboring particles remains 
fixed throughout the entire simulation.
Additionally, $\mathbf{B}^0_a $ is 
computed only once under the initial reference configuration. The momentum conservation in 
Eq. \eqref{eq:mechanical-mom} can be approximated
in the strong form  as 
\begin{equation}\label{eq:sph-mechanical-mom}
	\frac{\text{d}\mathbf{v}_a}{\text{d}t} = \frac{2}{\rho_a} \sum_b V^0_b \tilde{\mathbf{P}}_{ab} \nabla^0_a W_{ab} , 
\end{equation} 
where $\rho_a$ represents the density of particle $a$,
$\tilde{\mathbf{P}}_{ab}$ is the averaged first Piola-Kirchhoff stress of the particle pair $(a,b)$, 
stated as 
\begin{equation}
	\tilde{\mathbf{P}}_{ab} = \frac{1}{2} \left( \mathbf{P}_a \mathbf{B}^0_a + \mathbf{P}_b \mathbf{B}^0_b \right). 
\end{equation}
Note that the first Piola-Kirchhoff stress tensor is dependent on the deformation tensor $\mathbf{F}$, 
the time derivative of which is computed from 
\begin{equation}
	\label{rate_F}
	\frac{d\mathbf{F}_a }{dt} = \left( \sum_b V_b \left( \mathbf{v}_b - \mathbf{v}_a \right) \otimes \nabla^0_a W_{ab}  \right) \mathbf{B}^0_a ,
\end{equation}
where $\mathbf{v}_a$ and $\mathbf{v}_b$ denote the velocities 
of particles $a$ and $b$.
Considering the plastic response which may exist 
in the solid deformation, a return mapping algorithm 
is used to   obtain the stress-strain evolution.
\subsection{SPH discretization for fluid-structure interaction}
In the fluid-structure interaction model discretization, 
each particle carries the location 
$\mathbf{x}_n = \phi (\mathbf{X}, t_n)$  
at time $t_n$, along with an initial 
representative volume $V^0$ that partitions 
the initial domain of the macroscopic solid. 
The deformation 
gradient $\mathbf{F}_n$ of the solid phase 
is stored to update the solid current 
volume $V_n $ 
and  density $\rho^s_n$. 
Additionally, the fluid mass $m_n^l$, 
saturation $\widetilde{a}_n$, and  density-weighted 
velocity of the fluid relative to 
solid $\mathbf{q}_n$ are stored. 
The fluid mass equation 
Eq. \eqref{fluid_mass_der} of particle $ i $ 
is discretized as
\begin{equation} 
	\label{fluid_mass_sph}
	  \frac{\text{D} m_i^l}{\text{D} t} =  2 V_i \sum_j\frac{m_j}{\rho_j}(\mathbf{q}_{i} -\mathbf{q}_{j})\nabla_{i} W_{i j}.
\end{equation}
Note that with the equation \ref{eq:deformationtensor-displacement}, we have the relation of gradient kernel function in the total Lagrangian and updated Lagrangian  $ \nabla_{i} W_{i j} =\mathbf{F}^{-1} \nabla_{i}^0 W_{i j} $.
Once fluid mass is updated, the locally effective
fluid density $ \rho^l $ is obtained subsequently.
According to Eq. \eqref{fluid_density} and  Eq. \eqref{grad_q}, 
we update the fluid  
saturation $ \widetilde{a} $ and the fluid flux $ \mathbf{q} $ in 
the particle form 
\begin{equation} 
	\label{ralative_velocity_sph}
	\mathbf{q} =  -K\rho^l V_i \sum_j\frac{m_j}{\rho_j}(\widetilde{a}_{i} -\widetilde{a}_{j})\nabla_{i} W_{i j}.
\end{equation}
With the fluid flux and the stress in hand, we obtain discrete formulations for the momentum 
balance equation Eq. \eqref{totalmomentumupdate} as
\begin{equation}
	\label{totalmomentumupdatesph}
	\frac{D\mathbf{M}_i}{D t} = 2 \sum_j V_j (\mathbf{T}_i + \mathbf{T}_j) \nabla_i W_{ij}  - 2 \sum_j V_j ( {\mathbf{v}_i^{l} \otimes \mathbf{q}_i} + {\mathbf{v}_j^{l} \otimes \mathbf{q}_j}) \nabla_i W_{ij},
\end{equation}
where $\mathbf{T}_i$ and $\mathbf{T}_j$ are the stress 
tensors between particles $i$ and $j$.
We then compute the updated solid 
velocity $\mathbf{v}^s$ using the 
total momentum definition 
Eq. \eqref{P_equation}, where the total density 
of the mixture is the sum of the solid and 
fluid densities $ \rho = \rho^s+\rho^l $, written as  
\begin{equation}
	\label{solid_velocity_update}
	\mathbf{v}^{s} =\frac{\mathbf{M} -\mathbf{q}}{\rho} = \frac{\mathbf{M}-\mathbf{q}}{\rho^s+\rho^l }.
\end{equation}
Subsequently, the fluid velocity $\mathbf{v}^l$ is calculated  
using Eq. \eqref{defination_q} as 
\begin{equation}
	\label{fluid_velocity_update}
	\mathbf{v}^l =  \mathbf{v}^s-  \frac{\mathbf{q}}{\rho^l}.
\end{equation}  
\section{Multi-time step algorithm} 
 
In multi-time scale coupling involving solid dynamic problems, different time scales simultaneously exist.
A multi-time step algorithm using explicit scheme to match different time scale processes is introduced in this section.
In this paper,
the slow process, e.g., fluid diffusion is integrated with larger time step sizes, while the fast solid 
dynamics with smaller ones.
With small time step size, the solid dynamics evolves to a quasi-equilibrium state to update velocity, position and other solid information.
Further, in order to reduce the stress relaxation time
of solid dynamics, 
a damping scheme is applied to accelerate the equilibrium process.
For the following numerical simulations, 
stretch loading or fluid diffusion 
is performed 
with a larger time step size, while the 
dynamic stress relaxation coupled with 
a damping term  is executed 
with a smaller time step size. 
\subsection{Multi-time criteria} \label{multi-time}
Since the explicit integration 
operator is conditionally stable,  
a time step criterion  ${\Delta t_s} $ 
in solid simulation is required when using 
explicit scheme, stated as  
\begin{equation}\label{dts-advection}
	\Delta t_s   =  0.6 \min\left(\frac{h}{c_s + |\mathbf{v}_s|_{max}},
	\sqrt{\frac{h}{|\frac{\text{d}\mathbf{v}_s}{\text{d}t}|_{max}}} \right),
\end{equation}
where the artificial speed of sound of a solid structure  $c_{s} = \sqrt{K/\rho_s}$. 
In multi-time scale coupling problems, considering that the solid dynamic relaxation
process is comparatively fast,
$ \Delta t_s $ is usually limited under a small value. 
In comparison, the time step for internal diffusion  evolution or stretching  
is allowed to be  much larger.  
For the tensile test simulation,  
we divide the stretching process into $ N_S $ steps 
and the time 
step is   
\begin{equation}\label{dt-thermal_diffusion}
	\Delta t_{l}  = \frac{T_t}{N_S},
\end{equation}
where  $ T_t$ is the entire process time 
of the tensile test, $\Delta t_{l} $ accordingly the time step  for stretch loading.
Similarly, for the fluid-structure interaction,
according to the Fick's law, the maximum time 
step allowed for 
explicit time stepping is characterized as
\cite{cleary1999conduction}
\begin{equation}\label{dt-thermal_diffusion}
	\Delta t_{d}  = 0.5\frac{ h^2}{D},
\end{equation}
stating that the time step is mainly limited by the
diffusivity  constant $ D $ and the kernel smoothing length $ h $.  
To address the difference between these
time step sizes
of different time scale processes,
we present a multi-time step algorithm  to simulate these processes respectively
with a iterative scheme.
 
\subsection{Iterative scheme}
\begin{figure*}[htbp]
	\centering
	\includegraphics[trim = 0mm 0mm 0mm 0mm, clip,width=0.65\textwidth]{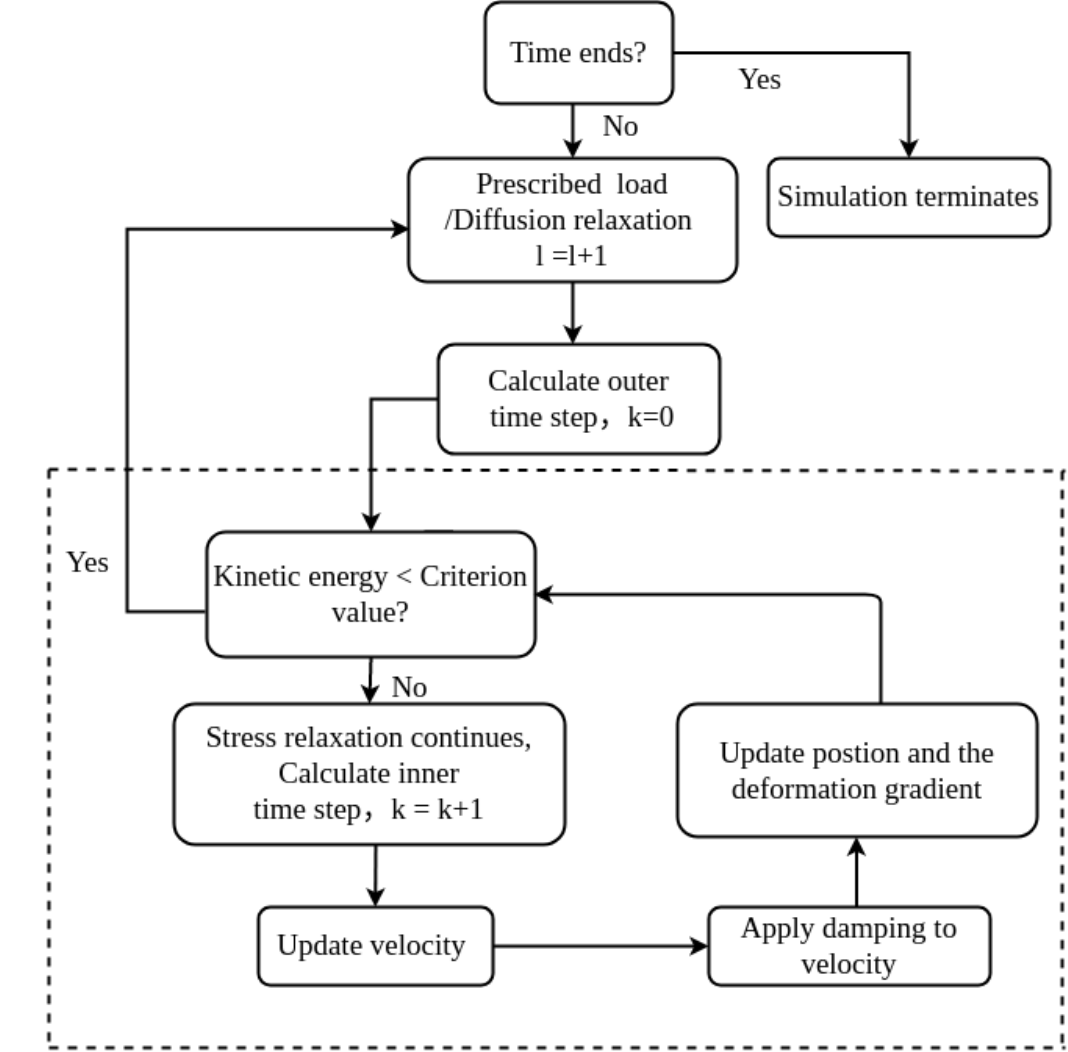}
	\caption{Flowchart of the iterative scheme in multi-time step algorithm.}
	\label{methodshceme}
\end{figure*}
Figure. \ref{methodshceme} shows the  iterative scheme 
of the proposed multi-time step algorithm  schematically.
It  can be seen this algorithm consists of two loops, 
where the outer loop  indicates that 
the entire dynamic progress is controlled by the prescribed  displacements 
or diffusion relaxation, 
which are executed incrementally with a subscript $ l $ denoting each increment. The inner loop 
describes the solid dynamics evolution 
with a subscript $ k $ signifying each
stress relaxation step.
The loading or diffusion criterion   
$ \Delta t_{l} $ or $\Delta t_d$ 
controls the external force exerting or the fluid diffusion process and
${\Delta t_s} $ determines the frequency of solid stress relaxation.
However, within one external loading 
time step $ \Delta t_{l} $ 
or diffusion time step   $\Delta t_d$, 
the time integration of structure 
should be computed as 
$ k_0 = [\frac{{\Delta t_{l/d}}}{\Delta t_s}]+ 1$ times.
With a limited ${\Delta t_s} $ and much larger $ \Delta t_{l} $  and $\Delta t_d$, 
$ k_0 $ is supposed to be very large and  
the computation of solid dynamics 
will  be trapped into
a meaningless iteration, 
increasing the unnecessary computation time. 

Since once solid dynamics achieves the static state, the inner loop can be finished to begin another outer loop.
Therefore, in order to save computation time, the inner loop is executed with a damping term to dissipate  the kinetic energy  and accelerate the relaxation of the transient response.
Solid governing equations with extra damping 
can be solved 
a small number of times $ k $
until the kinetic energy is reduced to
a sufficient small value $ E_k $.
Specific criteria values of the kinetic energy 
are given in different cases.
After the equilibrium state of the solid deformation   is  achieved in the inner steps,   
a new outer step begins and this procedure is performed 
once again until the physical computation time ends.

\subsection{Damping scheme}
As we mentioned before, 
obtaining equilibrium for a
dynamic system is  excessively time-consuming 
in SPH method with explicit time-stepping.
To address this issue, we apply 
a damping term into the stress relaxation
to dissipate the extra kinetic 
energy inside the system and
accelerate the convergence of 
stress relaxation process.
Following  Zhu et al. work \cite{zhu2022dynamic},
a viscous damping term $\mathbf{f}^{v} $ is added in the   solid momentum equation as
\begin{equation} \label{solid_deformation}
	\frac{d\mathbf{v}}{dt} =\mathbf{f}^s  +\mathbf{g} +  \mathbf{f}^{v},
\end{equation}
Where $ \mathbf{f}^s $  and $ \mathbf{g} $ represents the surface and body forces, 
the added damping term $ \mathbf{f}^{v} $ can be 
discreized in the total Lagrangian form as
\begin{equation}
	\label{damping-sph}
	\mathbf{f}^{v}_a =\frac{\eta}{\rho_a}  \nabla_a^2 \mathbf{v} = \frac{2 \eta}{m_a} \sum_{a} V_aV_b \mathbf{v}_{ab} \nabla_a^0 W_{ab},
\end{equation}
where $ \eta $ is the dynamic viscosity, given separately in different cases, and usually it depends on   the characteristic length scale of the problem and materials parameters.
$ \mathbf{v}_{ab} = \mathbf{v}_{a}-\mathbf{v}_{b} $ denotes 
the velocity difference between a particle pair 
$ (a,b) $. 
This viscous force can 
deduce  the  system oscillation  
caused by large velocity gradient and eliminates 
the extra kinetic energy. 
Therefore,
the solid stress is relaxed much 
faster to a equilibrium state where the kinetic 
energy decreases below a criterion value. 
Also, a pairwise splitting scheme is 
adopted to update
the velocity implicitly and locally,
keeping the conservation of momentum in 
each particle pair.
More detailed information can be referred to Zhu's work \cite{zhu2022dynamic}. 

\section{Numerical examples}
In this section,  several  tests including 
the stretching-necking and the fluid diffusion
coupled solid deformation in two  and
three dimensions, are simulated using 
the present method to show its accuracy and efficiency.
\subsection{Necking  of a two-dimensional bar}  
\begin{figure}[htp!]
	\centering
	\includegraphics[trim = 0mm 0mm 0mm 0mm, clip,width=0.55\textwidth]{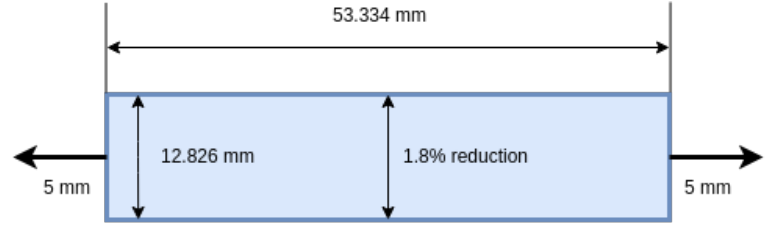}
	\caption {2D tensile necking: geometry and initial and boundary condition setup.}
	\label{2d-necking_setup}
\end{figure}

The standard tensile necking test simulation has been  previously studied in several papers \cite{simo2006computational, neto2005f, de2006orthogonal,elguedj2014isogeometric} with experimental and  numerical results to compare. 
\begin{table}[htb]
	\centering
	\caption{Necking test simulation: physical material parameters.}
			\renewcommand{\arraystretch}{1.1} 
			\begin{tabular}{cc} 
			\hline		
			Parameters & Value  \\  
				\hline	
			Shear modulus   & 80.1938 Gpa\\  
			Bulk modulus  & 164.21  Gpa  \\ 
			Initial flow stress 	&450 MPa \\ 
			Saturation flow stress	&715 MPa\\ 
			Saturation exponent 	& 16.93 \\   
			Linear hardening coefficient & 129.24 MPa\\ 
			\hline	
	\end{tabular}
	\label{stretching-parameter-table}
\end{table}
 With a length of 53.334 mm and a width of 12.826 mm, 
 the test sample  is stretched from the surface 
 under an increasing (uniaxial) load.
 A reduction in the width and thickness happens consistent with the elongation of this specimen.
 A slight imperfection of this sample (1.8\% reduction) 
  is imposed initially in the center part  as shown in Figure \ref{2d-necking_setup}
to trigger the necking phenomenon.
The specimen  is composed of a elastic  deformation  
depicted by the  Neo-Hookean law and a plastic response by the nonlinear isotropic hardening law.
 The material parameters are given in Table \ref{stretching-parameter-table}.
A total stretching of 10 mm is realized via a 
symmetric displacement  boundary conditions. 
 Here, $dp  = PH/50$=0.25652 mm.
Three layers of particle are imposed with 
the aforementioned boundary condition.
Consistent with the experimental time around 2 minutes,
the physical time in this simulation  is set to 
$ t= 100 $s,  with  stretching steps  $  N_S =$ 10000 the
 corresponding velocity  $v =0.5\times 10^{-4}$ m/s. 
This is different from  that 
 in reference papers where the velocity usually 
is increased to about 1 m/s  to reduce 
the physical time to $1.5 \times 10^{-3}$ s.
After each step of stretch loading, 
stress relaxation coupled with damping is performed. The damping ratio is set to an experienced value
 of $\eta = 1.0 \times 10^{4}$ based on the work of Zhu \cite{zhu2022dynamic}. 

\begin{figure*}[htbp]
	\centering
	\begin{subfigure}[b]{0.95\textwidth}
		\includegraphics[trim = 1mm 2mm 1mm 5mm, clip,width=0.99\textwidth]{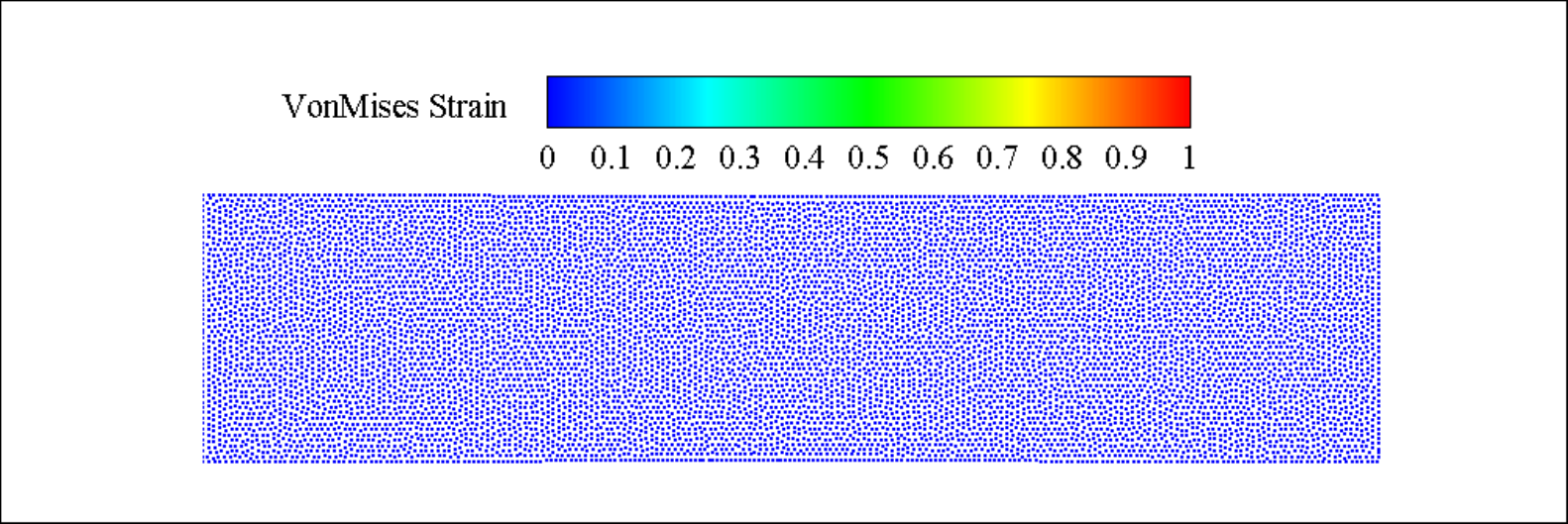}
		\caption {0s}
		\label{2dstrain-contour0}
	\end{subfigure}
	\begin{subfigure}[b]{0.95\textwidth}
		\includegraphics[trim =1mm 1mm 1mm 5mm, clip,width=0.99\textwidth]{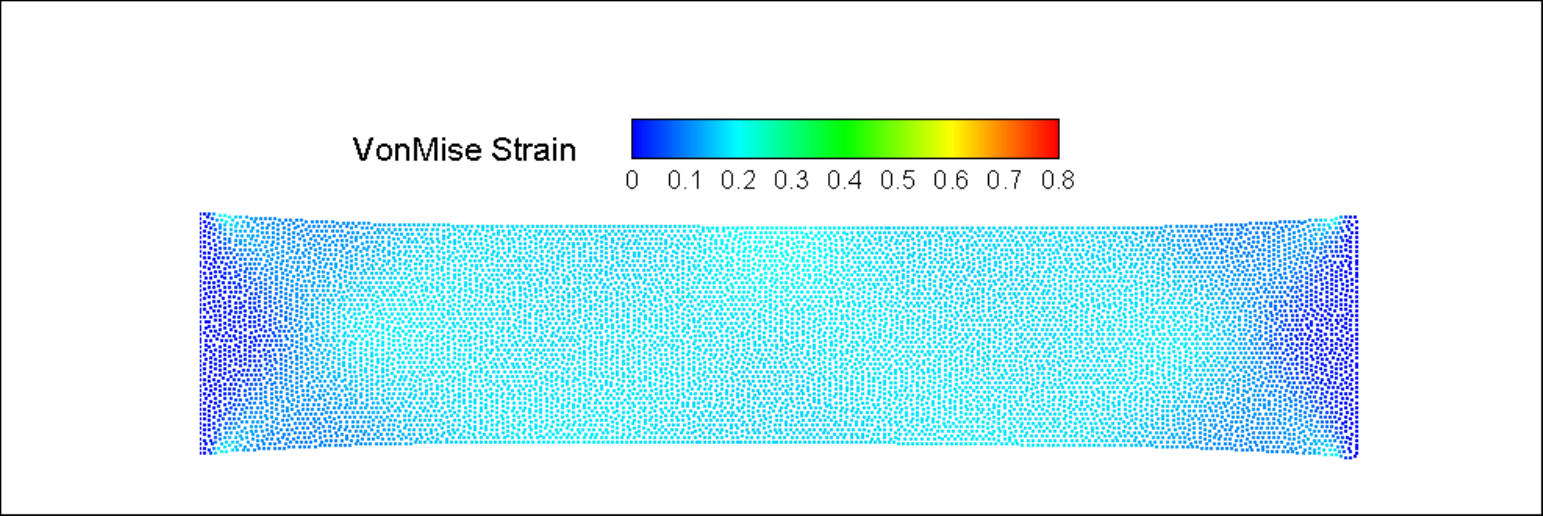}
		\caption {50s}
		\label{2dstrain-contour50}
	\end{subfigure}
	\begin{subfigure}[b]{0.95\textwidth}
		\includegraphics[trim = 1mm 2mm 1mm 5mm, clip,width=0.99\textwidth]{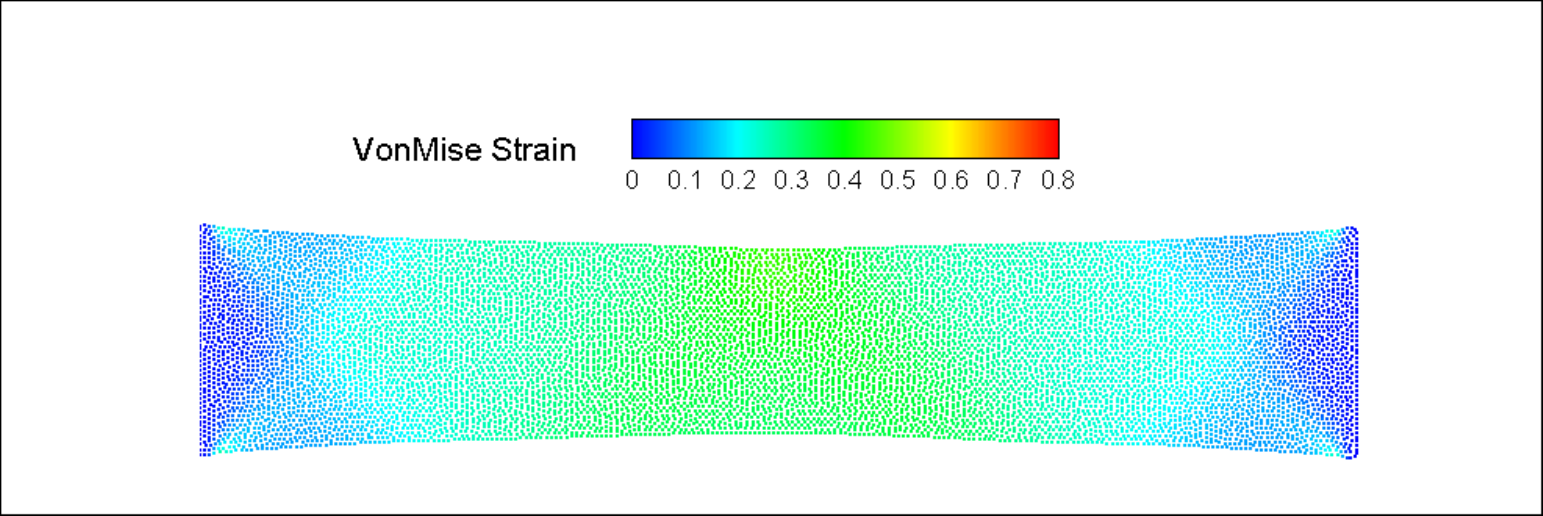}
		\caption {80s}
		\label{2dstrain-contour80}
	\end{subfigure}
	\begin{subfigure}[b]{0.95\textwidth}
		\includegraphics[trim =1mm 2mm 1mm 5mm, clip,width=0.99\textwidth]{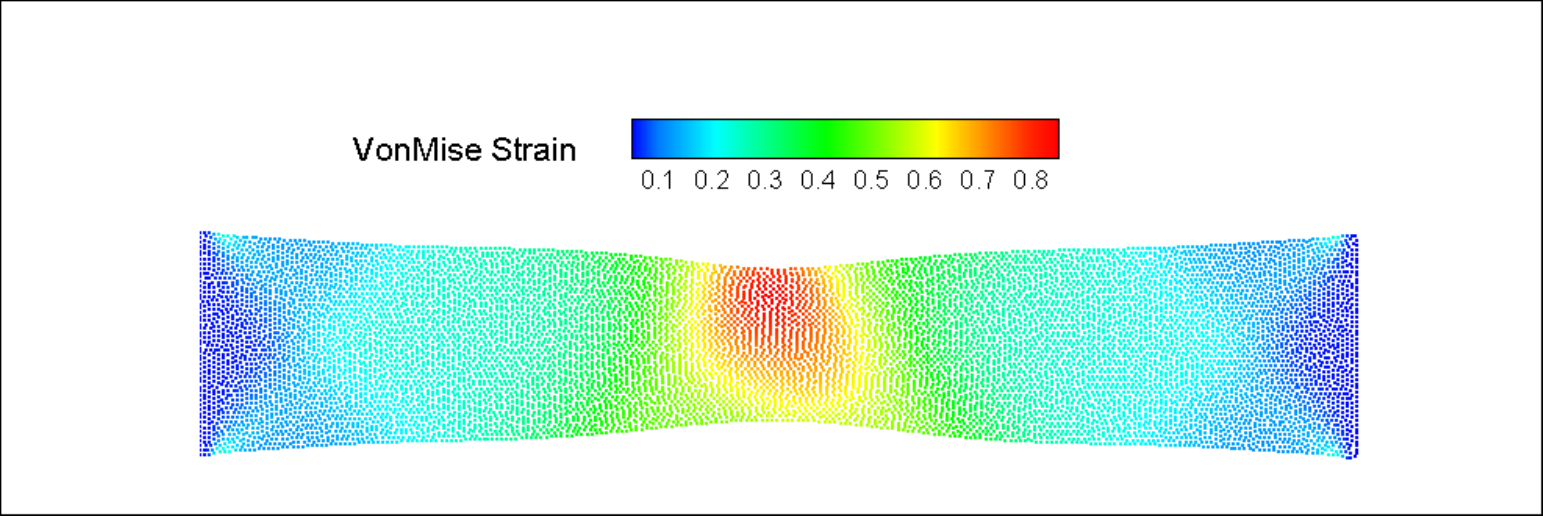}
		\caption {100s}
		\label{2dstrain-contour100}
	\end{subfigure}
	\caption{2D tensile necking: the deformation colored by von Mise Strain at different time instants. }
	\label{stretching-stress}
\end{figure*}

\begin{figure*}[htbp]
	\centering
	\includegraphics[trim = 2mm 2mm 2mm 2mm, clip,width=0.75\textwidth]{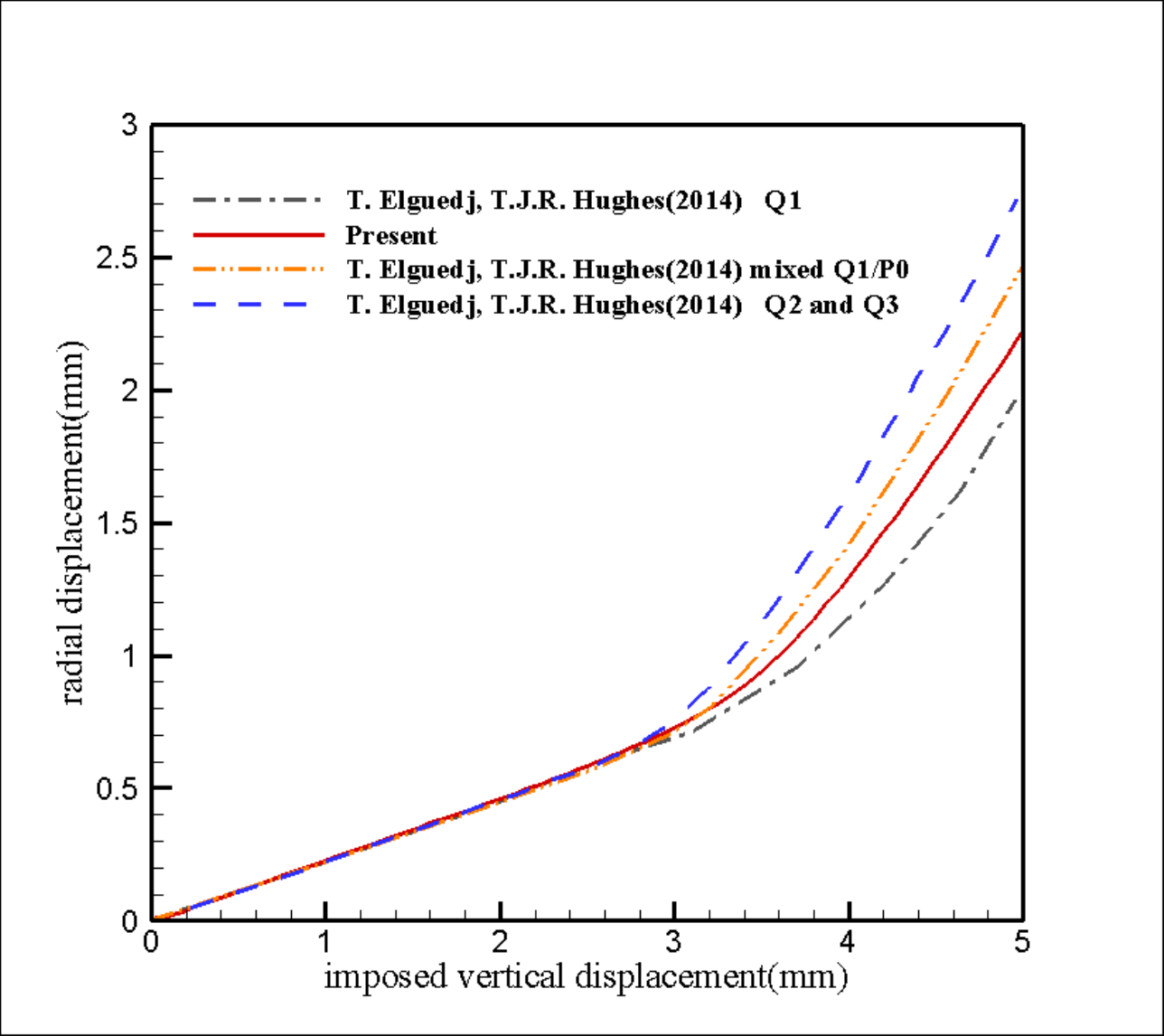}
	\caption {2D tensile necking: the evolution of the radial displacement as a function of the imposed vertical displacement of the central part.   }
	\label{2d-radial-displacement}
\end{figure*}

\begin{figure*} 
 \centering
	\includegraphics[trim = 2mm 2mm 2mm 2mm, clip,width=0.55\textwidth]{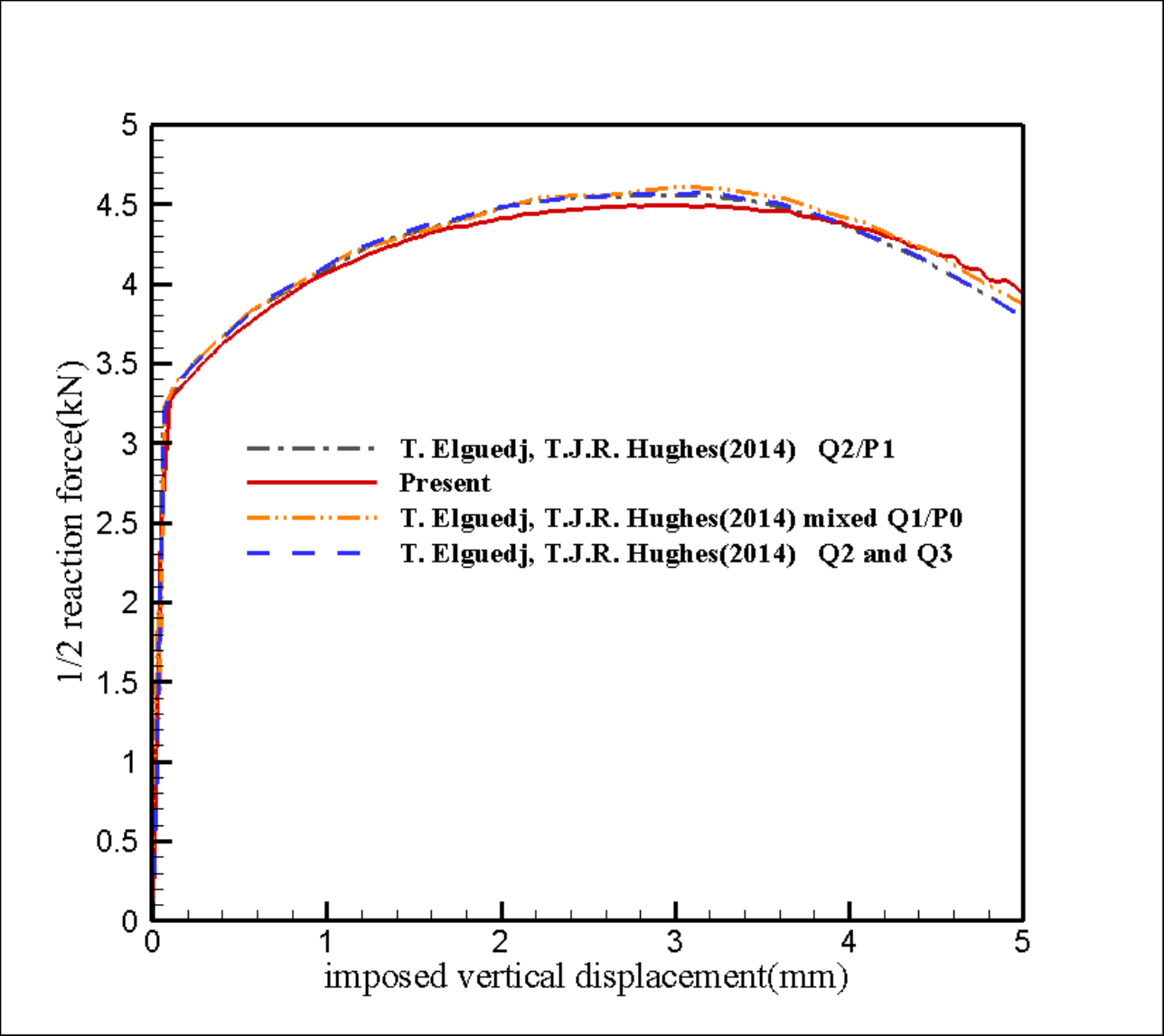}
	\caption {2D tensile necking: the evolution of 
		the reaction force versus the imposed vertical displacement. }
	\label{2dreaction-force} 
\end{figure*}

Figure \ref{stretching-stress} shows the deformation evolution 
colored by von Mise Strain at different time instants.  
A clear necking pattern is observed in the center of the specimen, 
which is consistent with that observed in both experimental and other numerical works \cite{neto2005f,elguedj2014isogeometric,de2006orthogonal}.
The specimen undergoes three distinct stages: elastic strain, followed by uniform plastic strain, and finally necking strain.
Figure \ref{2d-radial-displacement} plots the radius evolution of the central part 
where necking occurs as a function of
 the imposed stretching displacement.
 It is compared with the  results from 
 the reference Elguedj and Hughes \cite{elguedj2014isogeometric} where different 
 mesh discretization and element types Q1, 
 mixed Q1/P0, etc. are used to model this test.
 As time progresses and the sample elongates, 
 the radius displacement of the central 
 part increases linearly, 
 while after necking occurs, 
 it experiences a rapid increase.
Figure \ref{2dreaction-force} depicts the evolution of the reaction force as time progresses. 
After a short elastic response, 
represented by the initial straight line, 
the specimen enters the stage of uniform plastic deformation with a smooth increase of reaction force.
During this stage, plastic deformation spreads slowly
and shows a homogeneous state throughout the specimen.
Eventually, when the boundary displacement
reaches a certain value, necking occurs 
in the central part, 
and the reaction force reaches its peak value. Subsequently, the deformation changes to a mode
where the plastic effect is concentrated in the central zone,
resulting in a decreasing reaction force, 
which is more obvious 
in the following three dimensional case.

\begin{figure*}[htbp]
	\centering
	\begin{subfigure}[b]{0.45\textwidth}
	\includegraphics[trim = 1mm 2mm 1mm 2mm, clip,width=0.9\textwidth]{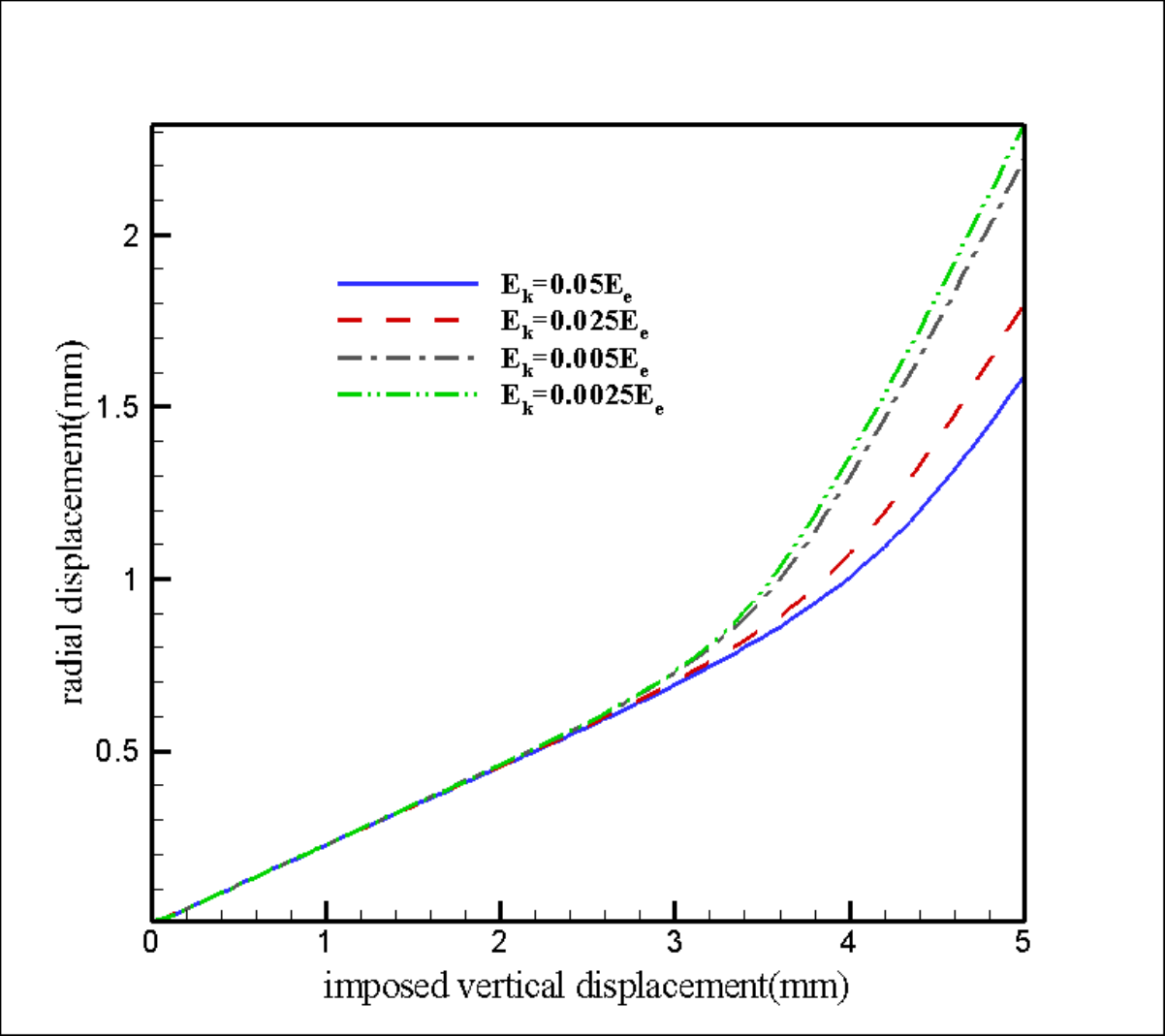}
	\caption {}
	\label{2d-streching-pos_convergence0}
\end{subfigure}
\begin{subfigure}[b]{0.45\textwidth}
	\includegraphics[trim =1mm 2mm 1mm 2mm, clip,width=0.9\textwidth]{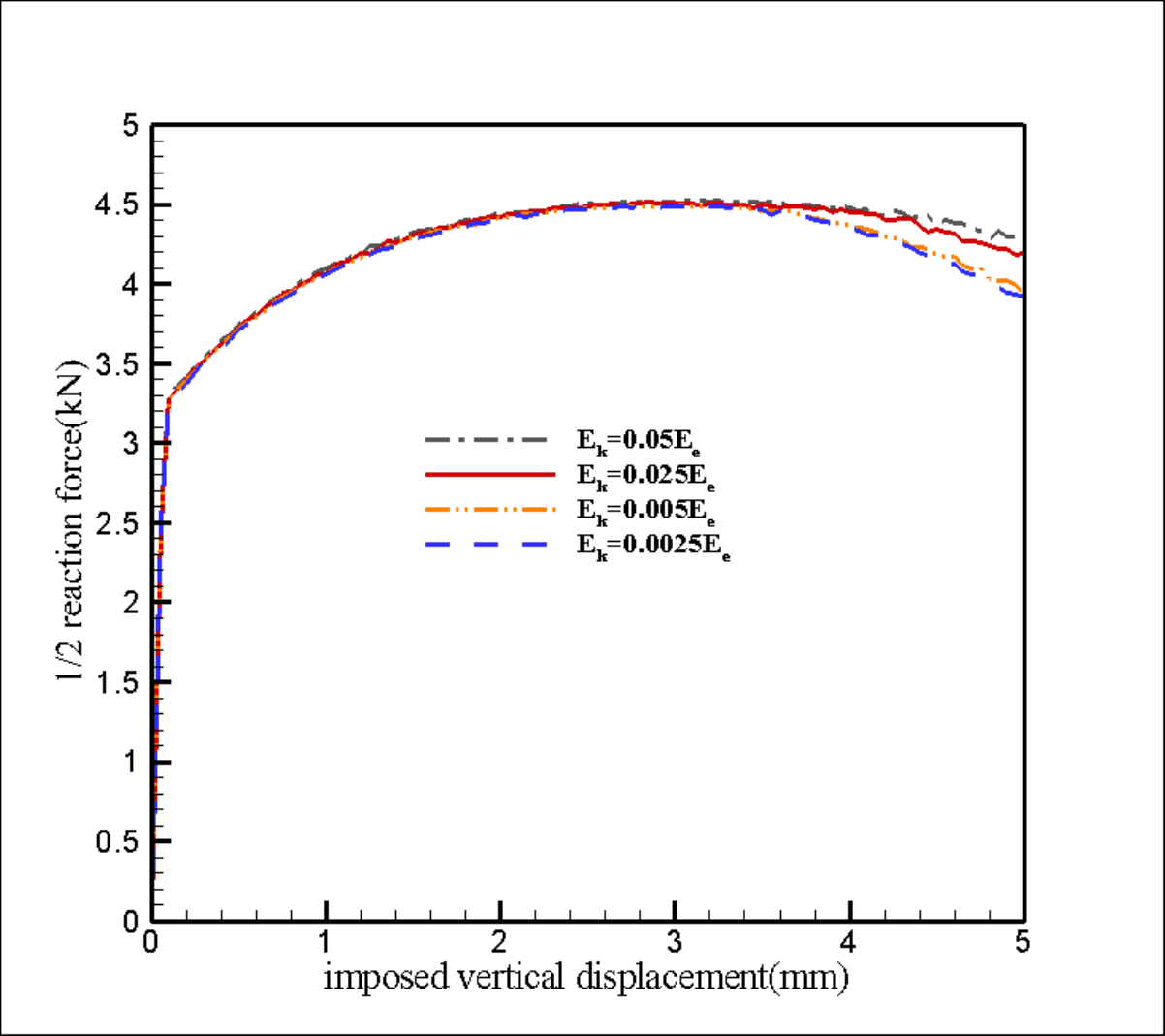}
	\caption {}
	\label{2d-streching-force_convergence}
\end{subfigure}
	\caption {2D tensile necking: radius displacement (a) and the loading force (b) convergence with different  kinetic energy criteria.}
	\label{2d-streching-convergence}
\end{figure*}

To determine when equilibrium is achieved,
we monitor the kinetic energy $E_k$ until
it is damped below a threshold value 
derived from the elastic energy $E_e$.
Here, $E_e$ is calculated using the formula $E_e = \frac{1}{2} F \Delta x$, 
where $F$ is the load force of $8000$ N
deduced from Figure	\ref{2dreaction-force}, 
and $\Delta x$ is the stretching length of $10$ mm. 
To investigate the effect of the kinetic energy threshold on the simulation results,
 we conducted a series of stretching simulations with varying criteria.  
Figure \ref{2d-streching-convergence} plots the
variation of the radius displacement and reaction force
for different kinetic energy criteria. 
Initially, we chose a larger criterion value of  $E_k =$  5\%$E_e$ 
and gradually decreased the criterion. 
The results reveal that when $E_k$ is set to be 5\%$E_e$, either the radius displacement and loading force evolution is not smooth enough, 
indicating that equilibrium is not achieved. 
This suggests that 5\%$E_e$ is too large as a criterion value. 
On the other hand, with too small criterion values, unnecessary calculation steps are performed, 
increasing computation time.
The results demonstrate that for this 2D case, 
the appropriate kinetic energy criterion value 
 is 0.5\%$E_e$. 

\begin{figure*}[htbp]
	\centering
	\includegraphics[trim = 2mm 2mm 2mm 2mm, clip,width=0.55\textwidth]{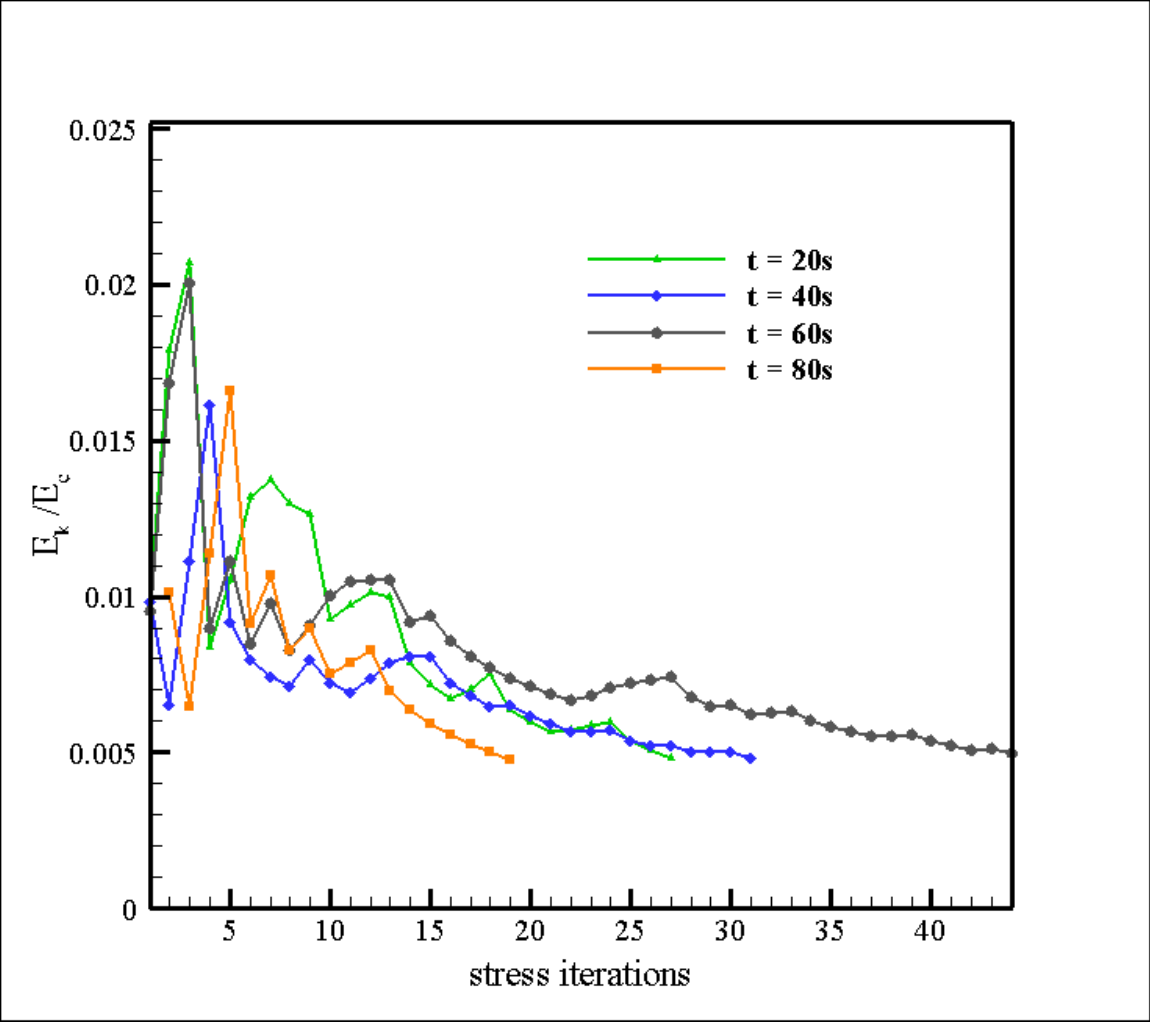}
	\caption {2D tensile necking: evolution of  kinetic energy  evaluated by the elastic energy after one stretching at different time. }
	\label{2d-streching-energy}
\end{figure*}

During the simulation, 
the evolution of the kinetic energy after one stretching at four different time instants,
as evaluated by the elastic energy $E_e$, 
is shown in Figure \ref{2d-streching-energy}. 
As expected, due to the stretching force,
there is a kinetic energy 
fluctuation. 
After each stretching event, 
the kinetic energy first increases, 
followed by a decrease to a certain criterion value of 0.5\%$E_e$, 
which is due to the damping effects. 
Throughout the simulation process, 
stress relaxation occurs with  viscous damping
immediately after each stretching. 
The relative kinetic energy at the end of each stretching step 
approaches 0.5\%$E_e$, showing that
the equilibrium is achieved.
 
With a  physical time in simulation $ t=100$s, 
due to the time step size limitation in explicit scheme,
the performed stretching times $ N_S $ and stress relaxation times $ N_s $
are supposed to be  $ N_S = N_s= t/	\Delta t_s  = 2.58\times 10^{9} $.
With this multi-time criteria algorithm, 
we firstly decrease the number of stretching time steps from $2.58 \times 10^9$ to $N_S=1.0\times 10^4$. 
Secondly, we decrease the stress relaxation times 
from $2.58\times 10^9$ to $N_s=3.26\times 10^5$ by coupling the damping term to accelerate the equilibrium obtaining.
Table \ref{2D-stretching-efficiency} lists the stress relaxation iterations performed in straightforward and multi-time step algorithms  respectively
and gives the quantitative efficiency 
of the present algorithm compared against the straightforward one 
in terms of stretching $N_S$ and stress relaxation iterations $N_s$ with the same total particle number $ N_p $.
It is obvious that the proposed algorithm yields a drastic reduction in computation time.
%%%%%%%%%%%%%%%%%%%%%%%%%%%%%%%%%%%%%%%%%%%%%%%%
\begin{table}[htb!]
	\centering
	\caption{2D tensile necking: quantitative validation of the efficiency of this multi-time step algorithm.}
		\renewcommand{\arraystretch}{1.1} 
	\begin{tabular}{ccccc}
		\hline
		algorithm   & $ N_p $  & $ N_S $& $ N_s $ & $N_{damping}$      \\ 	
		\hline
		straightforward algorithm  & 10788 & $  2.58e^{9} $ &  $ 2.58e^{9} $  &  -  \\
		multi-time step algorithm	&  10788  & $ 1.0e^{4} $ & $ 3.26e^{5} $ & $ 3.26e^{5} $  \\
		\hline	
	\end{tabular}
	\label{2D-stretching-efficiency}
\end{table}
%%%%%%%%%%%%%%%%%%%%%%%%%%%%%%%%%%%%%%%%%%%%%%%%
\subsection{Necking of a three-dimensional bar} 
Further, a  three-dimensional necking analysis of a cylindrical bar is carried out,
which has been studied by Simo and Armero \cite{simo1992geometrically, simo2006computational}, de Souza Neto et al. \cite{ neto2005f},   Elguedj  and Hughes \cite{elguedj2014isogeometric}.
The same geometry of radius 6.413 mm and length 53.334 mm with a
slight  reduction (1.8\%) in the center of the bar
as in the previous 2D case is considered.
Loading is imposed using displacement control, 
with a total vertical displacement of 
7 mm applied on both the top 
and bottom surface of the bar. 
The same  material properties in Table \ref{stretching-parameter-table} 
and elastic-plastic response  
as that applied in previous two-dimensional 
case are employed herein. 
In this work, initial particle spacing $dp$ = 0.3 mm
with a total particle number almost  $ N_p = 2.5e^5 $.
With  physical  time $ t= 100$s and stretching steps $N_S = 10000 $,  the
corresponding velocity  is $0.7\times 10^{-4}$  m/s, 
which allows problem  to be simulated in 
a real stretching rate. 
The damping ratio used here is $\eta = 1.0 \times 10^{4} $.
%which is necessary for problems concerning rate-dependent materials.
%For example, high deformation speeds in experiments could lead to an unacceptable heating of the sample  and thus also falsify the result.
%Also, in order not to distort the result, the deformation speeds and other details are specified in the corresponding standards. 
 %For steels, for example, the increase in stress must not exceed 30 $ N/mm^2$ per second.  }
 \begin{figure*}[htbp]
 	\centering
 	\begin{subfigure}[b]{0.48\textwidth}
 		\includegraphics[trim = 1mm 2mm 1mm 5mm, clip,width=0.95\textwidth]{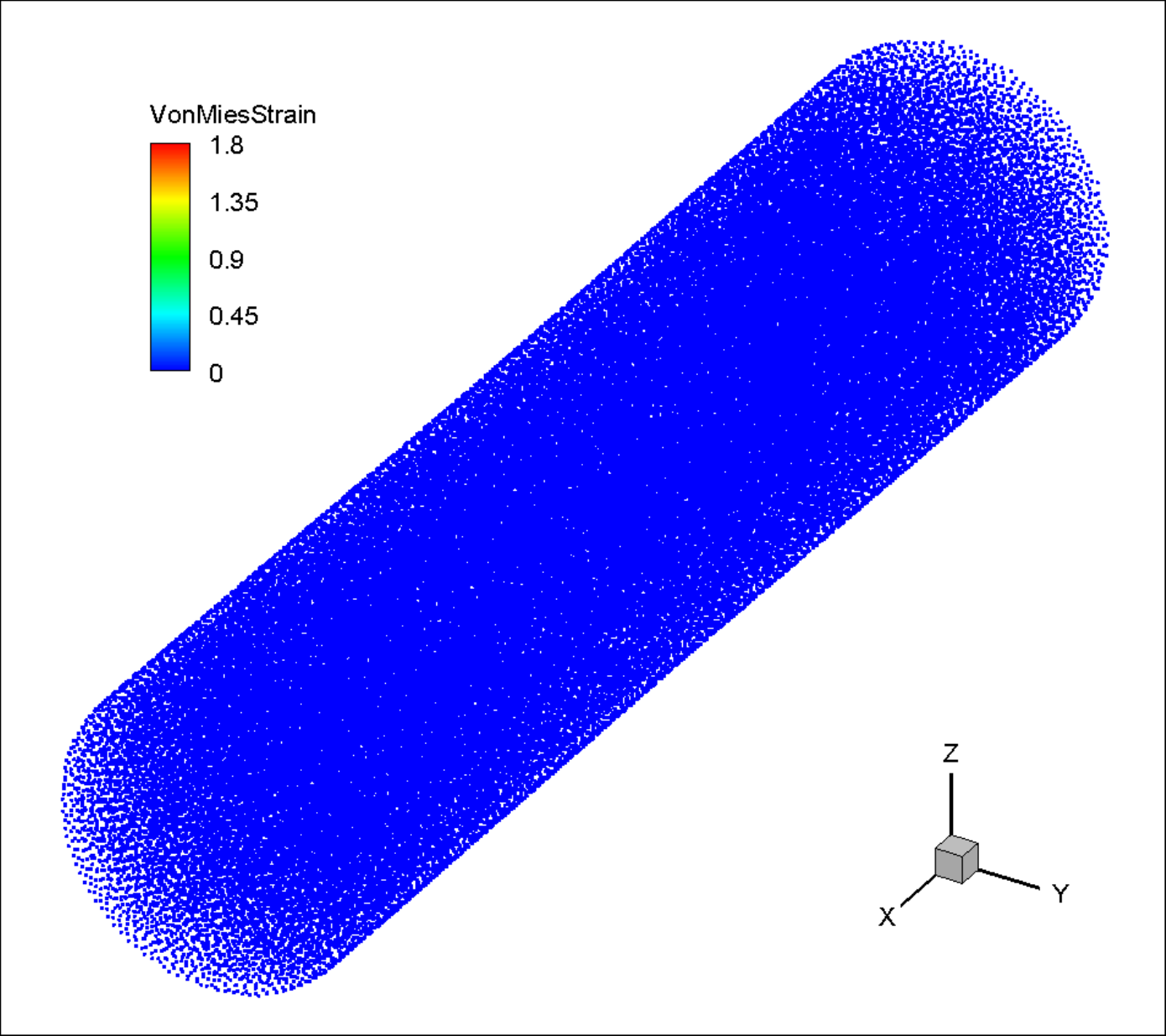}
 		\caption {0s}
 		\label{3dstrain-contour0whole}
 	\end{subfigure}
 	\begin{subfigure}[b]{0.48\textwidth}
 		\includegraphics[trim =1mm 1mm 1mm 5mm, clip,width=0.95\textwidth]{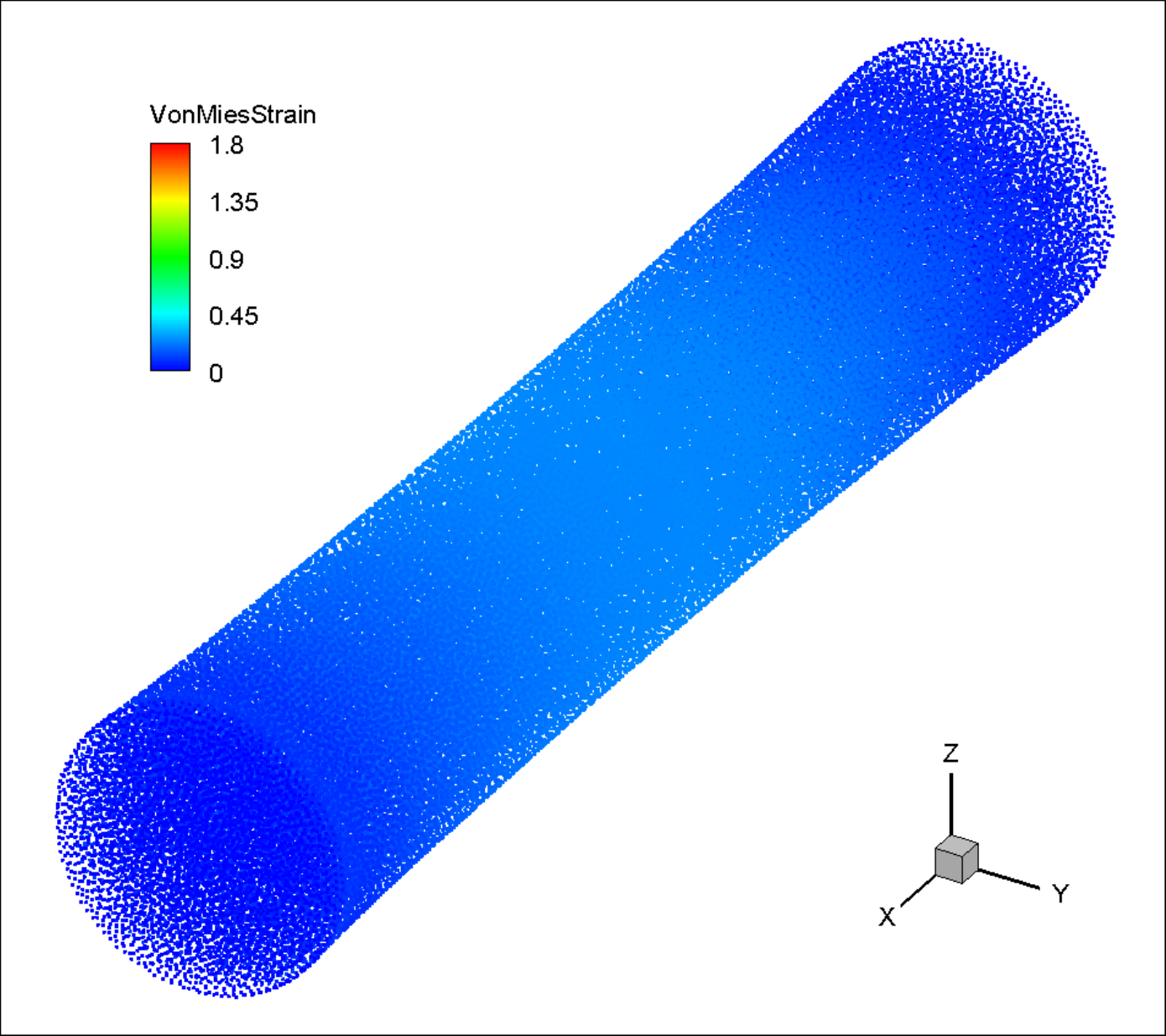}
 		\caption {50s}
 		\label{3dstrain-contour50whole}
 	\end{subfigure}
 	\begin{subfigure}[b]{0.48\textwidth}
 		\includegraphics[trim = 1mm 2mm 1mm 5mm, clip,width=0.95\textwidth]{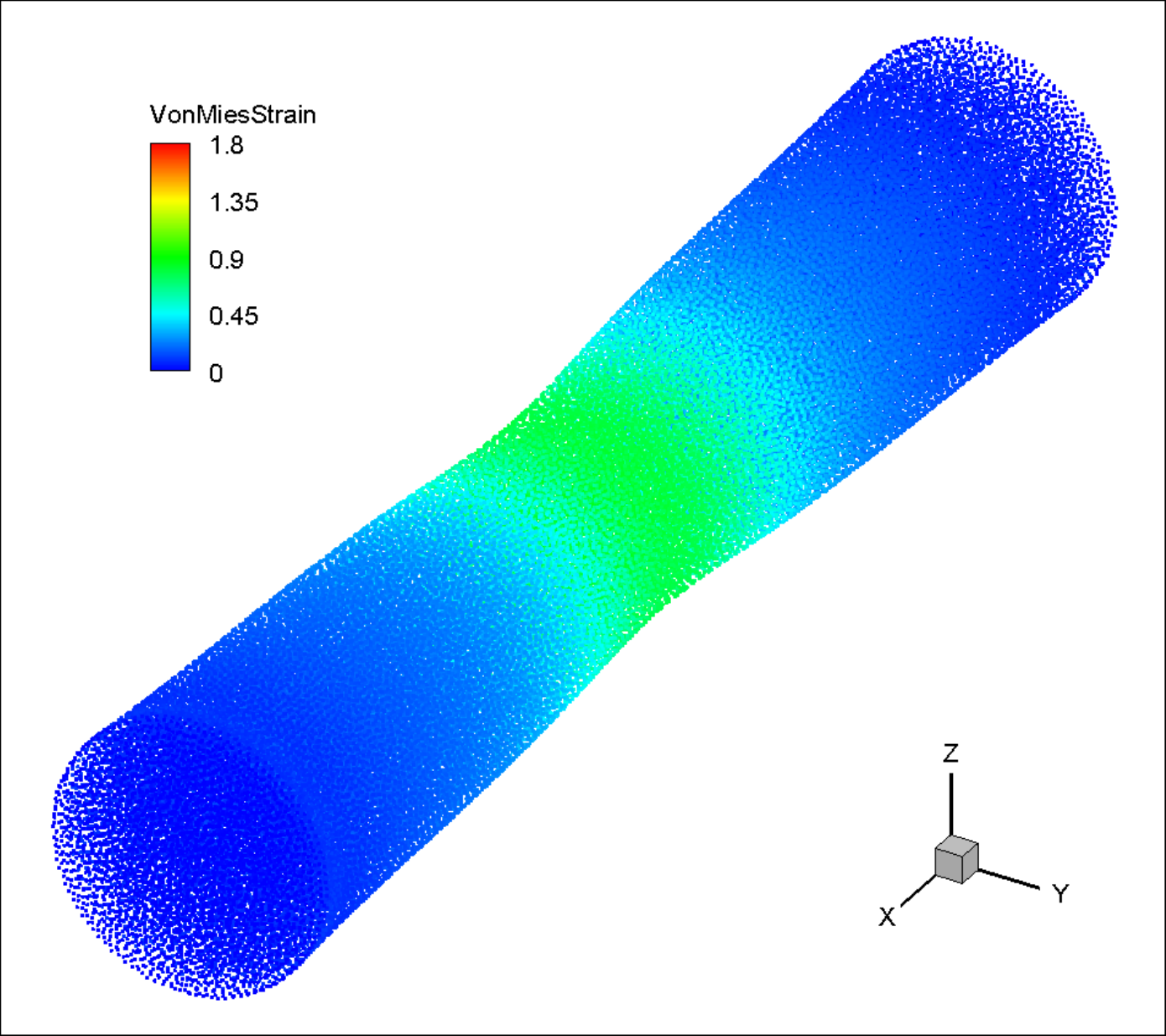}
 		\caption {80s}
 		\label{3dstrain-contour80whole}
 	\end{subfigure}
 	\begin{subfigure}[b]{0.48\textwidth}
 		\includegraphics[trim =1mm 2mm 1mm 5mm, clip,width=0.95\textwidth]{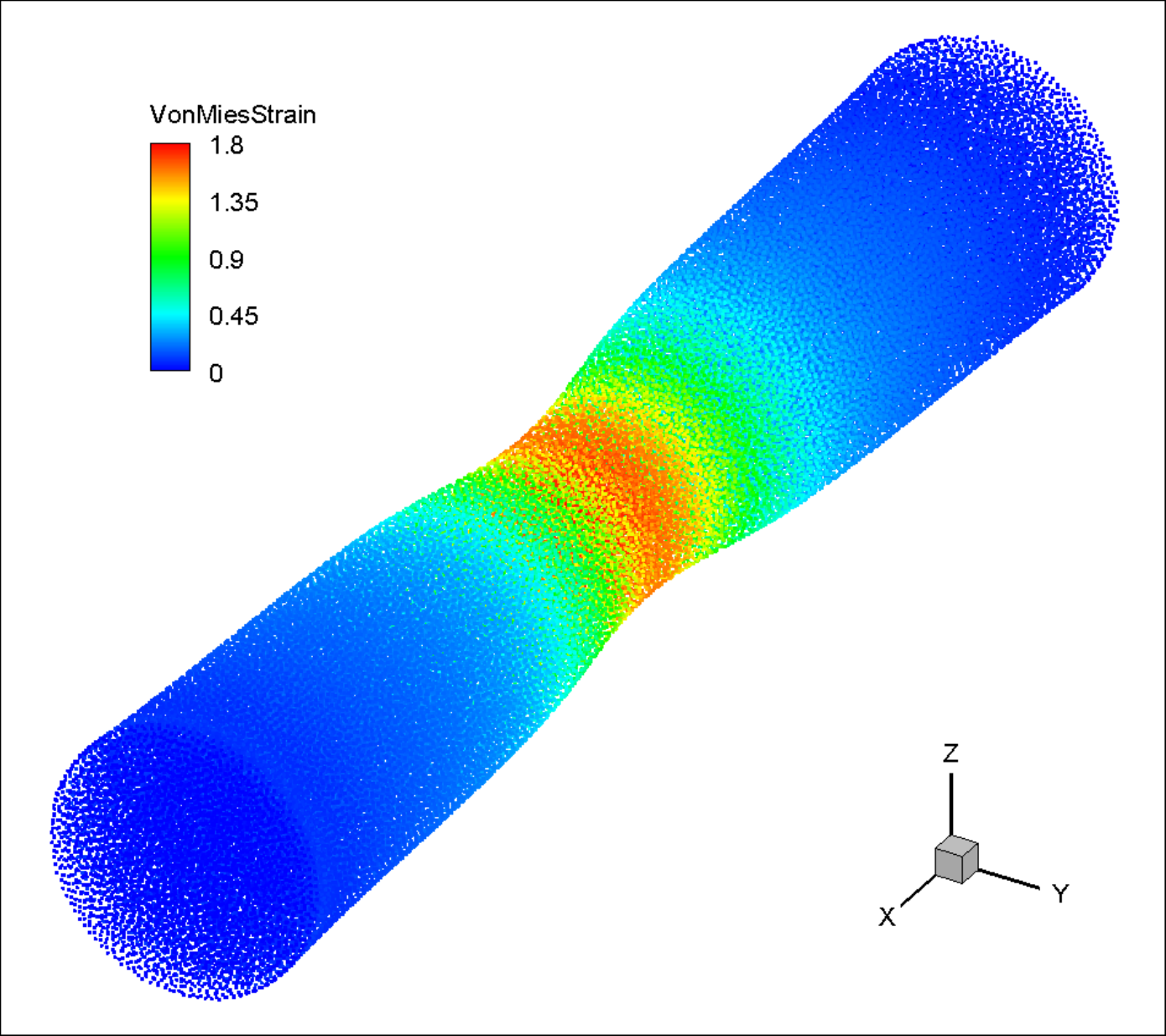}
 		\caption {100s}
 		\label{3dstrain-contour100whole}
 	\end{subfigure}
 	\caption{3D tensile necking: the deformation colored by von Mise Strain at different time instants (top side view). }
 	\label{3d-stretching-stresswhole}
 \end{figure*}

\begin{figure*}[htbp]
	\centering
	\begin{subfigure}[b]{0.45\textwidth}
		\includegraphics[trim = 1mm 2mm 1mm 5mm, clip,width=0.95\textwidth]{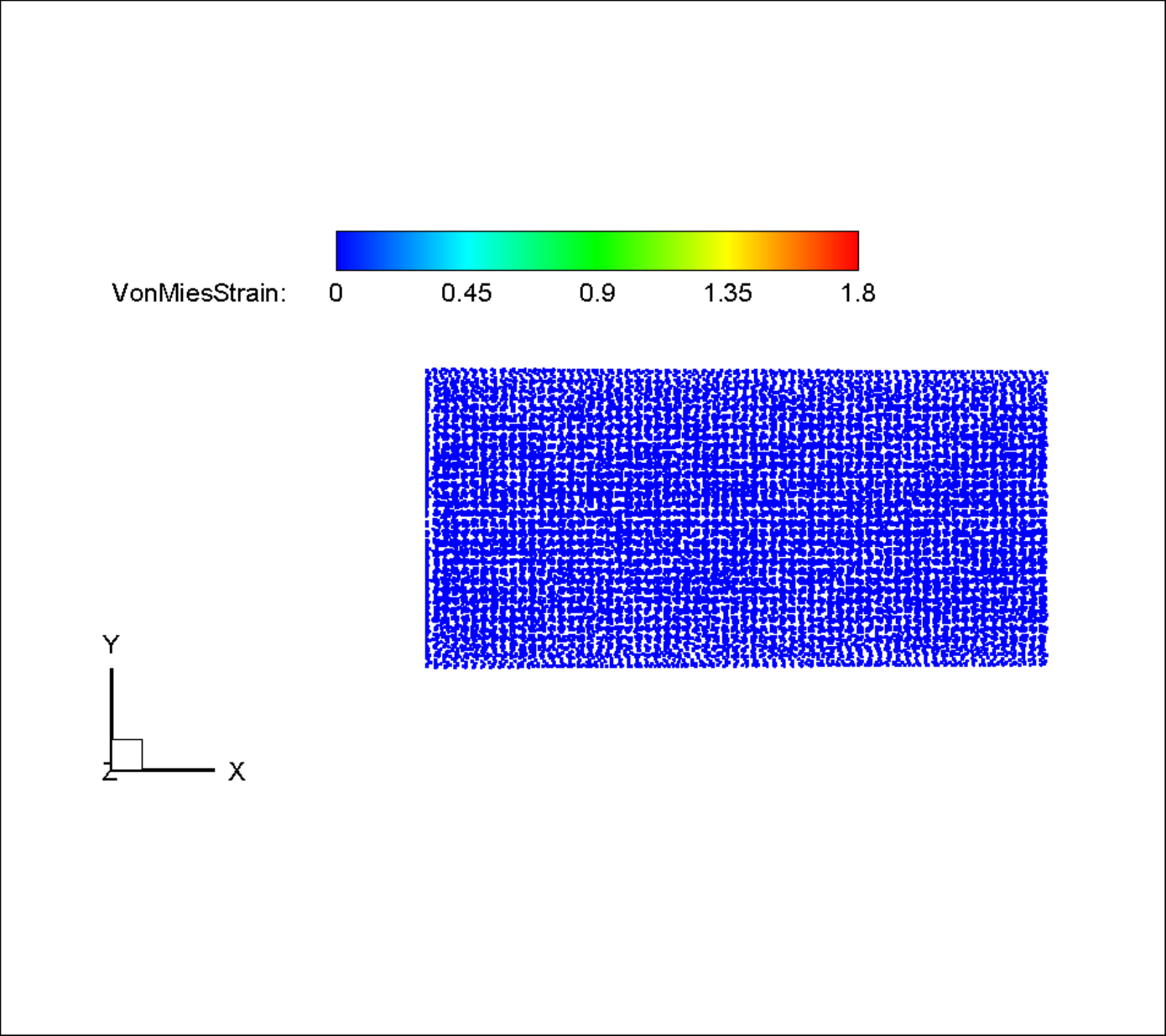}
		\caption {0s}
		\label{3dstrain-contour0}
	\end{subfigure}
	\begin{subfigure}[b]{0.45\textwidth}
		\includegraphics[trim =1mm 1mm 1mm 5mm, clip,width=0.99\textwidth]{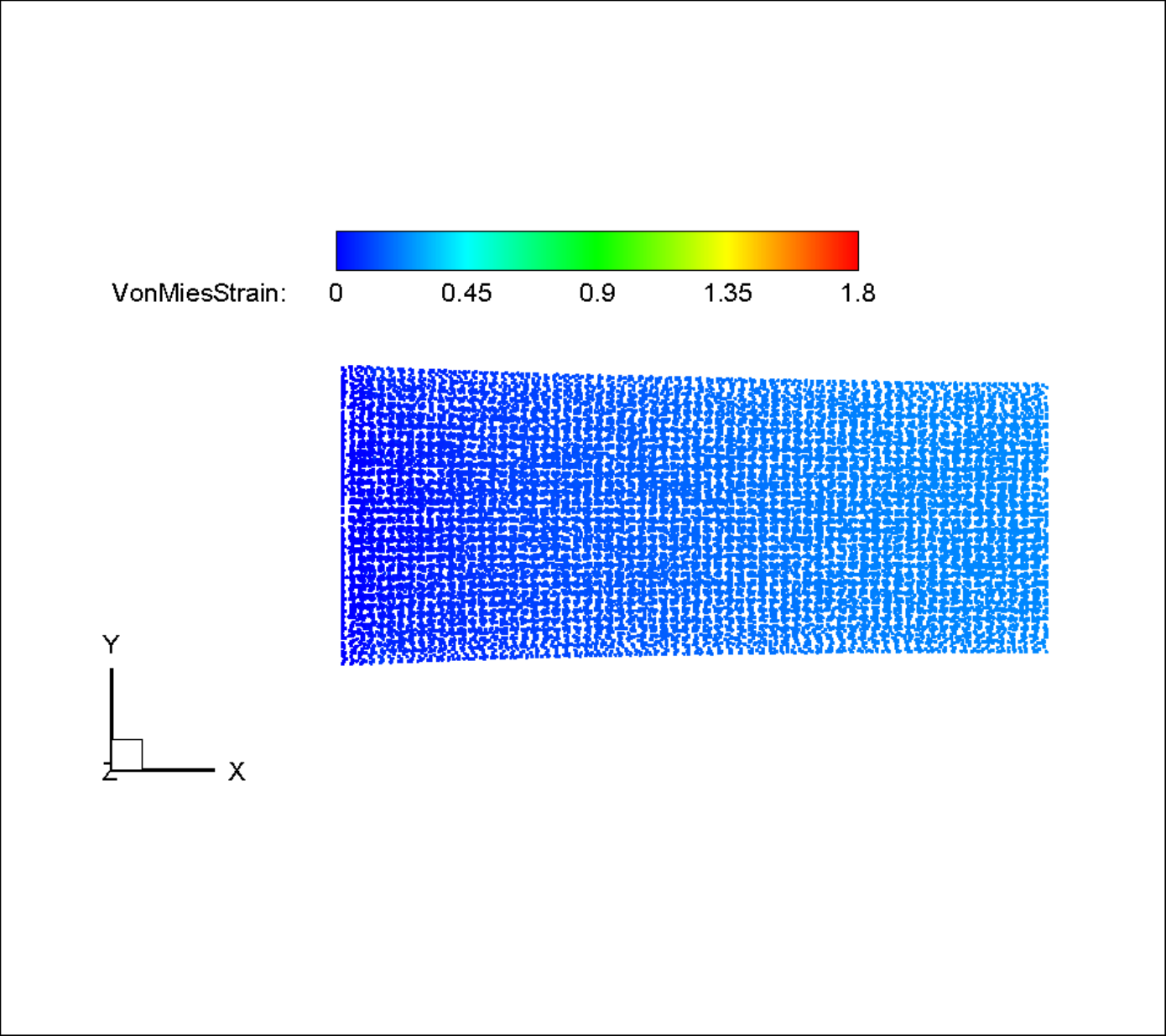}
		\caption {50s}
		\label{3dstrain-contour50}
	\end{subfigure}
	\begin{subfigure}[b]{0.45\textwidth}
		\includegraphics[trim = 1mm 2mm 1mm 5mm, clip,width=0.99\textwidth]{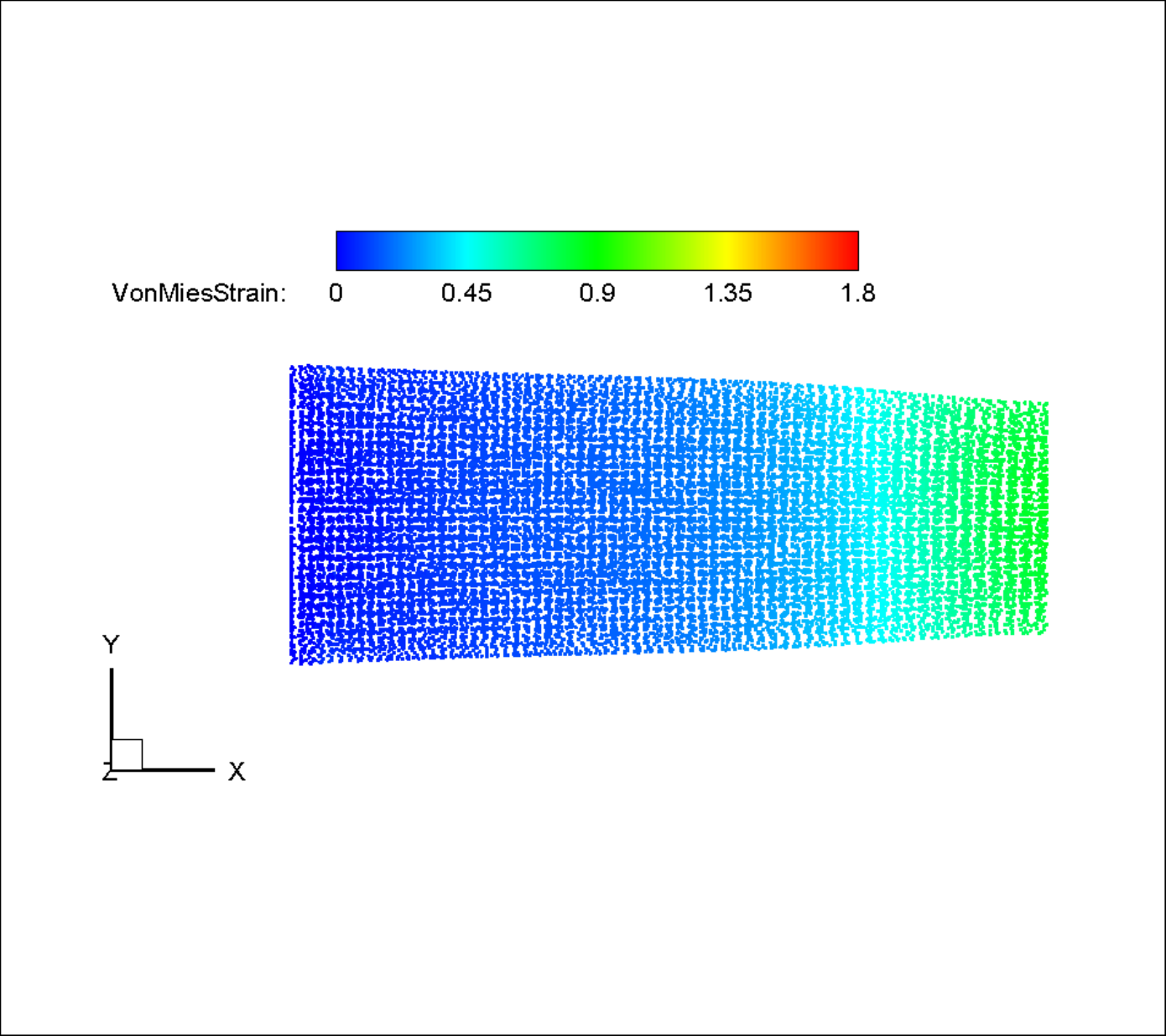}
		\caption {80s}
		\label{3dstrain-contour80}
	\end{subfigure}
	\begin{subfigure}[b]{0.45\textwidth}
		\includegraphics[trim =1mm 2mm 1mm 5mm, clip,width=0.99\textwidth]{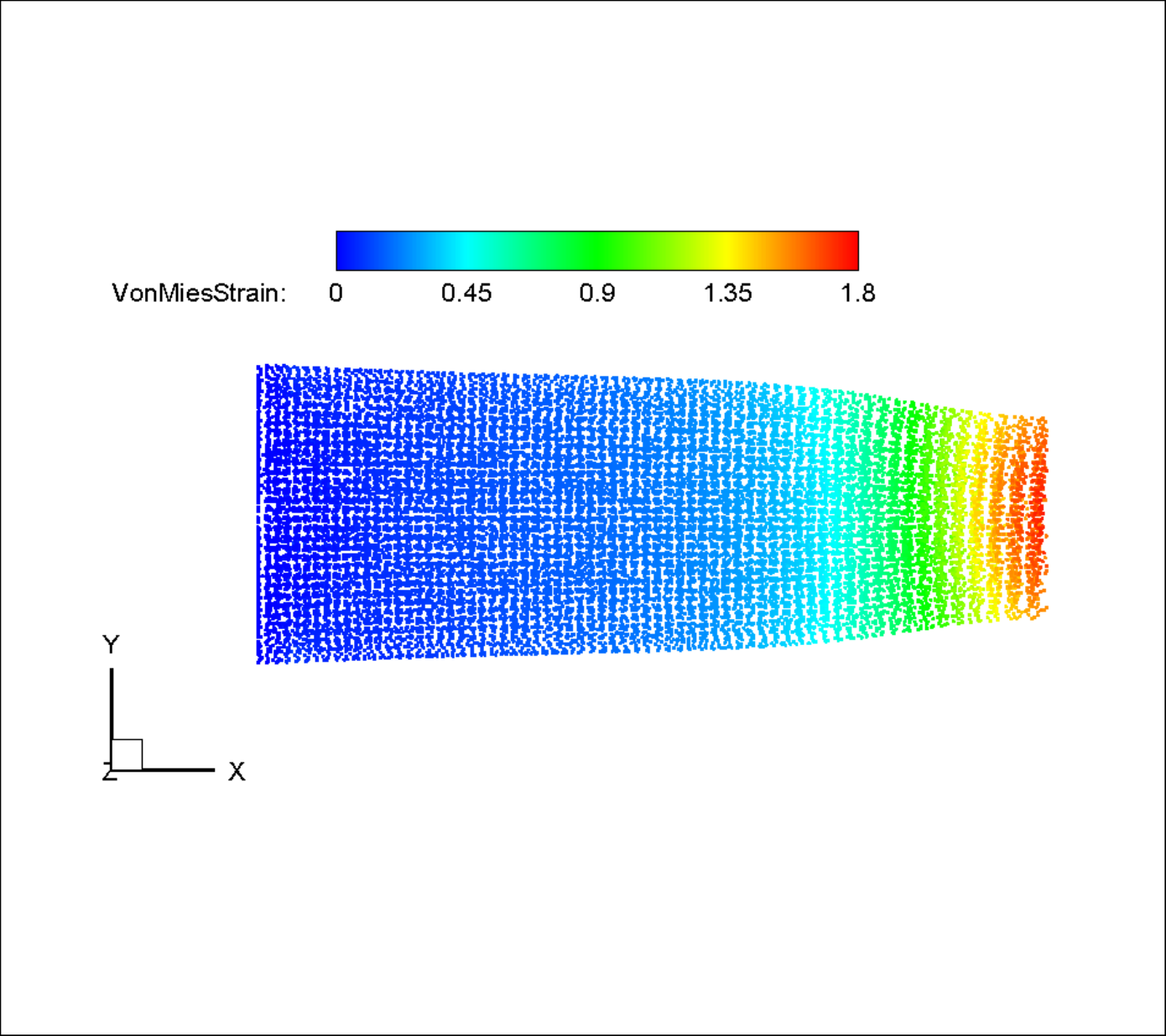}
		\caption {100s}
		\label{3dstrain-contour100}
	\end{subfigure}
	\caption{3D tensile necking: the deformation colored by von Mise Strain at different time instants (front view  with half the specimen). }
	\label{3d-stretching-stress}
\end{figure*}
\begin{figure*}[htbp]
	\centering
	\begin{subfigure}[b]{0.45\textwidth}
		\includegraphics[trim = 1mm 2mm 1mm 5mm, clip,width=0.95\textwidth]{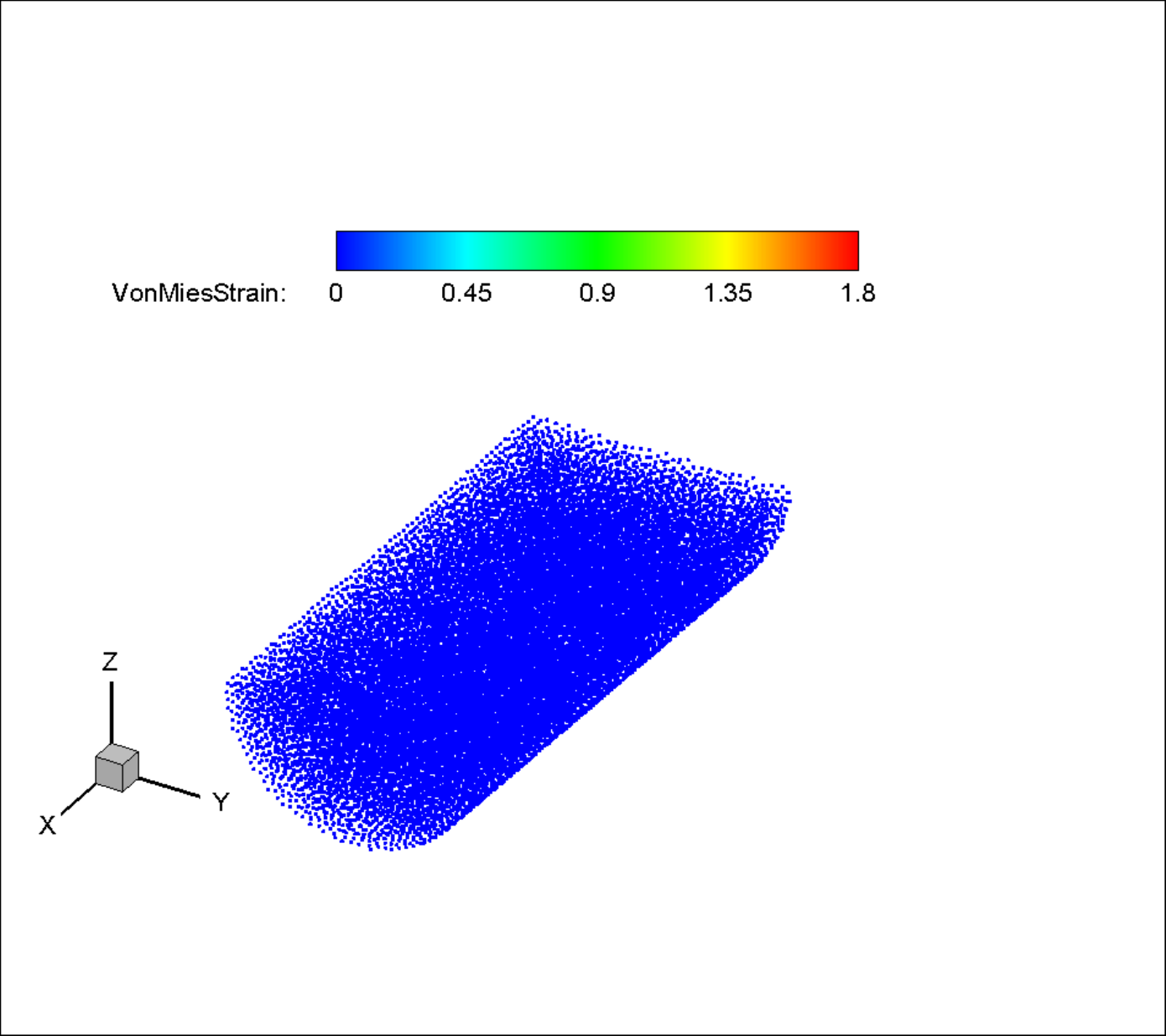}
		\caption {0s}
		\label{3dstrain-contour0xyz}
	\end{subfigure}
	\begin{subfigure}[b]{0.45\textwidth}
		\includegraphics[trim =1mm 1mm 1mm 5mm, clip,width=0.99\textwidth]{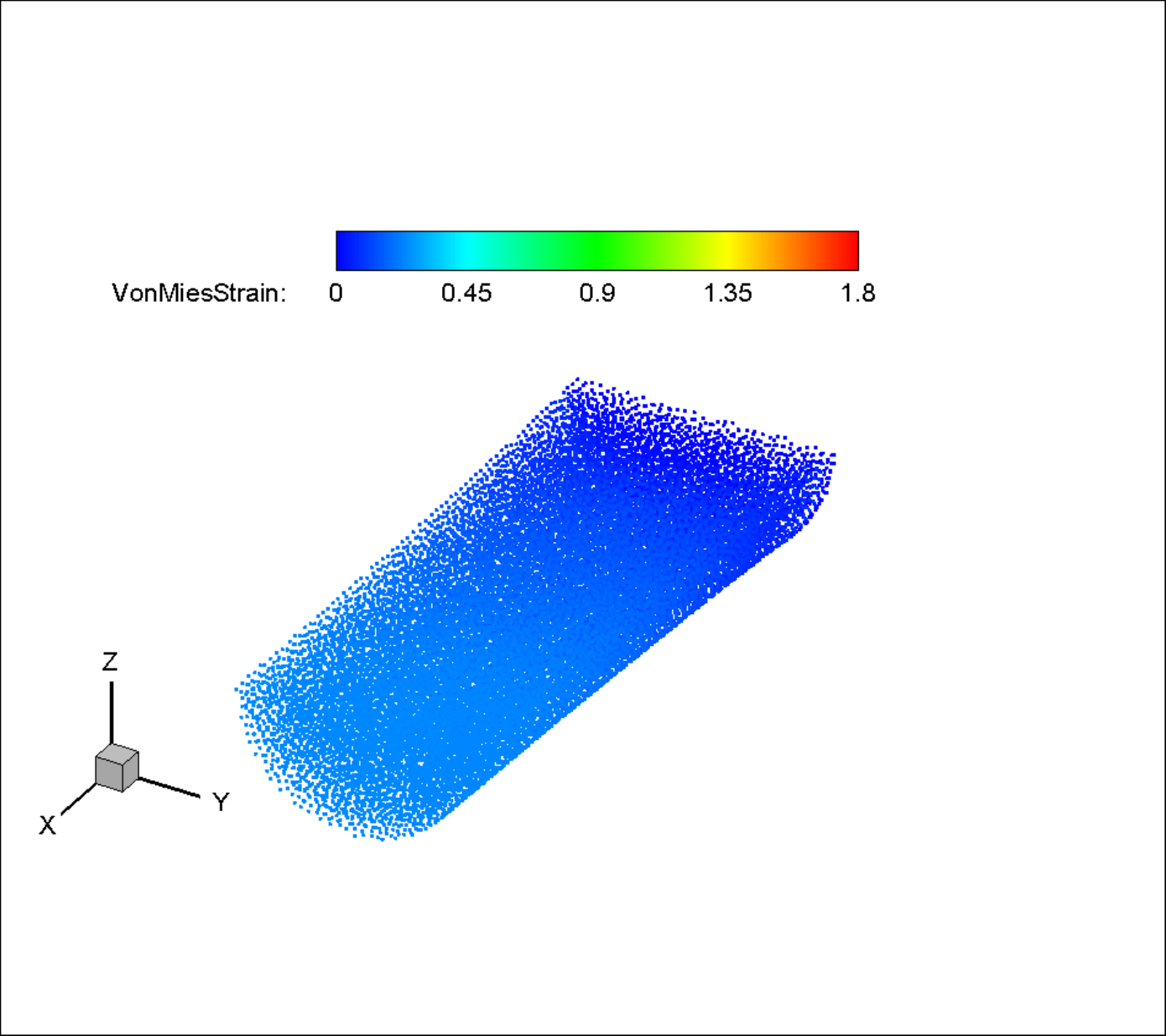}
		\caption {50s}
		\label{3dstrain-contour50xyz}
	\end{subfigure}
	\begin{subfigure}[b]{0.45\textwidth}
		\includegraphics[trim = 1mm 2mm 1mm 5mm, clip,width=0.99\textwidth]{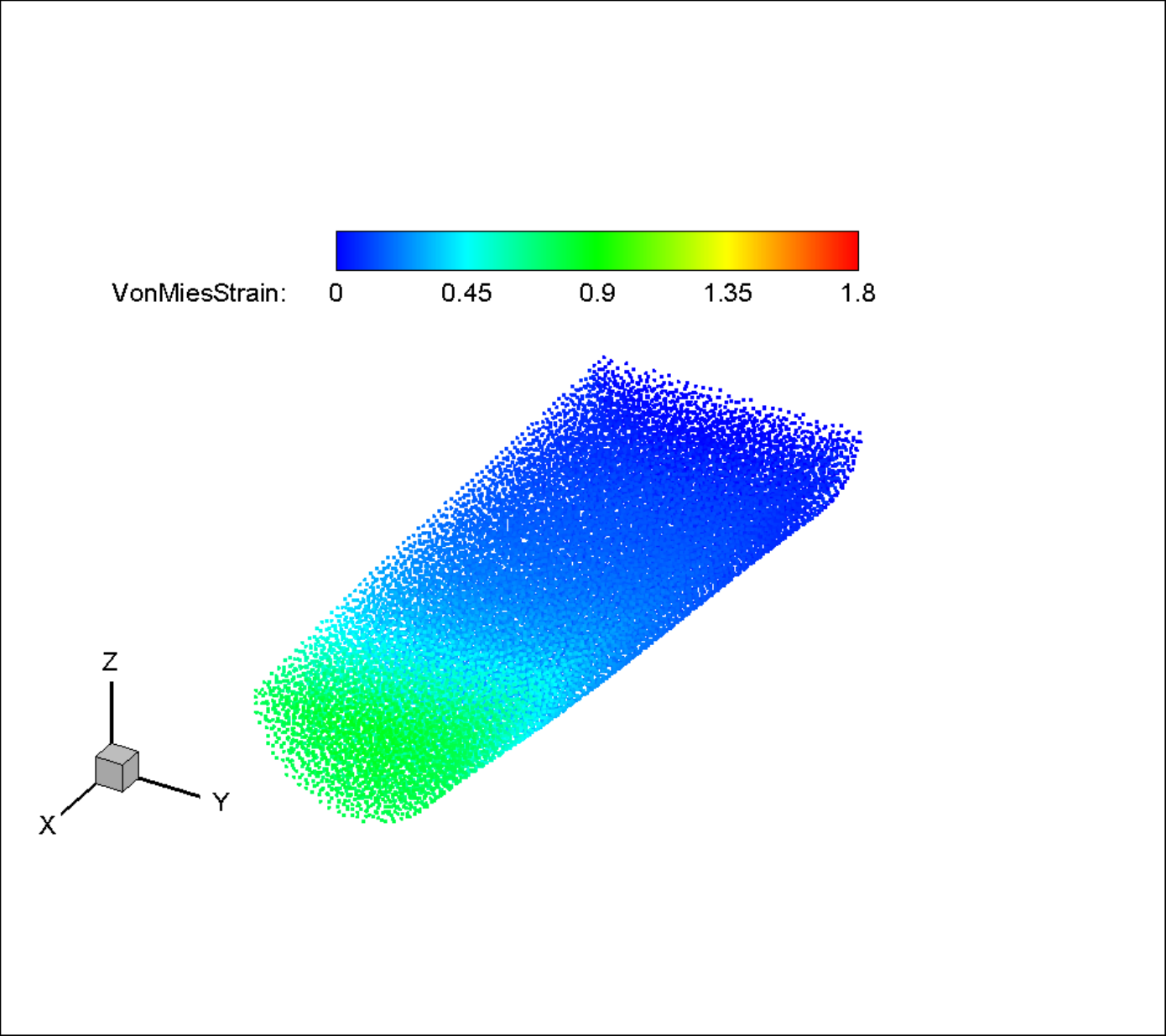}
		\caption {80s}
		\label{3dstrain-contour80xyz}
	\end{subfigure}
	\begin{subfigure}[b]{0.45\textwidth}
		\includegraphics[trim =1mm 2mm 1mm 5mm, clip,width=0.99\textwidth]{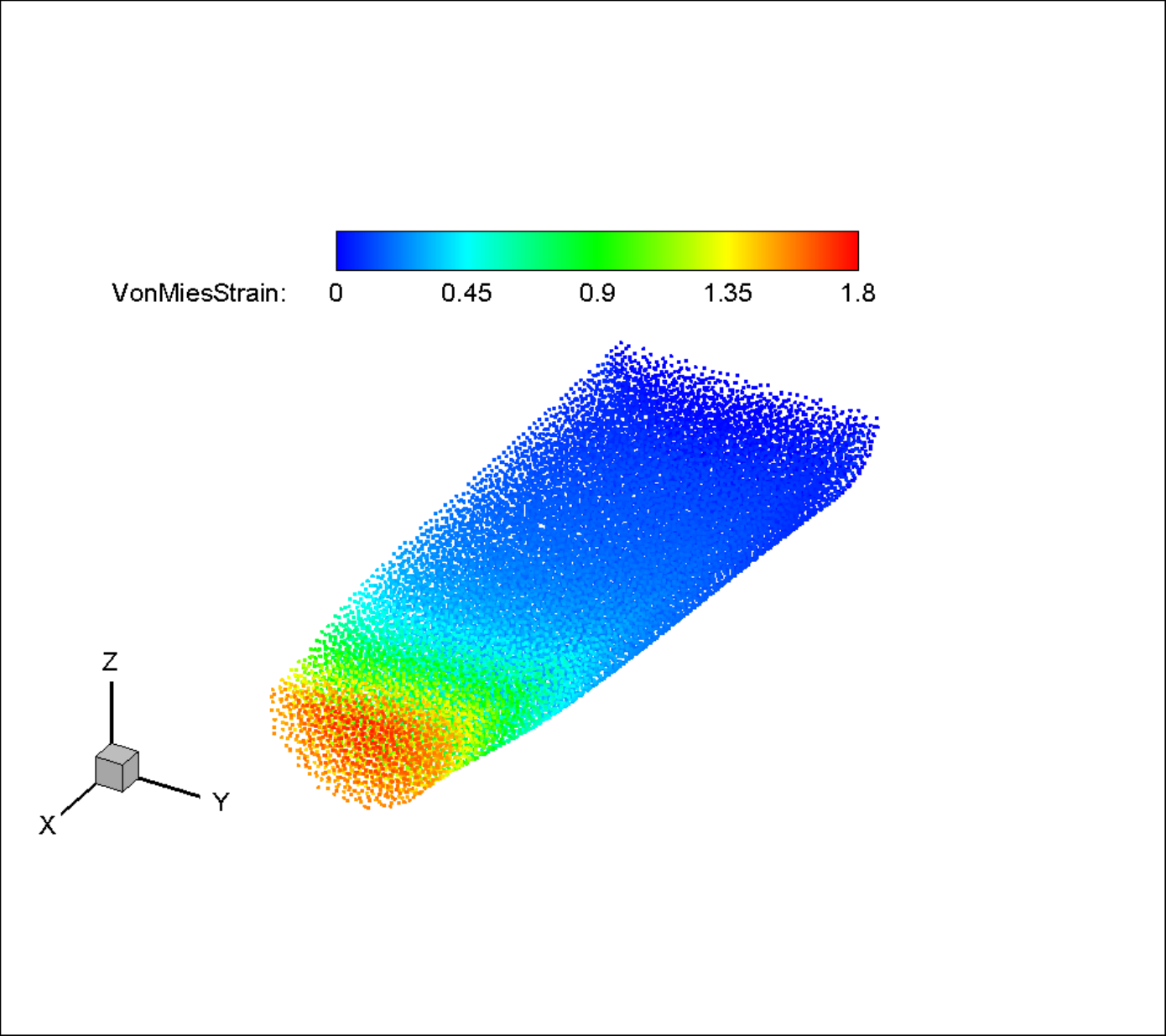}
		\caption {100s}
		\label{3dstrain-contour100xyz}
	\end{subfigure}
	\caption{3D tensile necking: the deformation colored by von Mise Strain at different time instants (top view with quarter the specimen). }
	\label{3d-stretching-stressxyz}
\end{figure*}
\begin{figure*}[htbp]
\centering
\includegraphics[trim = 2mm 2mm 2mm 2mm, clip,width=0.55\textwidth]{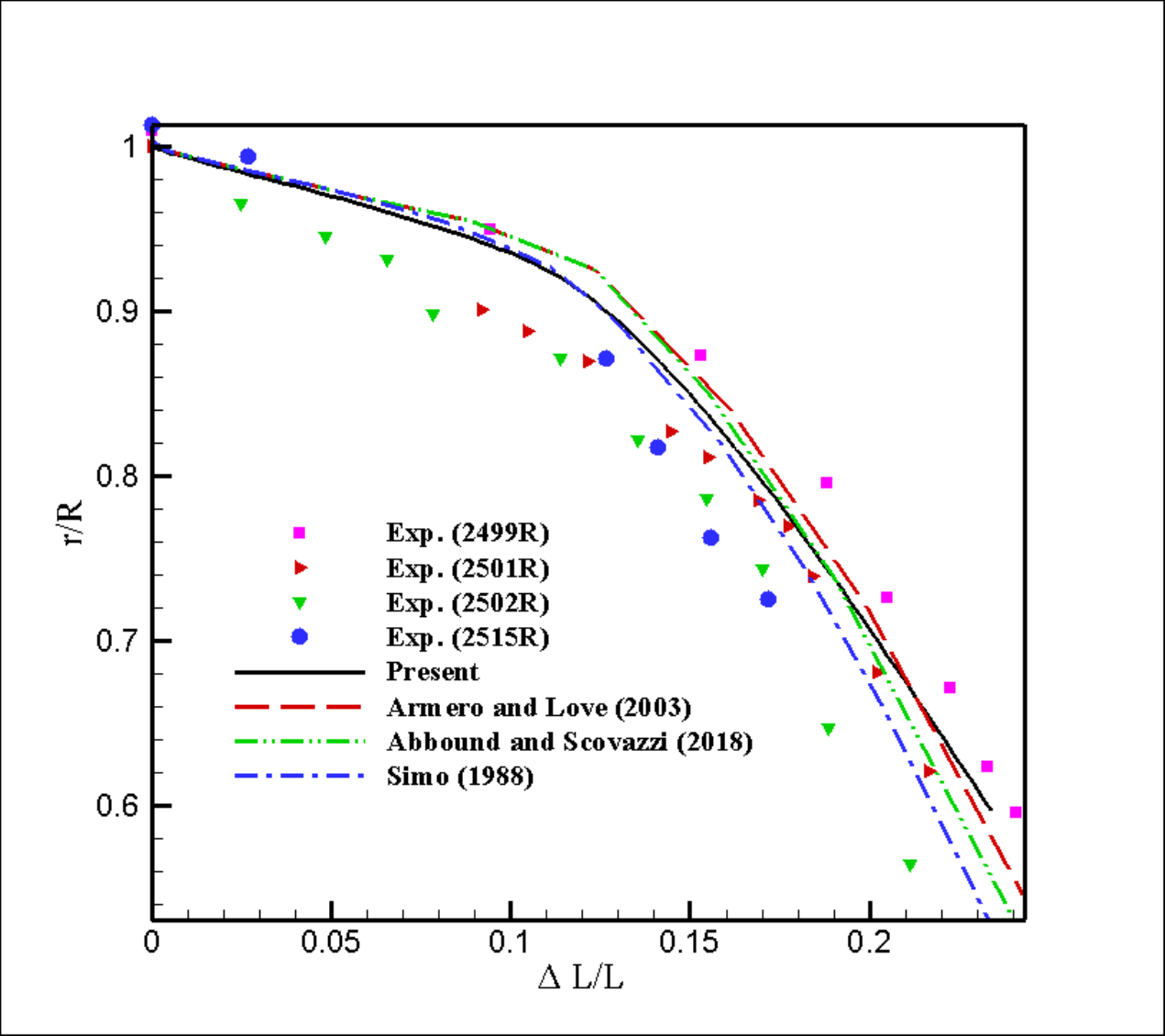}
\caption {3D tensile necking: the evolution of radial displacement of the central part compared with the reference \cite{elguedj2014isogeometric, rodriguez2002arbitrary, neto2005f}. }
\label{3D-stretching-pos}
\end{figure*}
\begin{figure*}[htbp]
	\centering
	\includegraphics[trim = 2mm 2mm 2mm 2mm, clip,width=0.55\textwidth]{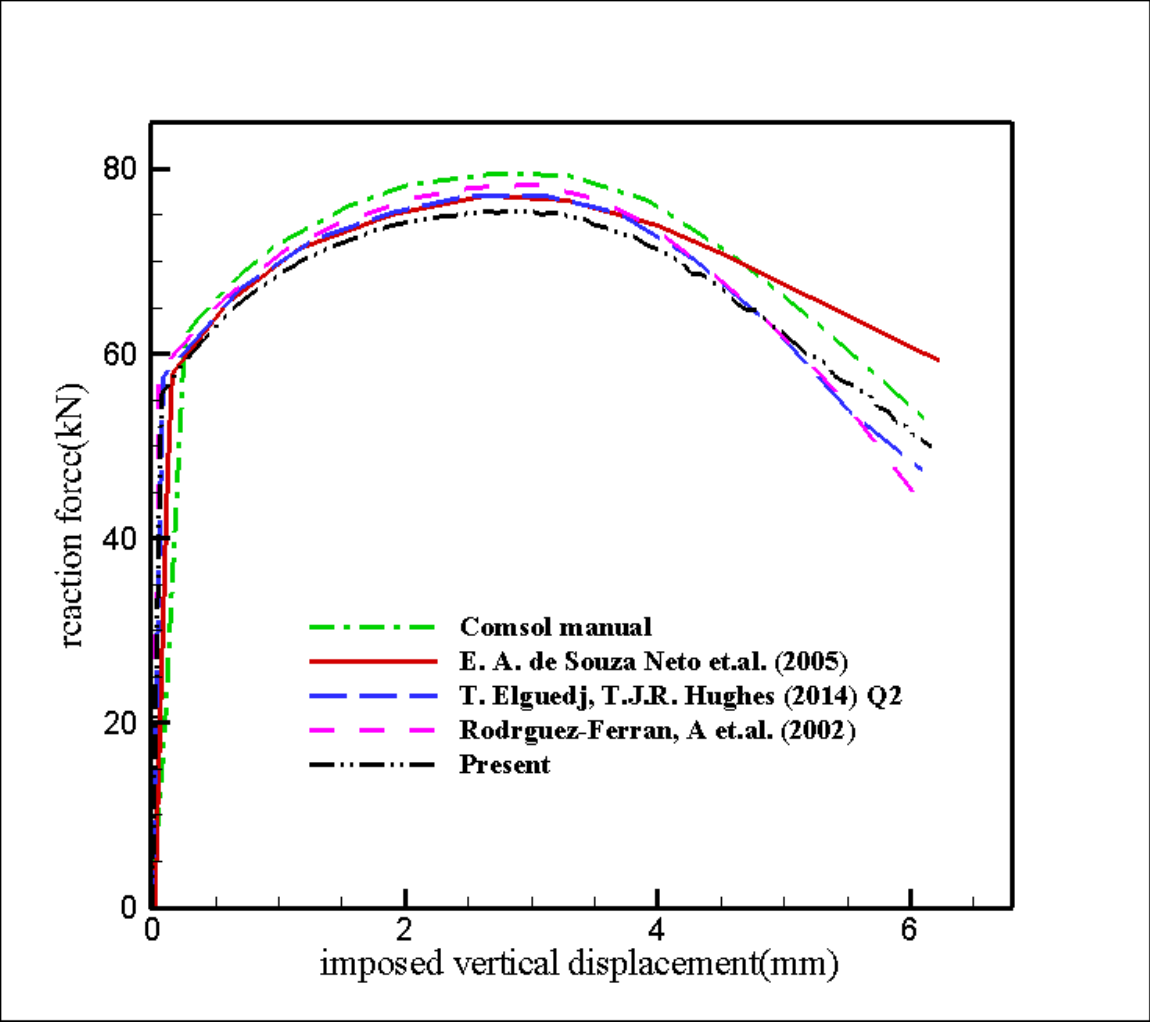}
	\caption {3D tensile necking: the overall evolution of 
		the reaction force versus the imposed vertical displacement compared with the reference \cite{elguedj2014isogeometric, rodriguez2002arbitrary, neto2005f}. }
	\label{3D-stretching-force}
\end{figure*}

Contour plots of the von Mise Strain at different time instants from different views are shown  in Figure \ref{3d-stretching-stresswhole}-\ref{3d-stretching-stressxyz}.
The last plots depict  the deformed shape of the specimen at the final stage of the simulation,
indicating the occurrence of a necking in the center of the specimen.
Based on these figures, 
we can deduce the deformation evolution of this specimen: 
initially, the boundary conditions enabled the specimen to maintain an  uniform elastic response in the short stage of loading history; 
subsequently, in the post-peak regime, a diffuse necking mode emerged, 
which eventually led to the formation of shear bands at high strain levels. 
These bands accumulated plastic deformations, 
ultimately leading to the final necking even failure of the specimen.
The evolution of this pattern is well-reproduced by the force and 
deformation data presented in Figures \ref{3D-stretching-pos} and \ref{3D-stretching-force}, which agrees well with experimental findings.

Specially, Figure \ref{3D-stretching-pos}  collected the numerical data of the radius displacement, 
normalized by the initial radius,  
versus the length displacement, 
normalized by the initial bar length, 
and the results are compared 
with experimental data and solutions
reported in other works  \cite{armero2003arbitrary, simo1988framework, de2022new,simo2006computational}.
The results obtained with the present method seems to 
be in good agreement with the experimental data of material 2501R.
The load displacement curve obtained from this 
 simulation is presented in Figure. \ref{3D-stretching-force} along with
the results from other numerical works.
Some differences can be observed in the later stages of 
the deformation and force variation curves,
which may be attributed to the hourglass mode 
and the limitations of particle discretization.
From the findings of de Saracibar  \cite{de2004orthogonal}, 
different mesh discretization and element types can 
result in distinct stabilization and locking patterns, 
as evidenced by the disparate peak and final values of force in 
load-displacement curves obtained using different mesh discretization strategies, 
including uniform and non-uniform mesh distributions
in the necking area.
A comprehensive comparison of these results, 
as well as a clear explanation, is provided in de Saracibar's work. 
Although this hourglass phenomenon depicted in Figure  \ref{3D-stretching-pos}
cannot be fully removed in this study,
 the same pattern is observed in prior research
that utilized standard 3D elements in FEM \cite{neto2005f}.
The current paper does not aim to address this issue comprehensively, 
but rather suggests that further improvements will be implemented in the future research.

%%%%%%%%%%%%%%%%%%%%%%%%%%%%%%%%%%%%%%%%%%%%%%%%%%%%%%%%%%%%%%%
%%%%%%%%%%%%%%%%%%%%%%%%%%%%%%%%%%%%%%%%%%%
%%%%%%%%%%%%%%%%%%%%%%%%%%%%%%%%%%%%%%%%%%
%%%%%%%%%%%%%%%%%%%%%%%%%%%%%%%%%%%%%%%%%%
As for determining the static state achieving,  similar with previous 2D case, the kinetic energy criterion 
is derived from the elastic energy  $ E_e = \frac{1}{2} F \Delta x$, 
while the load force $F = 80000N$, consistent with the value in Figure	\ref{3D-stretching-force}, 
and  the stretching length $\Delta x = 0.014$ m.
Using the same method applied in the 2D case, 
we find that the results are converged when the kinetic energy is reduced to 0.5\% of the elastic energy.
Finally, to check the efficiency of the present algorithm,
we computed the relaxation iterations  and the results are summarized in   Table 	\ref{3D-stretching-efficiency}. Evident reduction is obtained in stress relaxation times,
saving the computation time significantly.
 
\begin{table}[htb!]
	\centering
	\caption{3D tensile necking:  quantitative validation of the efficiency of this  multi-time step algorithm.}
		\renewcommand{\arraystretch}{1.1} 
	\begin{tabular}{ccccc}
		\hline
		algorithm   & $ N_p $ & $ N_S $& $ N_s $ & $N_d$      \\ 	
		\hline
		straightforward algorithm  & 250852 & $  2.21e^{9} $ &  $ 2.21e^{9} $  &  -  \\
		multi-time step algorithm	&  250852 & $ 1.0e^{4} $ & $ 3.21e^{5} $ & $ 3.21e^{5} $  \\
		\hline	
	\end{tabular}
	\label{3D-stretching-efficiency}
\end{table}
%%%%%%%%%%%%%%%%%%%%%%%%%%%%%%%%%%%%%%%%%%%%%%%%
\newpage
\subsection{Two-dimensional fluid-structure interaction}  
\begin{figure*}[htbp]
	\centering
	\includegraphics[trim = 0mm 0mm 0mm 0mm, clip,width=0.55\textwidth]{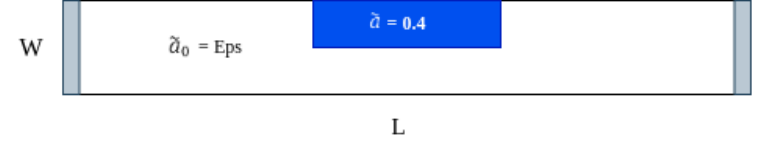}
	\caption {2D fluid-structure interaction: physical configuration of the thin porous beam.}
	\label{configuration}
\end{figure*}

In this section, we  perform a two-dimensional simulation  of  fluid diffusion 
coupling with porous solid deformation and  
the model  is described in Section \ref{section_2}, to verify the efficiency of the presented method. 
As Figure \ref{configuration} shows,
a thin porous beam with a length of $ L $ = 10.0 mm and
width of $ W $ = 0.125 mm is considered, 
with the left and right sides being constrained 
to prevent any curling or movement.
The simulation starts with a fluid droplet
 contacting the center part of the beam 
with a length of 0.3$ L $, 
and this contact continues for 10 seconds 
while the total physical time is 100 seconds.
Given the thin nature of the beam, 
we assume all pores in the upper half part 
are filled with fluid initially.
As we stated before,
the relationship between fluid
saturation $\widetilde{a}$
and solid porosity  ${a}$ is
$0 \leq \widetilde{a} \leq {a} < 1$.
For this 2D and 3D cases discussed later, 
we assume a solid porosity of ${a} = 0.4$,
meaning that the fluid saturation $ \widetilde{a} $ 
in the central part($0.5W \times 0.3L$) 
is constrained to $ \widetilde{a} = {a} = 0.4$
for the initial 10 seconds,
while in other regions $ \widetilde{a}_0 = 0.0$.
\\

In accordance with the experimental setup,  
the solid material is considered as 
a porous and elastic Nafion membrane, 
with water  serving as the fluid.
The physical properties and  material parameters 
of this membrane  
are listed in Table \ref{parameter-table}.
The pressure coefficient C has been calibrated to fit the experimentally measured flexure curves,
 while other   parameters are obtained from previous research papers \cite{motupally2000diffusion,goswami2008wetting}.
\begin{table}[htb!]
	\centering
	\caption{Fluid-structure interaction: physical material parameters value of Nafion film. 
 Data estimated from Motupally  and Goswami \cite{motupally2000diffusion,goswami2008wetting}.}
	  
\resizebox{\textwidth}{!}{
	\renewcommand{\arraystretch}{1.1} 
		\begin{tabular}{cccccc} 
		\hline
			\centering
		Parameters & $\rho$  $ {\rm(kg/m^3)}$ & K $ {\rm(m^2/s)} $ & Pressure coefficient C  $ {\rm(Pa)}$& Young modulus $ {\rm(Pa)}$ & Poisson ratio \\ 
		\hline
			\centering
		Value & 2000  & $ 1.0e^{-10} $	& $ 3.0e^{6} $   & $ 8.242e^{6} $& 0.2631\\
		\hline	
	\end{tabular}}
	\label{parameter-table}
\end{table}
In the simulation, eight particles are placed in the vertical direction, 
with a particle spacing  of $dp_y = W/8 = 1.5625 \times 10^{-2}$ mm. 
However,  due to the high aspect ratio of the beam, 
using the same particle spacing $ dp$ in
horizontal $ x $ and vertical $ y $ directions
 would require a large number of particles, 
thus increasing the computation time.
To address this issue, an anisotropic kernel algorithm is employed, 
with an anisotropic ratio of 4.0,
meaning   $dp_x = 4dp_y  = 0.0625$ mm.
In this simulation, an  experienced damping ratio of $ \eta = 1.0e^3$ in the damping 
term is utilized. 
\begin{figure}[htbp]
	\centering
	\begin{subfigure}[b]{0.9\textwidth}
		\includegraphics[trim = 1mm 25mm 1mm 130mm, clip,width=1.0\textwidth]{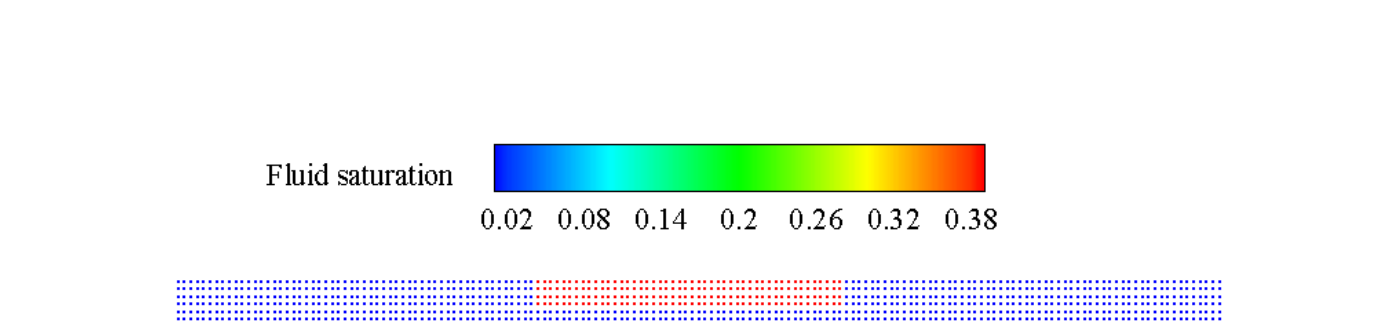}
		\caption {0s}
		\label{2d-saturaion-0}
	\end{subfigure}
	\begin{subfigure}[b]{0.9\textwidth}
	\includegraphics[trim = 1mm 25mm 1mm 100mm, clip,width=1.0\textwidth]{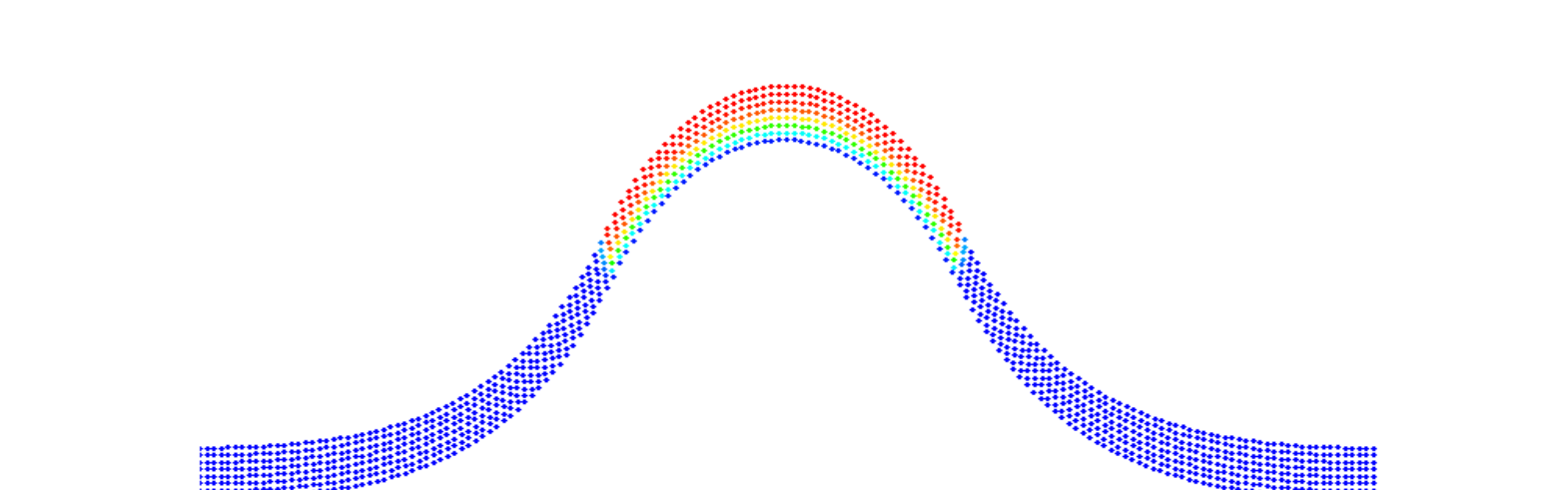}
	\caption {10s}
	\label{2d-saturaion-10}
\end{subfigure}
	\begin{subfigure}[b]{0.9\textwidth}
		\includegraphics[trim = 1mm 25mm 1mm 100mm, clip,width=1.0\textwidth]{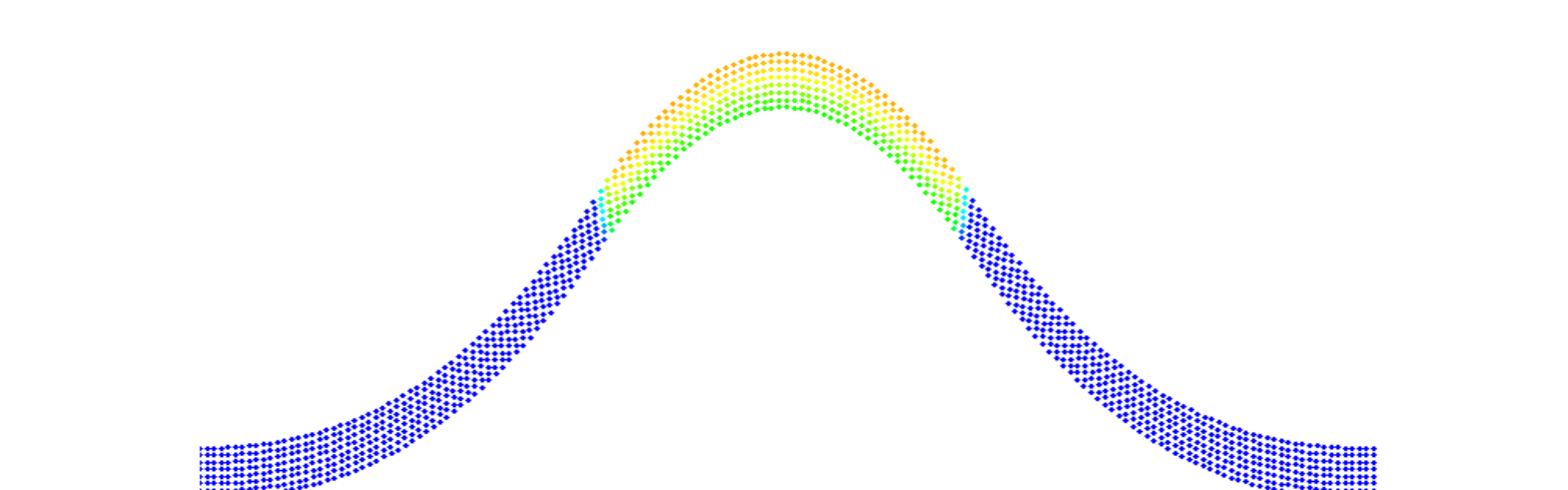}
		\caption {50s}
		\label{2d-saturaion-50}
	\end{subfigure}
	\begin{subfigure}[b]{0.9\textwidth}
		\includegraphics[trim =1mm 25mm 1mm 100mm, clip,width=1.0\textwidth]{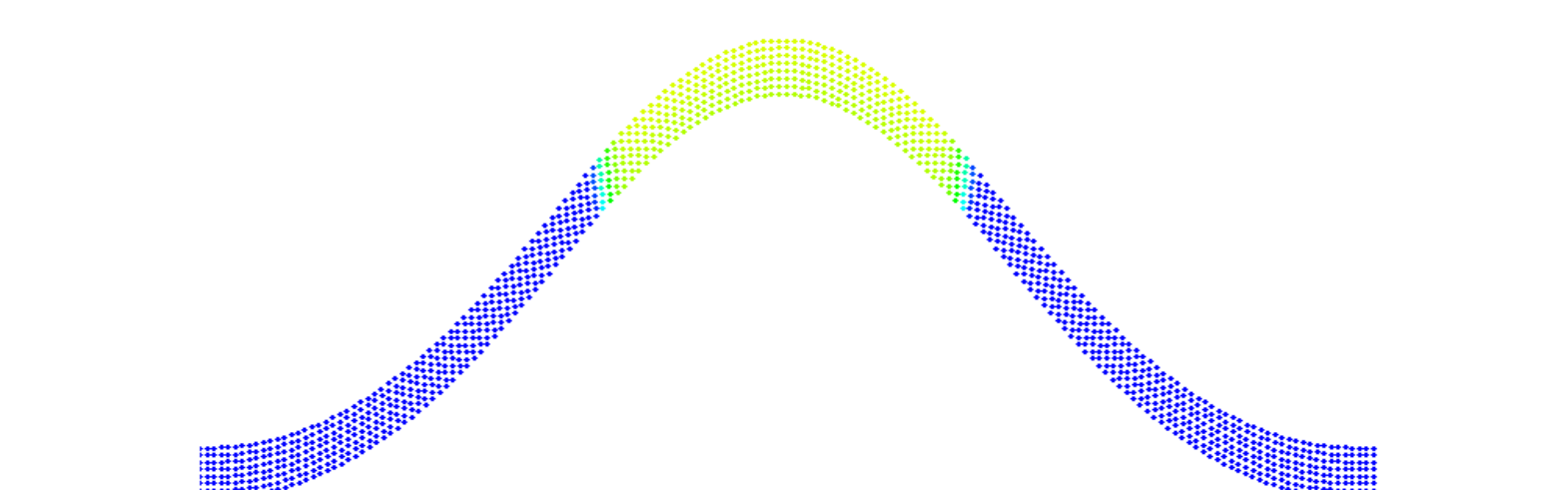}
		\caption {100s}
		\label{2d-saturaion-100}
	\end{subfigure}
	\caption{2D fluid-structure interaction: the deformation colored by fluid saturation at different time instants. }
	\label{deformation coupling saturation}
\end{figure}

\begin{figure*}[htbp]
	\centering
	\includegraphics[trim = 1mm 1mm 1mm 1mm, clip,width=0.55\textwidth]{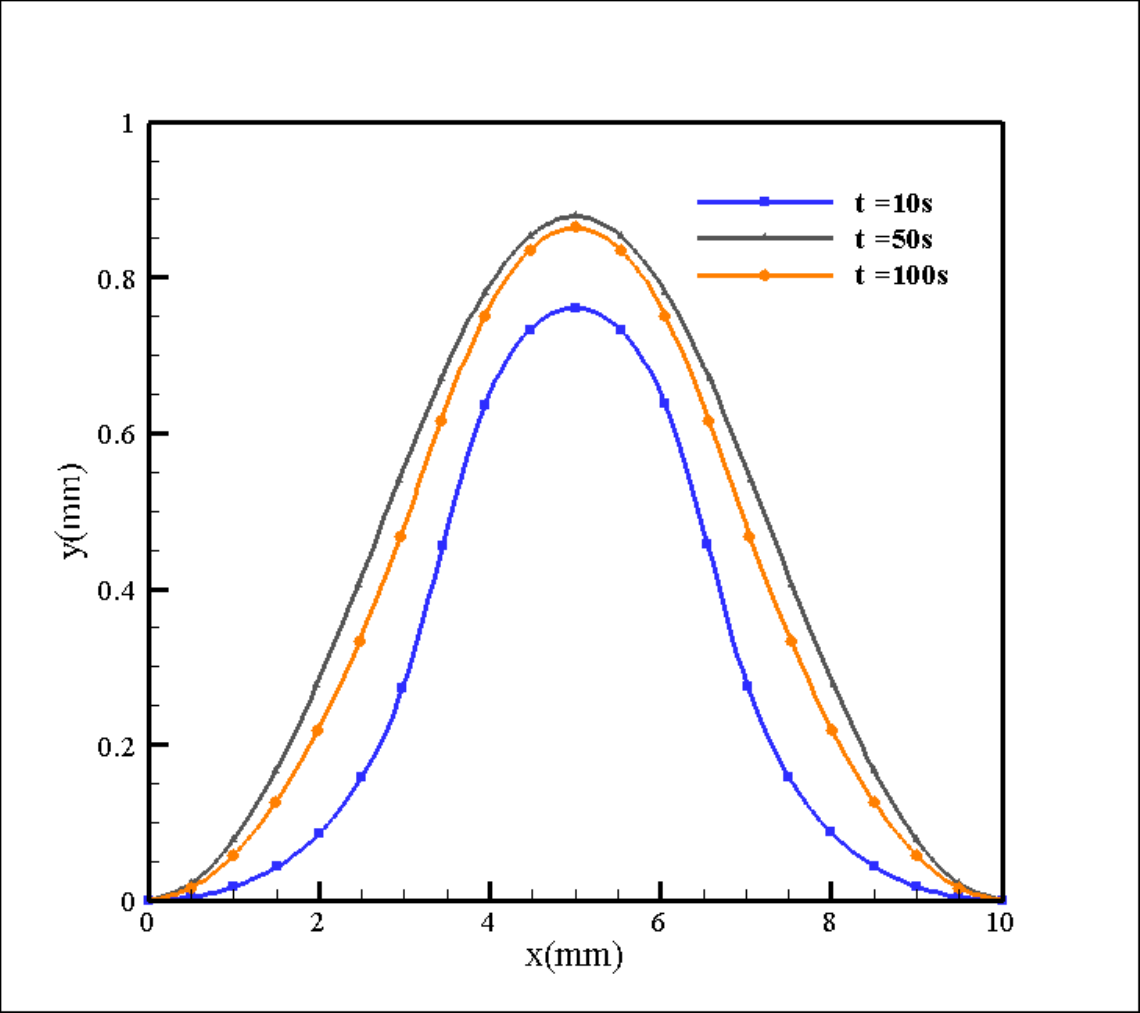}
	\caption {2D fluid-structure interaction: bending amplitude of the beam at different time instants.}
	\label{2dpoint_position}
\end{figure*}
With the conditions given above, the simulation produces a  deformed
configuration colored by fluid saturation, 
as shown in  Figure \ref{deformation coupling saturation}.
Initially, the presence of a water droplet in the upper 
central region generates a fluid pressure, 
as explained in Eq. \ref{fluid_pressure},
leading to a localized bending in the central region. 
As time progresses, the  saturation difference drives
water diffusing continuously, 
and the total water amount within 
the porous solid increases,
causing a rising flexure.
This is also depicted in Figure \ref{2dpoint_position}, 	
which records the  vertical position $ y $ 
versus the horizontal $ x $ position of the beam at different 
time instants.
After the contact finishes,
 no more water is added into the beam, 
 and the central water flows slowly into the side areas.
 Clearly, the fluid saturation shows 
a smooth transition from the center to the surrounding area  in  Figure \ref{deformation coupling saturation}.
Accordingly, a more uniform pressure distribution is developing, 
 resulting in a  more smooth  flexure of the beam  
as shown in Figure \ref{2dpoint_position} in the later period.

\begin{figure*}[htbp]
	\centering
	\includegraphics[trim = 1mm 1mm 1mm 1mm, clip,width=0.65\textwidth]{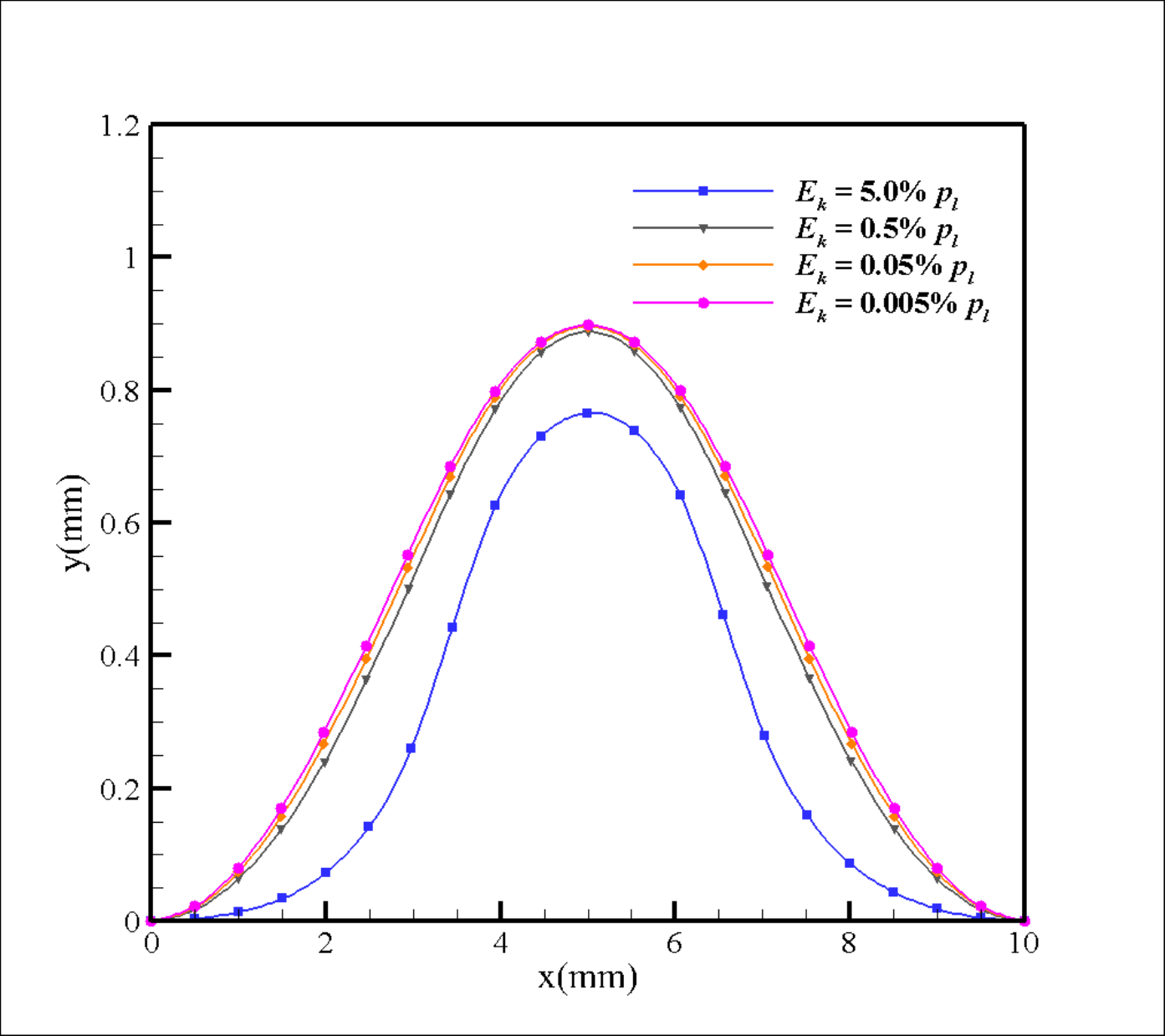}
	\caption {2D fluid-structure interaction: bending amplitude convergence with different density kinetic energy criteria.}
	\label{2dpoint_position_energy}
\end{figure*}

\begin{figure*}[htbp]
	\centering
	\includegraphics[trim = 1mm 1mm 1mm 1mm, clip,width=0.65\textwidth]{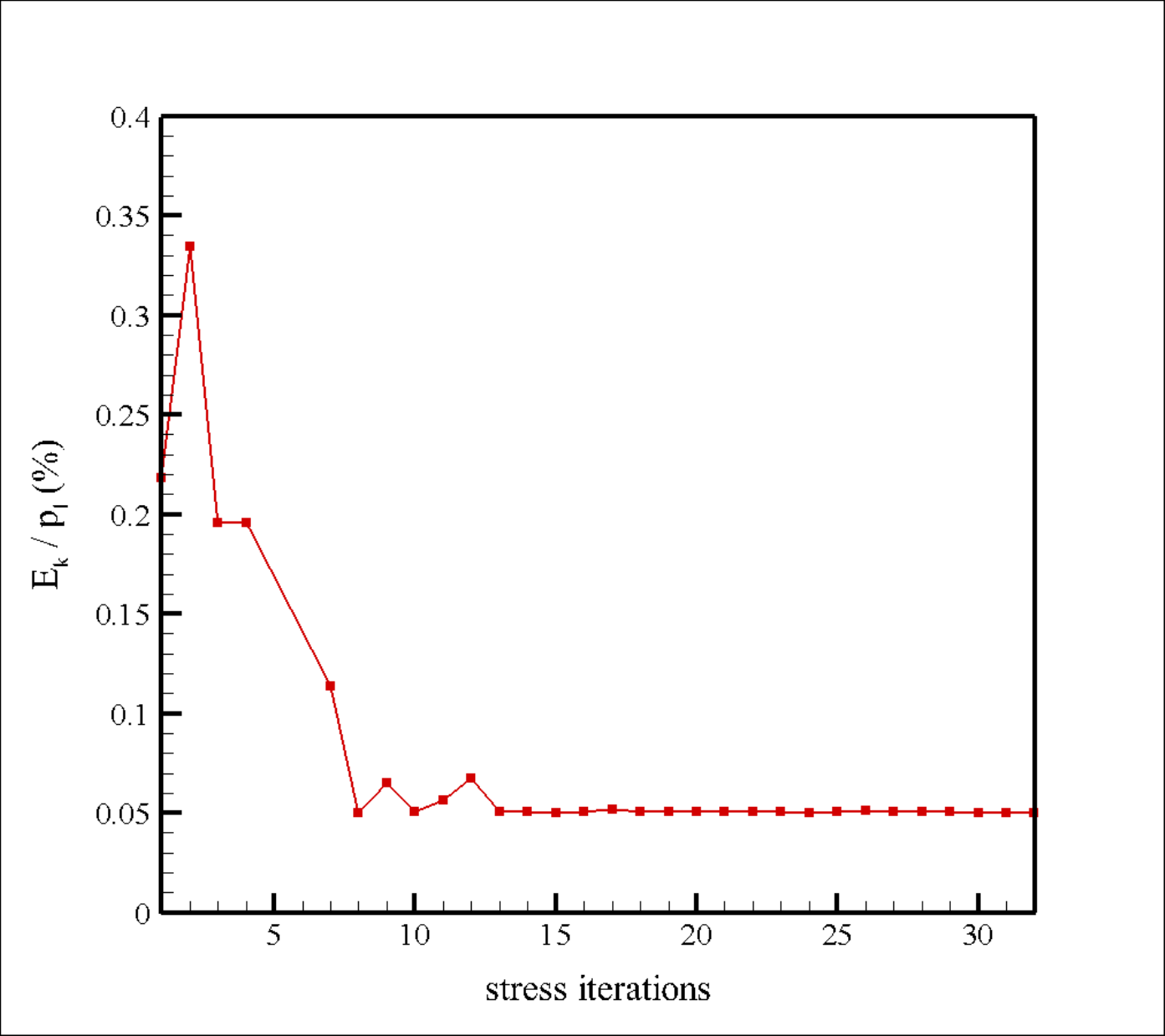}
	\caption{2D fluid-structure interaction: the density kinetic energy variation within the diffusion period when $ t = 20 $s valuated by the water pressure $ p^l$.}
	\label{kinetic-energycomparision}
\end{figure*}
For determining the density kinetic energy criterion $ E_k $, 
we use the pressure from water, $ p^l $, stated in Eq. \ref{fluid_pressure},
as the reference since the  fluid pressure induces the beam swelling.
 To evaluate the effect of the relative density kinetic energy threshold on the simulation results, 
a series of simulations are conducted using various criteria $ E_k$. 
The time evolution of the bending amplitude with different kinetic energy
criteria is presented in Figure \ref{2dpoint_position_energy}.
With a relatively large criterion value of $ E_k $ = 5\%$ p^l$,
it is observed that the equilibrium state is not achieved and the energy is not fully eliminated 
with a relatively light deformation.
On the other hand, 
using a very small criterion value leads to 
unnecessary calculation steps,
increasing computation time.
Therefore, it can be concluded that the 
appropriate density kinetic energy criterion value 
for this 2D case is 0.05\%$ p^l$.

Referring to Figure  \ref{kinetic-energycomparision}, 
the  evolution of the density kinetic energy 
within the diffusion period when $ t = 20$s, 
evaluated by the water pressure $p^l$, is presented. 
Due to the water pressure, 
the density kinetic energy firstly experiences a peak
 after one diffusion performance,  
then followed by a decrease to a certain criterion 
value of 0.05\% $p^l$ we set before, 
which is attributed to the damping effects.
Throughout the simulation process, 
the stress relaxation takes place 
accompanying with viscous  damping
immediately after each diffusion relaxation event.
The relative density kinetic energy at the end of each diffusion step approaches
0.05\%$ p^l$,
indicating that the velocity almost vanishes. 
This signifies that equilibrium is achieved 
at the end of each diffusion time step.

% By testing different thresholds, we aimed to investigate the relationship between the kinetic energy threshold and the speed at which balance is achieved. This allowed us to gain a better understanding of the influence of the kinetic energy criterion on the rate of convergence towards equilibrium.
%The result in Figure shows that the damping ratio have no impact on the result.  Note that much larger ratio does not show a big enhancement in the balance reaching speed. According to our experience, 5 for 2d and 500 for 3d are appropriate values.
 The efficiency of the proposed approach is demonstrated through Table \ref{2D-fluid-structure-interaction-efficiency}, 
 which presents a quantitative comparison of the algorithm against the straightforward approach
 in terms of diffusion and stress relaxation iterations $N_D$, $N_s$ with a total particle number $ N_p $.
The results reveal a great reduction 
in computation iterations,
thus demonstrating the significant improvement 
in efficiency achieved by the proposed approach.
\begin{table}[htb!]
	\centering
	\caption{2D fluid-structure interaction:  quantitative validation of the efficiency of this multi-time step algorithm.}
		\renewcommand{\arraystretch}{1.1} 
	\begin{tabular}{ccccc}
		\hline
		algorithm   & $ N_p $ & $ N_D $ & $ N_s $ & $ N_d $ \\ 	
		\hline
	straightforward algorithm  & 1336 & 1.58$ e^7 $  &  1.58$ e^7 $  & -\\
		\hline
	multi-time step algorithm	& 1336 &  125 & 2.76$ e^5  $   & 2.76$ e^5  $ \\
		\hline	
	\end{tabular}
	\label{2D-fluid-structure-interaction-efficiency}
\end{table}
 
\subsection{Three-dimensional fluid-structure interaction} 
Next, we consider the fluid diffusion coupling swelling in a three-dimensional film,
specifically the diffusion of water within a porous Nafion membrane.
This system has been previously studied numerically by Zhao \cite{zhao2013modeling}
and experimentally by Goswami \cite{goswami2008wetting}.  
This reference thin porous body is in the form
of a polymer film  with a x-y plane of dimensions $L_x = 10.0$ mm,  $L_y = 10.0$ mm and a height of $L_z = 0.125$ mm. 
Four boundary sides are  constrained to prevent any curling or movement.
The physical parameters are taken to be the same as those listed in Table \ref{parameter-table}.
The initial conditions are similar to those used in the two-dimensional case.
The central square part of the membrane in contact with water occupies 
a region of dimensions $0.3L_x \times 0.3L_y \times 0.5 L_z$, 
and this contact lasts for 450 seconds, 
while the total physical time is 2500 seconds. 
No fluid is allowed to diffuse out from the membrane.
The fluid saturation $\widetilde{a} $ 
in the central square part is constrained to 
$\widetilde{a}={a} = 0.4$ for the initial 450 seconds,
while in other regions $ \widetilde{a}_0 = 0.0$. 
Similar with that in the previous two-dimension case, an anisotropic kernel algorithm  to reduce the total particle number evolving in this membrane simulation is used. Specifically, 8 particles are set in the vertical $ z $  direction, meaning that the particle spacing $dp_z = W/8 = 1.5625 \times 10^{-2}$ mm. Here, the anisotropic ratio is 8.0, meaning the  $dp_x = dp_y = 8dp_z  = 0.125 $ mm.
In the stress relaxation process of the simulation, 
the experienced	 damping ratio is set to  $\eta= 1.0e^4 $.
In terms of the convergence study of density
kinetic energy criteria,
by using the same method with that in 2D case,
the 3D case has a converged criterion 
value of  $E_d$ = 0.1\% $p^l$. 
\begin{figure*}[htbp]
	\centering
	\begin{subfigure}[b]{0.45\textwidth}
		\includegraphics[trim = 1mm 2mm 2mm 2mm, clip,width=0.9\textwidth]{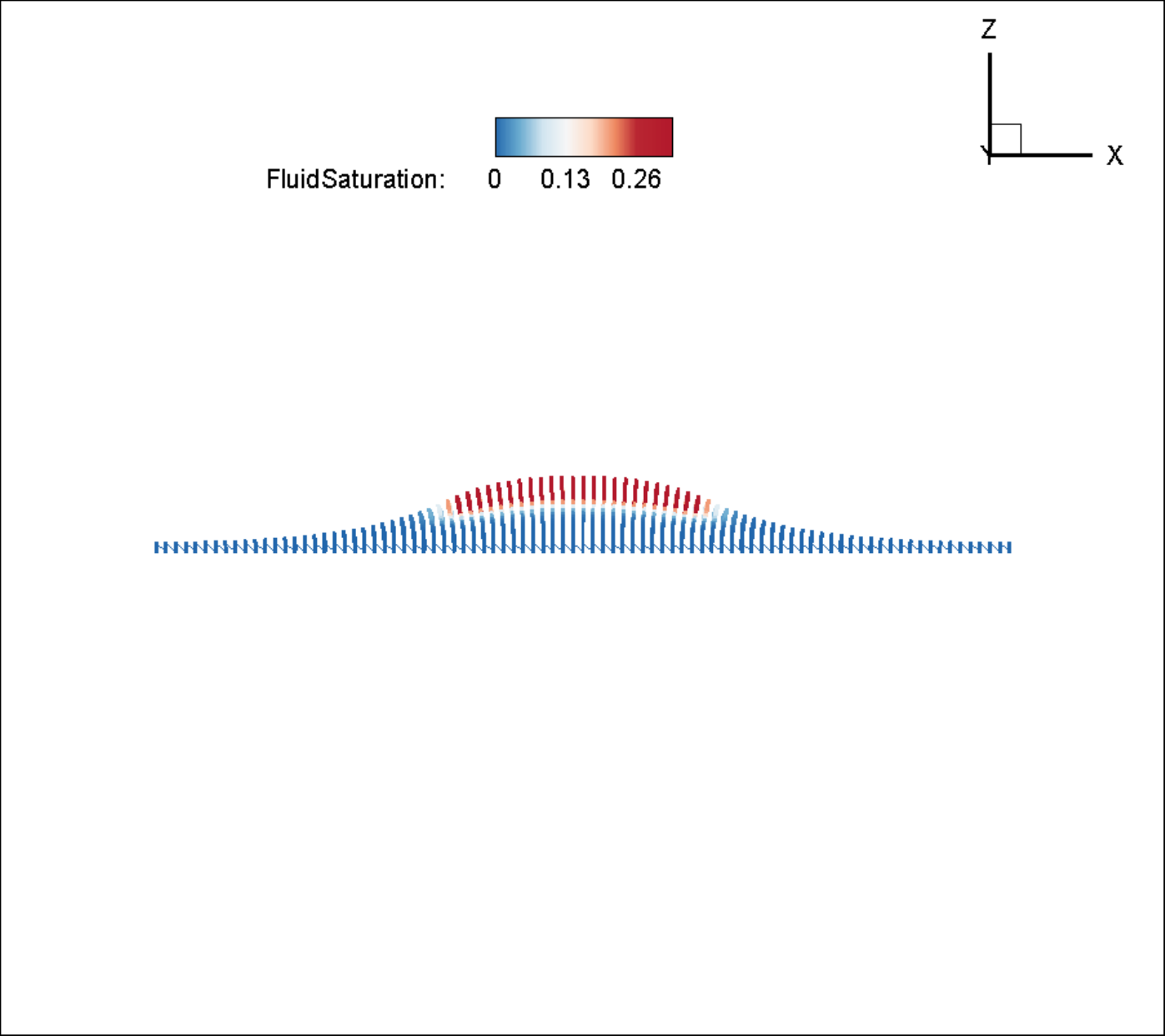}
		\caption {t = 450s, front view}
		\label{3d-saturaion-450-3}
	\end{subfigure}
	\begin{subfigure}[b]{0.45\textwidth}
	\includegraphics[trim =  1mm 2mm 2mm 2mm, clip,width=0.9\textwidth]{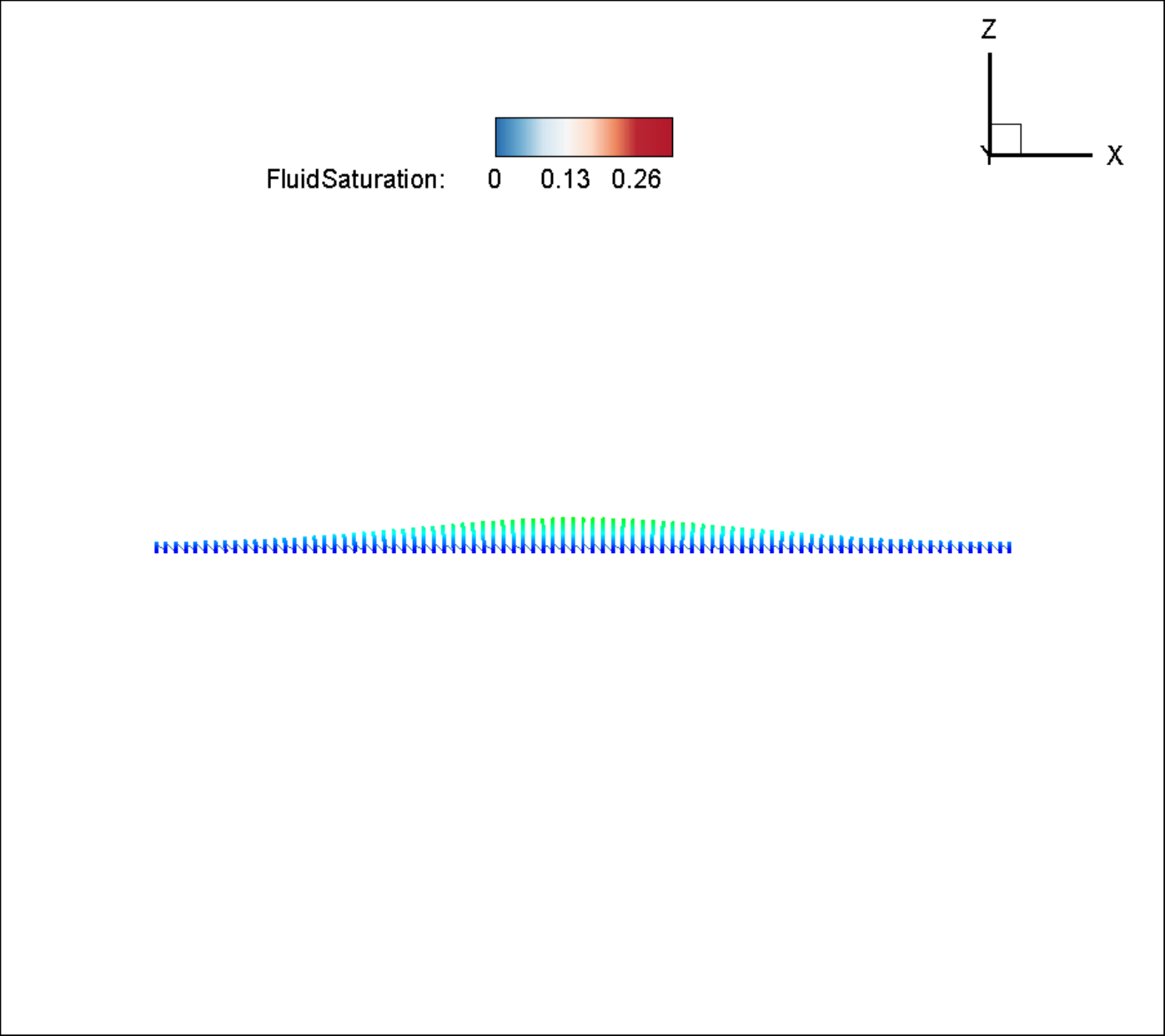}
	\caption {t = 1500s, front view}
	\label{3d-saturaion-1500}
\end{subfigure}
	\begin{subfigure}[b]{0.45\textwidth}
		\includegraphics[trim = 2mm 2mm 2mm 2mm, clip,width=0.9\textwidth]{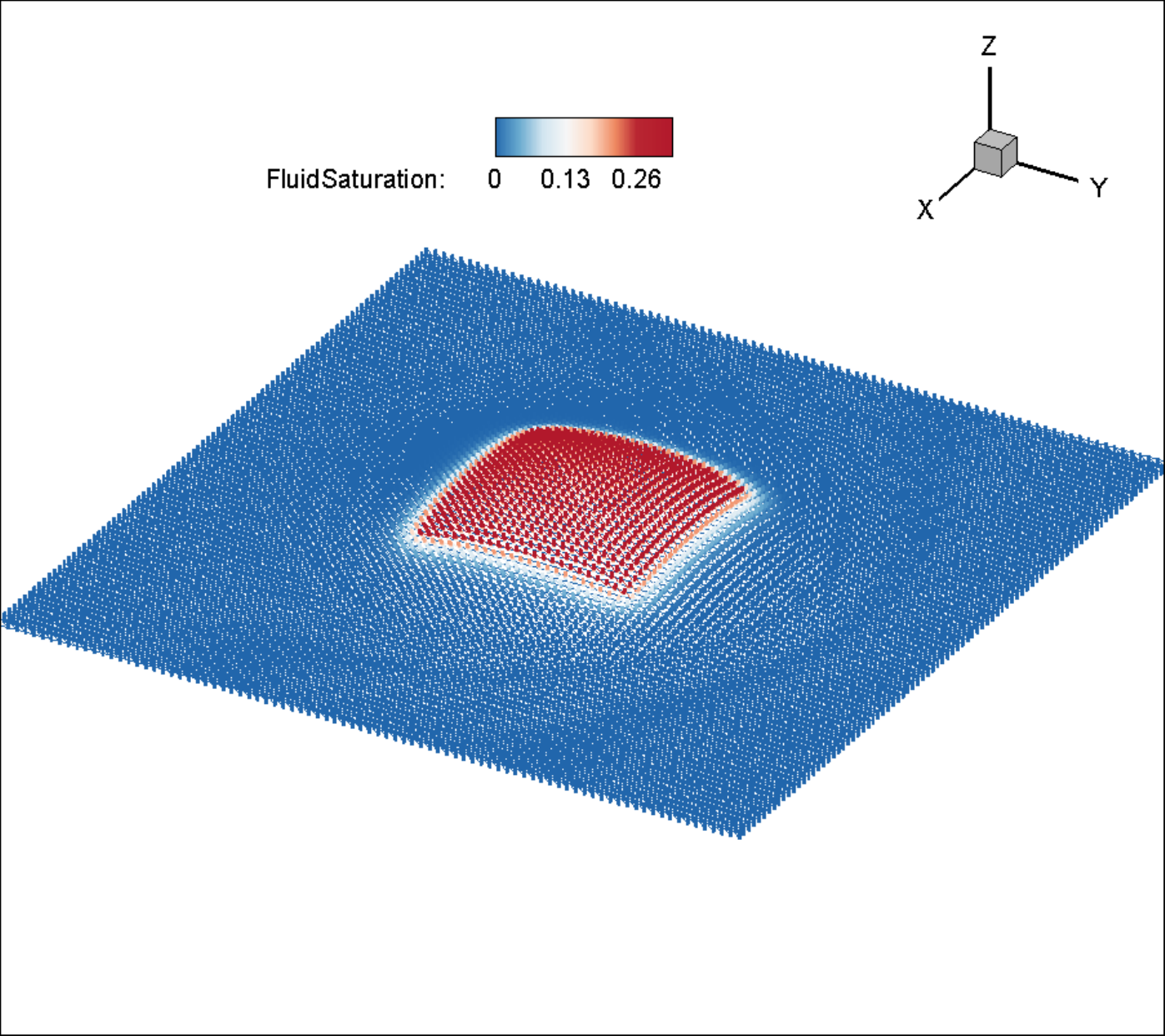}
		\caption {t = 450s, top side view}
		\label{3d-saturaion-450}
	\end{subfigure}
	\begin{subfigure}[b]{0.45\textwidth}
		\includegraphics[trim = 2mm 2mm 2mm 2mm, clip,width=0.9\textwidth]{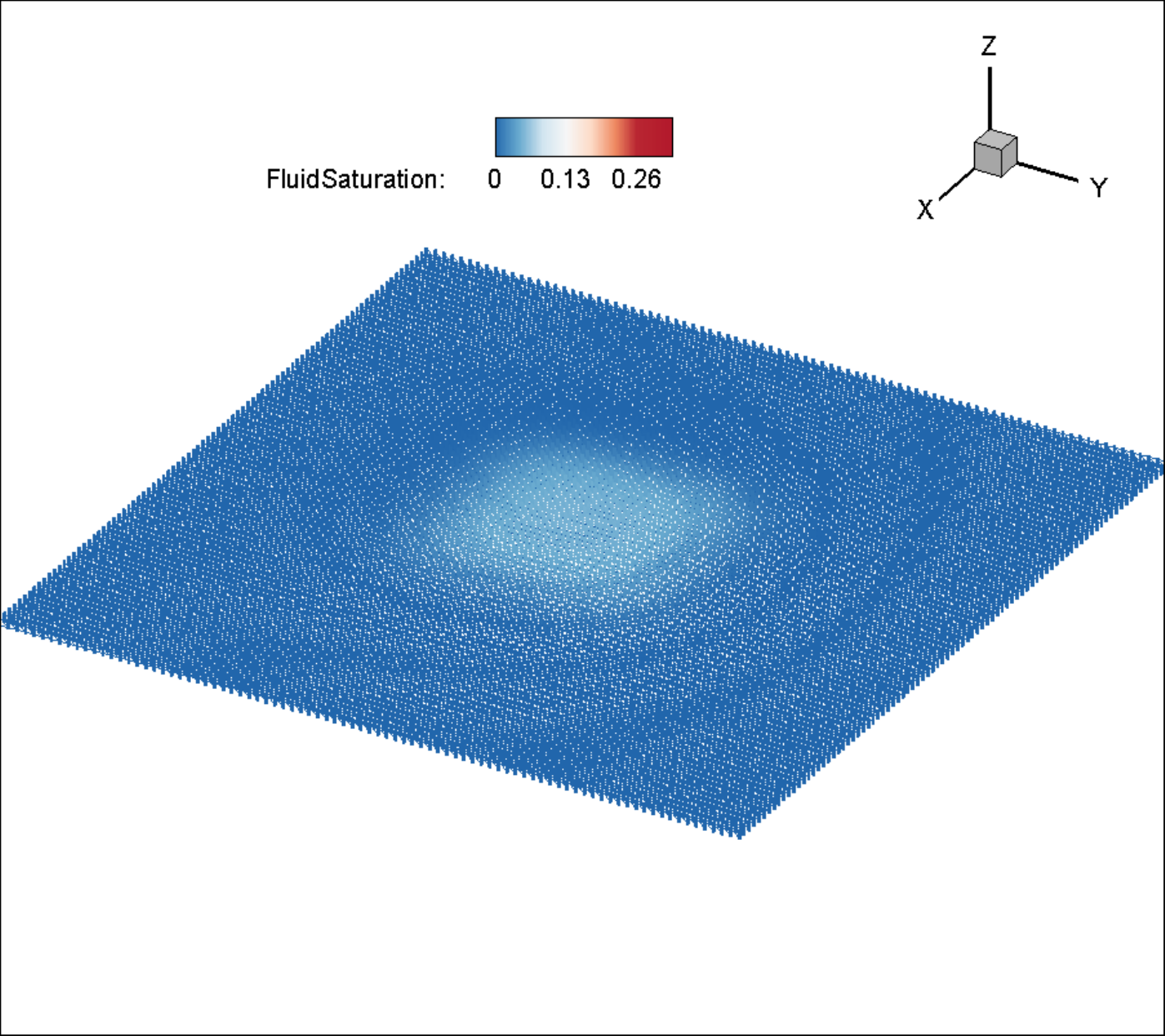}
		\caption {t = 1500s, top side view}
		\label{3d-saturaion-100}
	\end{subfigure}
	\begin{subfigure}[b]{0.45\textwidth}
	\includegraphics[trim = 1mm 2mm 1mm 2mm, clip,width=0.9\textwidth]{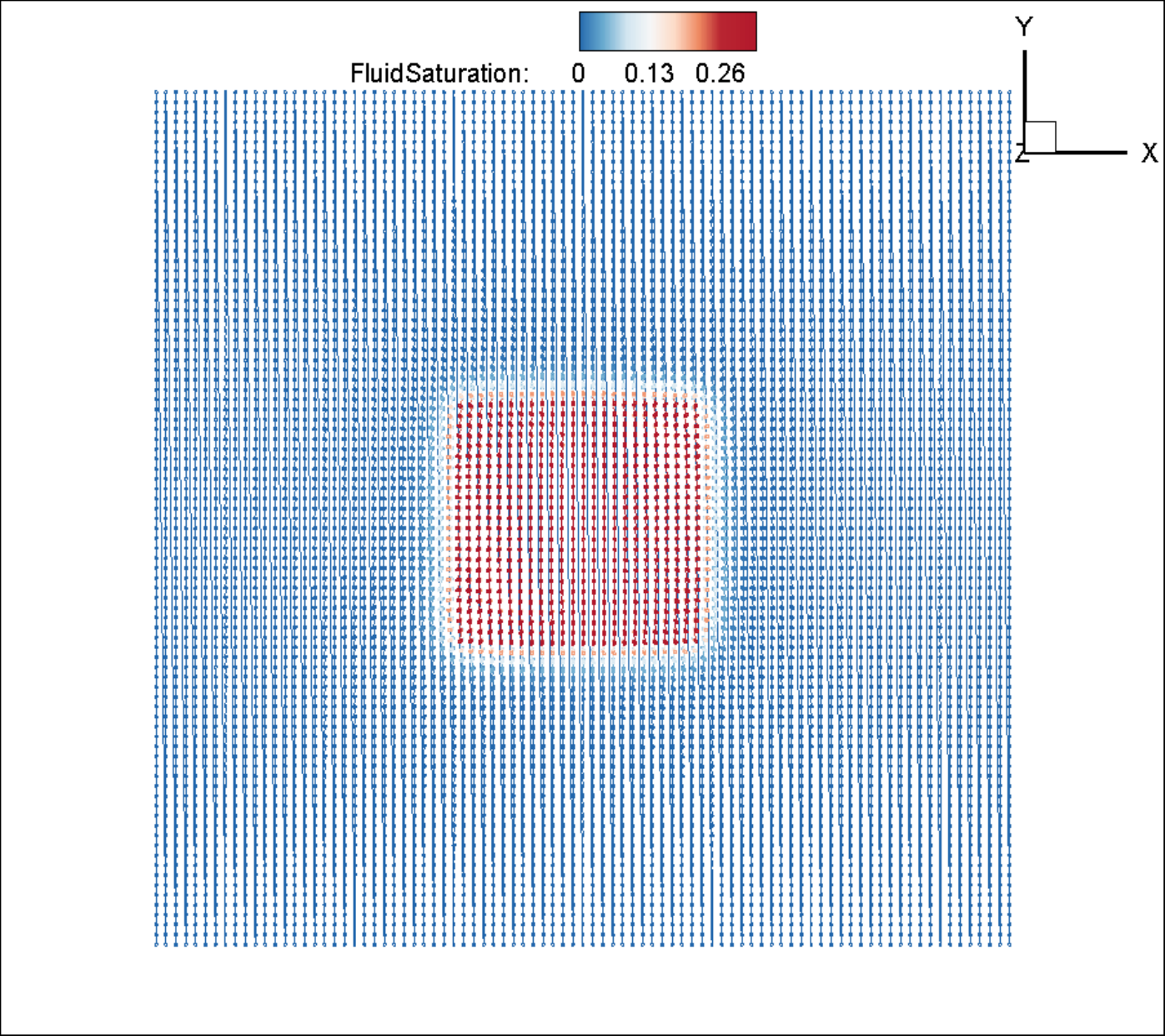}
	\caption {t = 450s, top view}
	\label{3d-saturaion-450-2}
\end{subfigure}
	\begin{subfigure}[b]{0.45\textwidth}
	\includegraphics[trim = 1mm 2mm 1mm 1mm, clip,width=0.9\textwidth]{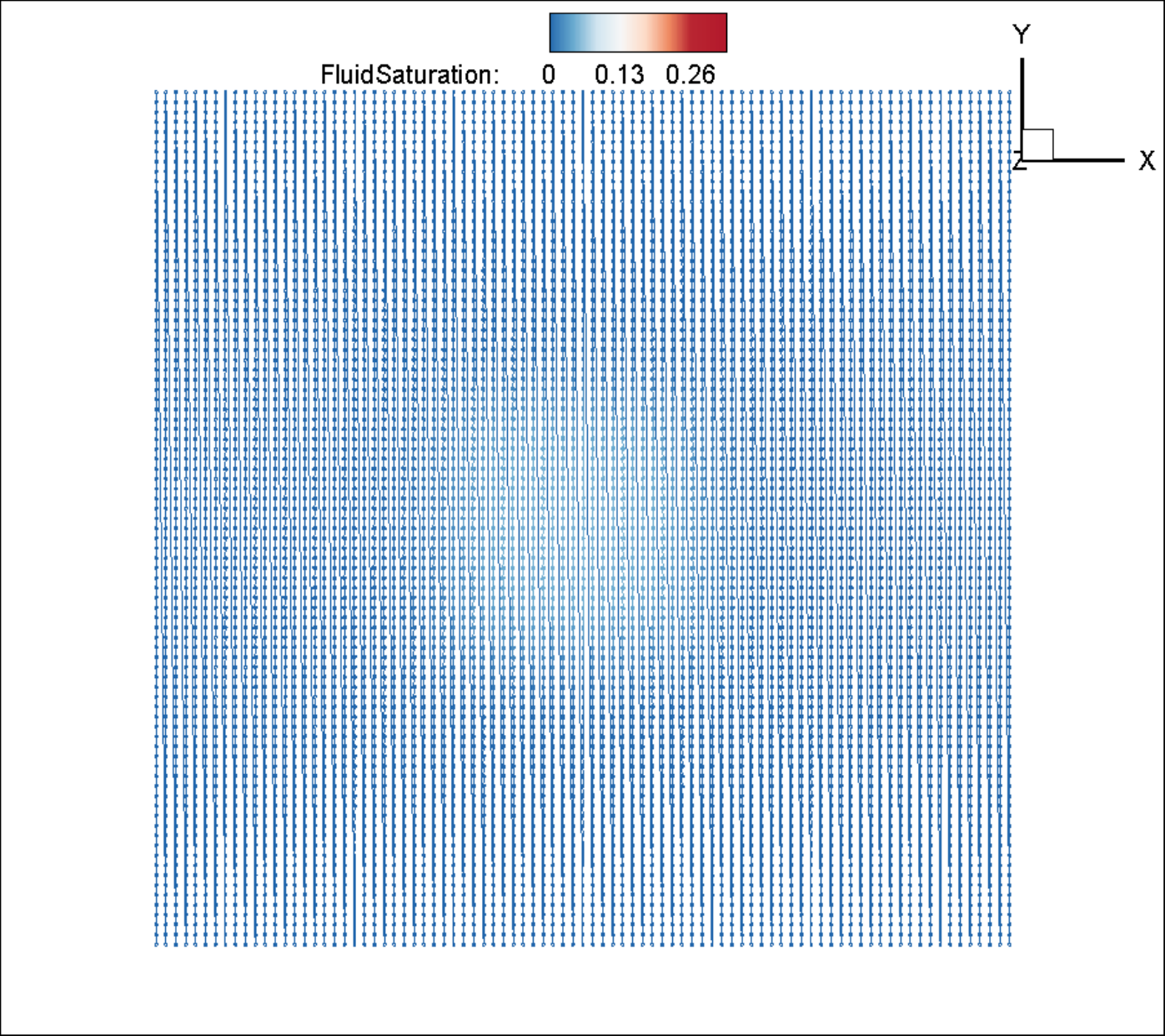}
	\caption {t = 1500s, top view}
	\label{3d-saturaion-1500-3}
\end{subfigure}
	\caption{3D fluid-structure interaction: the deformation colored by water saturation at different time instants. }
	\label{3ddiffusion}
\end{figure*}
In order to provide a more accurate representation of the experiment, 
 the evaporation process is taken into consideration,
i.e., the water loses as time progresses. 
Deformation flexure occurs during the initial period, 
and later as the mass of fluid loses from the membrane, 
it eventually returns back to the original shape.

\begin{figure*}[htbp]
	\centering
	\includegraphics[trim = 2mm 2mm 2mm 2mm, clip,width=0.55\textwidth]{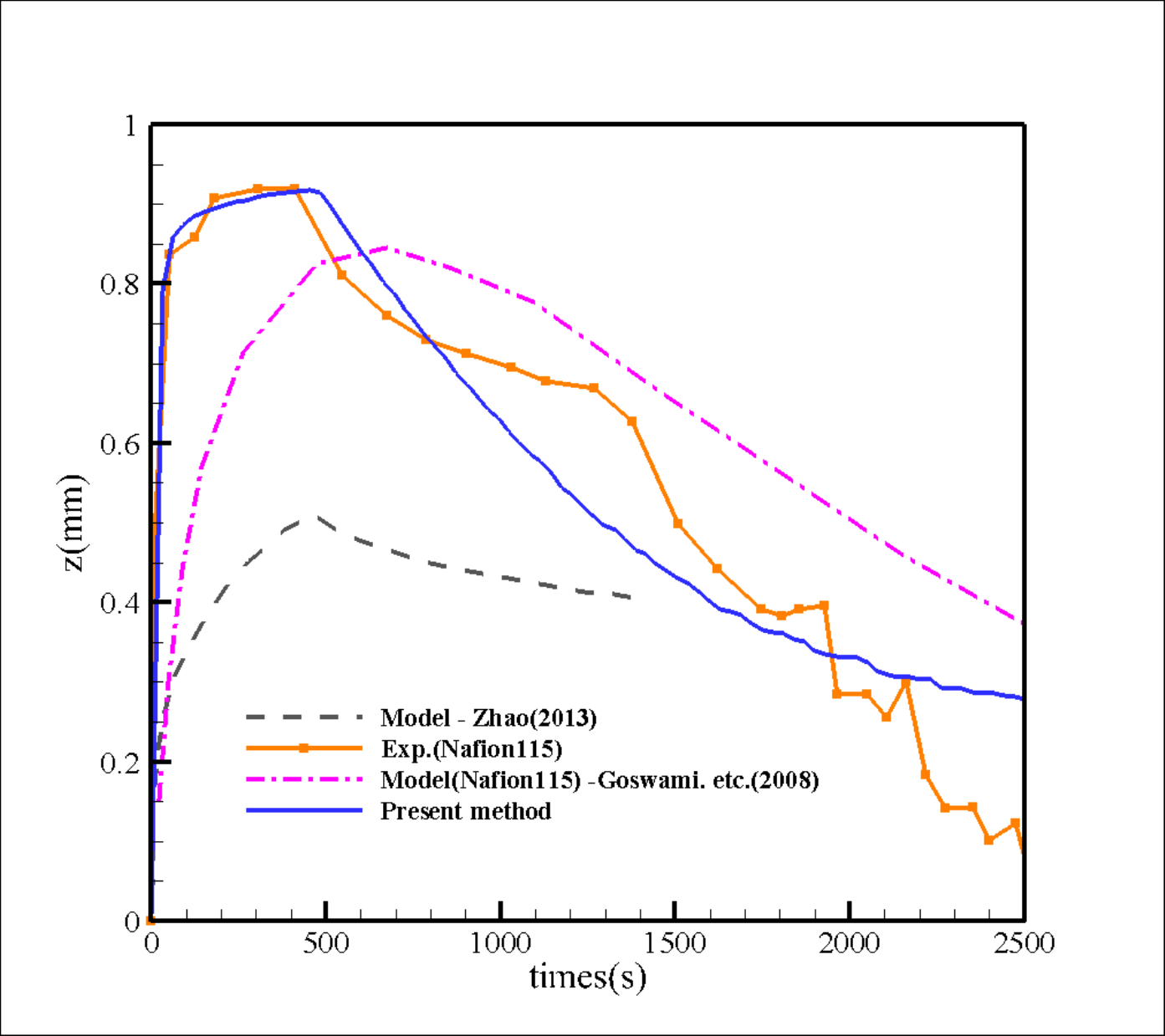}
	\caption {3D fluid-structure interaction: bending amplitude of the center point compared with experimental data and results from other numerical models.}
	\label{3dpoint_position}
\end{figure*}

Figure \ref{3ddiffusion} shows the membrane deformation colored by water saturation at different time instants.
In the first 450 seconds, 
water amount continues to increase as time progresses,
leading to a rising flexure 
as depicted in Figure \ref{3dpoint_position}, which records the time history of the height $z$ of the central point.
Once the contact period finishes,
 no further water is added into the beam,
and the central water flows slowly into the side areas.
At the same time, water evaporates from the membrane,
resulting in a rapid decrease of water pressure  
 and a corresponding decrease of the flexure,
as shown by the blue line in Figure \ref{3dpoint_position} beyond 450 seconds.
Figure \ref{3dpoint_position} also includes corresponding 
data points measured experimentally by Goswami \cite{goswami2008wetting} 
and results from other numerical models
for the swelling degree of the very center point
 versus different time instants.
Clearly, the present numerical simulation results 
exhibit good agreement with experimental results 
in terms of  
the deformation amplitude pattern, 
reproducing the increasing flexure during the water contact period 
and the subsequent decrease after the contact finishes, 
consistent with the saturation variation. 

\begin{figure*}[htbp]
	\centering
	\includegraphics[trim = 1mm 1mm 1mm 1mm, clip,width=0.55\textwidth]{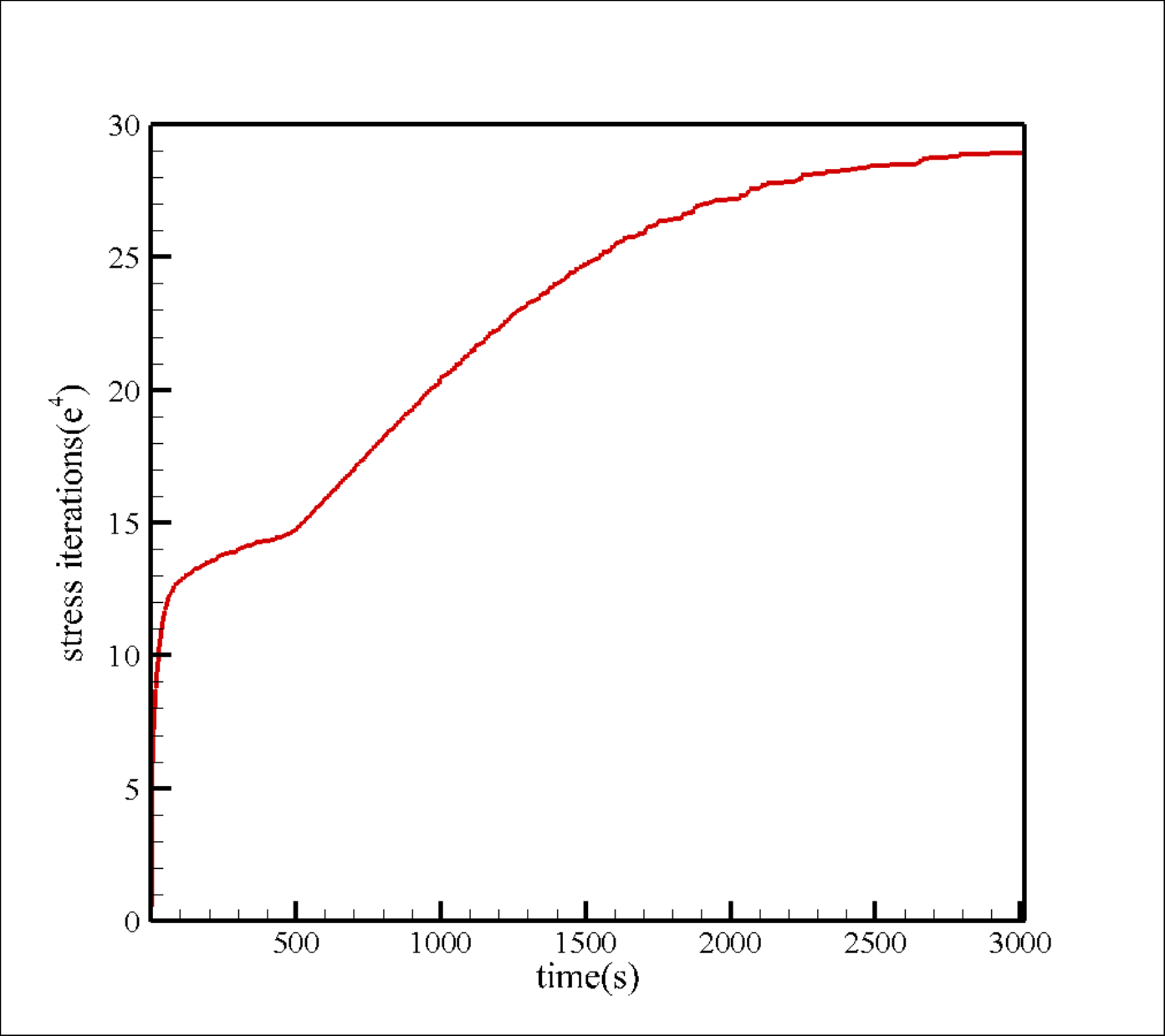}
	\caption {3D fluid-structure interaction: the stress iterations history during the whole simulation.}
	\label{3d-time-reduce}
\end{figure*} 

Drawing from the previous discussion, 
the optimal large outer time step is determined by the diffusion constant
and the smoothing length, 
while the small inner time step is dictated by the material properties of the solid. 
Ideally, the outer time step  allowed in principle is hundreds or
thousands of times larger than the inner time step size allowed.
However, in the standard explicit algorithm, 
the time step is limited to the smaller one, 
resulting in the execution of numerous stress 
relaxation steps
and consuming a substantial amount of time.
In the presented method,
first, diffusion is performed with the larger time step, 
while stress relaxation is executed multiple times
with damping effects until a kinetic energy is reached.
Our approach saves time in two ways. 
Firstly, the number of diffusion relaxation times is reduced
since multi-time step algorithm allows diffusion 
to be performed  with its own time step as the outer loop.
Secondly, once the kinetic energy criterion is satisfied, 
we consider the equilibrium achieved, 
and the inner loop is halted accordingly, 
avoiding unnecessary stress relaxation calculations.
Figure \ref{3d-time-reduce} indicates the stress iterations
 $ N_s $ during this 3D simulation.
There is  an increase 
in the initial 450 seconds
when the fluid is in contact with the film,
and then a slower increase in the later stages.
Table \ref{3D fluid-structure interaction efficiency} presents the quantitative efficiency of our new algorithm compared to the straightforward one,
by listing the diffusion iterations $ N_D $ and $ N_s $
separately.
As shown in the table, both two 
 iterations are obviously reduced,
representing a significant improvement in saving computation time.

\begin{table}[htb!]
	\centering
	\caption{3D fluid-structure interaction:  quantitative validation of the efficiency of this multi-time step algorithm.}
	\renewcommand{\arraystretch}{1.1} 
	\begin{tabular}{ccccc}
		\hline
		 algorithm   & membrane  & $ N_D $ & $ N_s $ & $ N_d $ \\ 
		\hline
straightforward algorithm		& 60552	& 1.5$ e^{10} $   &  1.5$ e^{10} $     & -\\
\hline
multi-time step algorithm	& 60552  & 1.25$ e^{5} $     &  2.89$ e^{6}$    & 2.89$ e^{6}$  \\
		\hline	
	\end{tabular}
	\label{3D fluid-structure interaction efficiency}
\end{table}

\section{Conclusion}
This paper proposed an approach employing a multi-time step algorithm to solve multi-time 
coupling problem involving solid dynamics.
In this algorithm,  the explicit scheme in time integration is used to simplify
the equation system solving. 
Inner and outer loops with different time step sizes 
are carried out to match different time scale process.
Another crucial feature of this algorithm 
is the utilization 
of a kinetic energy criterion to ascertain the attainment of equilibrium of solid dynamics
 and a damping term 
to accelerate this equilibrium attainment process, 
thereby enabling the earlier 
termination of the inner 
loop of solid stress relaxation
and avoiding redundant computations.
Two types of multi-time coupling problem, 
including a nonlinear hardening bar stretching 
and a fluid diffusion in porous media 
coupling solid deformation are 
simulated to test the performance of this algorithm.
Results demonstrate  the accuracy and 
a significant decrease in computation time.
Further, the application of  this algorithm
in practical fluid diffusion coupling hydrogel deformation
paves the way 
for  simulating complex multi-physics 
problems of multi-time scales in the field of complex chemistry reaction.

\textbf{Authorship contribution statement}

Xiaojing Tang  made the methodology,  designed the research,  developed code and tested the present library components, performed the visualization and validation, and wrote the original draft of the manuscript.   
Dong Wu  investigated the topic, made the methodology,  developed code and tested the present library components, conducted the formal analysis, modified the draft.
Zhentong Wang developed code and tested the present library components,  and revised the manuscript. 
Oskar Haidn and Xiangyu Hu made the conceptualization, supervised and administered the project, and revised the manuscript.

\textbf{Statements and Declarations}

The authors have no known competing financial interests or personal relationships that could
have appeared to influence the work reported in this paper.

\textbf{Acknowledgments} 

Xiaojing Tang was partially supported by the China Scholarship Council (Grant No. 201906120034).
Dong Wu was partially supported by the China Scholarship Council (Grant No. 20190613018).
Xiangyu Hu would like to express his gratitude to
Deutsche Forschungsge meinschaft (DFG)  
for their sponsorship of this research (Grant No. DFG HU1527/12-4).

\newpage
\bibliographystyle{IEEEtran}
\bibliography{mybio}
\newpage
\begin{appendices}
\section{Plasticity theory and nonlinear hardening plastic model}\label{appendixA}
In this appendix, we present the J2 plasticity theory coupling with a hardening elastic-plastic model to determine the plastic deformation. 
In A1 we describe the multiplicative decomposition technique for the material deformation.  A2 presents the constitutive relation in this model. To describe the strain-stress evolution, the flow rule and a hardening plastic model is stated in A3. Then a return mapping algorithm is given in A4 to explain
the time  integration.

\subsection{Multiplicative decomposition technique}
To describe the elastoplastic model, 
we adopted the flow plasticity theory where total 
strain can be multiplicatively decomposed into an  
elastic part and a plastic part \cite{simo2006computational, yue2015continuum}.
Using this technique,
$\mathbf{F}$ can be written as the product  
of its elastic 
volumetric part $\mathbf{F}^e$ and plastic 
deviatoric part $\mathbf{F}^p$:
\begin{equation} \label{eq:deformationtensor-split}
	\mathbf{F} = \mathbf{F}^e \mathbf{F}^p.
\end{equation}
Similarly, $\mathbf{b}^e $,  the elastic part of the left 
Cauchy-Green tensor  $ \mathbf{b} = \mathbf{F} \mathbf{F}^{T}$,
is defined as $\mathbf{b}^e = \mathbf{F}^e \mathbf{F}^{e\,T}$. 
When strains are within the elastic range,  
$ \mathbf{F} = \mathbf{F}^e $ and  
$ \mathbf{b} = \mathbf{b}^e $.
For plasticity analysis, 
the plastic Lagrangian tensor $\mathbf{C}^p$ 
is introduced as
\begin{equation} \label{eq:plastic-Lagrangian-tensor}
	\mathbf{C}^p = \mathbf{F}^{p\,T} \mathbf{F}^p.
\end{equation}
The relationship between $\mathbf{b}^e$ and 
$\mathbf{C}^p$ can be described as 
\begin{equation} \label{eq:plastic-tensor}
	\mathbf{b}^e = \mathbf{F}  \mathbf{C}^{p\,-1}  \mathbf{F}^T,
\end{equation}
which is used in the subsequent hardening plastic model. 
Additionally, to adhere to the volume 
preserving assumption in plasticity, 
we assume that the determinant of the plastic deformation part, det($\mathbf{F}^p$) $ = 1$.
\subsection{Constitutive relation}
According to the theoretical framework 
proposed by Simo and Hughes \cite{simo2006computational}, 
with an isotropic stress response assumption,
the elastoplastic constitutive model   
incorporates a nonlinear elastic strain energy 
function which is decomposed into volumetric 
and deviatoric parts:
\begin{equation} \label{eq:energy_total}
	W = W_v(J) + W_s(\overline{\mathbf{b}}^e),
\end{equation}
where $\overline{\mathbf{b}}^e$ is the 
volume-preserving left-Cauchy Green tensor.
The volumetric part weighted by the bulk 
modulus $K$, is given by 
\begin{equation} \label{eq:energy_volume}
	W_v(J) = \frac{1}{2}K[\frac{1}{2}(J^2-1)- \ln J].
\end{equation}
The deviatoric part related to 
the shear modulus $\mu$, is obtained by
\begin{equation} \label{eq:energy_shear}
	W_s({\overline{\mathbf{b}}^e}) = \frac{1}{2} \mu [\operatorname{Tr}(\overline{\mathbf{b}}^e)-D].
\end{equation}
Here,  $D =$ \{1,2,3\} depends on the dimension 
of the problem.
With $E$ denoting Young's modulus and $\nu$ the Poisson ratio,
the bulk and shear moduli are interconnected 
through the relationship:
\begin{equation}\label{relation-modulus}
	E = 2 \mu \left(1+\nu\right) = 3K\left(1 - 2\nu\right).
\end{equation}
With the energy function Eq. \eqref{eq:energy_total} in hand, 
the  Kirchhoff stress tensor, 
which characterizes the stress response, 
can be expressed as
\begin{equation}\label{kirchoff_stress_equation}
	\boldsymbol{\tau} =\frac{\partial W}{\partial \mathbf{F}^{e}} \mathbf{F}^{e T}=\frac{\kappa}{2}\left(J^{2}-1\right) \mathbf{I}+\mu \operatorname{dev}(\overline{\mathbf{b}}^{e}),
\end{equation} 
where the two parts account for the 
volumetric and shear stresses.

Note that in the equations above, the expression
\begin{equation}
	\overline{\mathbf{T}}=[\operatorname{det}(\mathbf{T})]^{-1/3} \mathbf{T}
\end{equation}
indicates the volume preserving 
treatment of a tensor $ \mathbf{T} $. Additionally,
\begin{equation}
	\operatorname{dev}(\mathbf{T})= \mathbf{T} -\frac{\operatorname{Tr}(\mathbf{T})}{3} \mathbf{I}
\end{equation}
represents the trace free part of the tensor 
$ \mathbf{T} $, i.e., 
$ \operatorname{Tr}[\operatorname{dev}(\mathbf{T})] =0 $ and  
$\operatorname{dev}(\mathbf{I}) =0 $.

\subsection{Flow rule and hardening plasticity model}
With the flow plasticity theory, a flow rule is needed to determine the orientation and magnitude of plastic deformation.
In this paper, the  classical $ J_2 $ flow theory,  
also known as the Mises–Huber yield condition 
proposed  by Hube and von Mises
\cite{mises1913mechanik},  is used 
to model plasticity stress-strain evolution.
This theory states that the plastic behavior
is governed by the  deviatoric part of the 
Kirchhoff stress tensor $\boldsymbol{\tau}$, represented by the 
second term   $ \mu \operatorname{dev}(\overline{\mathbf{b}}^{e}) $ in Eq.  \eqref{kirchoff_stress_equation}. 
For simplicity,  we define $\mathbf{s}=\operatorname{dev}(\boldsymbol{\tau})=\mu\operatorname{dev}(\overline{\mathbf{b}}^{e})$.
With a  Frobenius norm  $ \|.\|_{F} $,
the magnitude scalar $s=\|\mathbf{s}\|_{F}$ 
is used to compare with the yield criterion 
to determine the onset of plasticity, 
and the normalized tensor of $ {\mathbf{s}} $  is given by 
$\hat{\mathbf{s}}=\mathbf{s}/s$.

In  mechanical engineering, isotropic 
work hardening plastic 
behavior is commonly observed. 
To incorporate this behavior, 
a scalar yield function 
$f(\boldsymbol{\tau}, \alpha)$ that depends on 
the hardening function $k(\alpha)$ is introduced, 
where $\alpha$ represents the equivalent plastic strain.
The yield function is formulated as 
\begin{equation}
	\label{trial_function}
	f (\boldsymbol{\tau}, \alpha)=\|\operatorname{dev}(\boldsymbol{\tau})\|_{F} - \sqrt{\frac{2}{3}} k(\alpha)= s - \sqrt{\frac{2}{3}} k(\alpha) \leq 0,
\end{equation} 
where $k(\alpha)$ is defined by a nonlinear 
isotropic hardening law, as proposed by Simo et al. \cite{simo2006computational, elguedj2014isogeometric}:
\begin{equation}
	k(\alpha)=\sigma_{0}+\left(\sigma_{\infty}-\sigma_{0}\right)[1-\exp (-\delta \alpha)]+H \alpha,
\end{equation} 
where $ \sigma_{0} $ represents the initial flow stress,
also called yield stress,
$ \sigma_{\infty} $  the saturation flow stress, 
$ \delta $ the saturation exponent $\delta>0 $, 
and $ H $  the linear hardening coefficient. 
The yield function $f$ defines the yield surface
(when $ f = 0 $), which classifies the purely
elastic response when $f<0$.
When the yield condition is violated ($f>0$), 
the stress response to deformation consists of 
both plastic and elastic components.
Once the deformation enters the plastic regime, 
the material cannot go back to its original shape, with a permanently plastic deformation.
\subsection{Return mapping algorithm}

To integrate the material deformation over time, 
the  return mapping algorithm, which  has been explained 
and widely applied  in literature \cite{wilkins1963calculation,ponthot2002unified, dunne2005introduction,yu2006generalized, simo1988framework},
is applied here. 
In this algorithm, the deviatoric part 
of the Kirchhoff stress tensor 
$ \mathbf{s}^{pre} =\|\operatorname{dev}(\boldsymbol{\tau})\| = \mu \operatorname{dev}(\overline{\mathbf{b}}^{e})$ 
is obtained using the predicted update of 
$\mathbf{b}^e$, considering only the elastic strain. The yield condition is then 
checked using Eq. \eqref{trial_function} 
to determine if plastic deformation occurs. 
If the yield condition is not satisfied, 
the strain in the current step remains elastic, 
and the predicted update $\mathbf{b}^e$ is 
considered acceptable. Otherwise, plastic 
correction (returning map) is introduced to 
obtain the final $\mathbf{b}^e$ and $\mathbf{s}$ 
for the next step. The framework of 
this algorithm is presented in Agorithm \ref{al:algorithm1}.

\section{Fluid-structure interaction model}\label{appendixB}
In this appendix, referring to Zhao's \cite{zhao2013modeling} algorithm, 
we briefly discuss
the porosity assumption and the corresponding relations, including porosity and fluid saturation(\ref{appendixB1}), and stress relations(\ref{appendixB3}).  
In this simplified mixture model, 
the solid and fluid densities $\rho_s$ and $\rho_l$, solid velocity $\mathbf{v}^s$, and fluid saturation $\widetilde{a}$ are treated as state variables, 
enabling the fluid velocity to be calculated 
referring to solid velocity, rather than being an independent variable. This approach is practically significant because it significantly reduces the complexity of the system, as it eliminates the need for two sets of equations to describe the fluid and solid separately.

\subsection{Porosity and fluid saturation}\label{appendixB1}
Considering  a representative volume element $dV$,
the macroscopic porosity $a$ is defined as 
the ratio of the total volume of the 
pores $dV^p$ to $dV$, yielding $a = \frac{dV^p}{dV}$. 
Note that $ 0 < {a} < 1 $ holds for all cases.
 
When the porous solid is partially saturated by fluid,  the fluid volume in the representative element $ dV $ is denoted by $ dV^l $. The fluid saturation level $ \widetilde{a} $ can be defined as
\begin{equation}
	\label{s_defination}
	\widetilde{a} =\frac{dV^l}{dV}.
\end{equation}
Clearly, $\widetilde{a}$ is always less than or equal to the maximum possible saturation ${a}$, i.e., $\widetilde{a} \leq {a} $.
The locally effective fluid density $ \rho^l $, 
defined as the mass of the fluid per unit volume, 
varies depending on the extent of fluid saturation 
and can be expressed as 
\begin{equation}
	\label{fluid_density}
	\rho^l =  \frac{dm^{l}} {dV } = \frac{dm^{l}} {dV^l }\frac{dV^{l}} {dV } =\rho^l_0 \widetilde{a} ,
\end{equation} 
where $dm^l$ represents
the mass of the fluid within   
a representative volume element $dV$, 
$ {\rho}^l_0$ the 
initial density of the fluid which 
is assumed to be a constant for 
incompressible fluids. 
 
\subsection{Effective  stress on solid}\label{appendixB3}
Following \cite{gawin1995coupled,korsawe2006finite,ghaboussi1973flow, atkin1976continuum}, 
the total stress acting on the solid 
is the sum  of Cauchy stress $ \boldsymbol{\sigma}^s $ and 
the  pressure  stress due to the presence of the fluid phase $ \boldsymbol{\sigma}^l $, written as:
\begin{equation}
	\boldsymbol{\sigma} =\boldsymbol{\sigma}^s +\boldsymbol{\sigma}^l = \boldsymbol{\sigma}^s - p^l \mathbf{I}.
\end{equation}
where $ p^l $ is fluid pressure. 
For a hyper-elastic material, 
the constitutive equation for the 
solid component is given by
\begin{equation}
	\label{cauchy-stress}
	\boldsymbol{\sigma}^s = 2\mu \mathbf{e} + \lambda \text{tr}(\mathbf{e})\mathbf{I},
\end{equation}
where  the Eulerian-Almansi finite strain tensor  $\mathbf{e}$  
can be evaluated  by
\begin{equation}
	\mathbf{e} = \frac{1}{2} ( \mathbf{I} - \mathbf{F}^T \mathbf{F}).
\end{equation}
The Lam$\acute{e}$ parameters $ \lambda $ can be calculated via  
shear modulus $\mu$  and bulk modulus K as 
$ \lambda =K - \frac{2\mu}{3}$.  

The excess fluid pressure simply depends on 
the fluid saturation level within the porous solid element,
with a function $ p^l = p^l (\widetilde{a}) $. 
The relationship between fluid diffusion and 
the solid deformation satisfies a fundamental 
principle: 
when fluid flows out of  a given zone,  
the saturation level decreases, 
resulting in a drop in pressure, and consequently, the material tends 
to contact inwardly. 
Conversely, when fluid penetrates a porous
solid area, 
there exists a higher saturation level corresponding 
to a stronger pressure force, 
leading to a material expansion. 
In the present model, this behavior is described mathematically using a linear relation, taking the form
\begin{equation}
	\label{fluid_pressure}
	p^l = C (\widetilde{a} -   \widetilde{a}_0),
\end{equation} 
where   $ C $ is a material constant,  $\widetilde{a}_0$ the  initial  saturation. Details can be referred to   \cite{atkin1976continuum}.

\begin{algorithm}
	\footnotesize
	Update deformation tensor
	\begin{equation}
	\mathbf{F}_{n+1}= \Delta t  \frac{d\mathbf{F}}{dt},  \quad \overline{\mathbf{F}}_{n+1}=[\operatorname{det}(\mathbf{F}_{n+1})]^{-1/3} \mathbf{F}_{n+1} \nonumber 
	\end{equation}
	
	Predict the elastic
	\begin{equation} \overline{\mathbf{b}}_{n+1}^{e, \mathrm{pre}}=\overline{\mathbf{F}}_{n+1} \overline{\mathbf{C}}_{p}^{n} \overline{\mathbf{F}}_{n+1}^{T}, \quad    	
	\mathbf{s}_{n+1}^{\mathrm{pre}}=\mu \operatorname{dev}(\overline{\mathbf{b}}_{n+1}^{e, \mathrm{pre}}) \nonumber \end{equation}
	
	Check the von Mises criterion
	\begin{equation}
	f_{n+1}^{\mathrm{pre}}= s_{n+1}^{\mathrm{pre}}-\sqrt{\frac{2}{3}} k(\alpha_n) \nonumber
	\end{equation}
	
	\eIf{$ f_{n+1}^{\mathrm{pre}}\leq 0 $} 
	{ Elastic state, Set $ (\textbf{.})_{n+1}=(\textbf{.})_{n+1}^{\mathrm{pre}} $, go to \ref{Update_stress}. }
	{Perform \ref{return_mapping} (the return mapping). }

	Compute the normlized shear modulus\label{return_mapping}
	\begin{equation}
	\tilde{\mu}=\frac{1}{3} \operatorname{Tr}(\overline{\mathbf{b}}_{n+1}^{e, \mathrm{pre}}) \mu \nonumber
	\end{equation}
	Initiate
	\begin{equation} 
	\Delta \gamma  = 0\nonumber 
	\end{equation}
	Compute $ \Delta \gamma $ so that 
	\begin{equation}
	\hat{f}(\Delta \gamma)=\left\|\mathbf{s}_{n+1}^{\text {trial }}\right\|-\sqrt{\frac{2}{3}} k\left(\alpha_{n}+\sqrt{\frac{2}{3}} \Delta \gamma\right)-2 \bar{\mu} \Delta \gamma=0 \nonumber
	\end{equation}
	$$ \hat{\mathbf{s}} =\mathbf{s}_{n+1}^{\mathrm{pre}} / s_{n+1}^{\mathrm{pre}} $$
	Return map
	\begin{equation}
	\mathbf{s}_{n+1} = \mathbf{s}_{n+1}^{\mathrm{pre}} -2 \tilde{\mu} \Delta \gamma \hat{\mathbf{s}}, \quad
	\alpha_{n+1} = \alpha_{n} + \sqrt{\frac{2}{3}} \Delta \gamma \nonumber
	\end{equation}
	
	Update stress\label{Update_stress}
	\begin{equation}
	J_{n+1} = det(\mathbf{F}_{n+1}), \quad
	\boldsymbol{\tau}_{n+1} =\frac{\kappa}{2}\left(	J_{n+1} ^{2}-1\right) \mathbf{I}+	\mathbf{s}_{n+1}, \quad
	\mathbf{P}_{n+1} =  \boldsymbol{\tau}_{n+1} \mathbf{F}_{n+1}^{-T}\nonumber
	\end{equation}
	
	Update local configuration
	\begin{equation}
	\overline{\mathbf{b}}_{n+1}^{e}=  \frac{1}{\mu} \mathbf{s}_{n+1}+\frac{1}{3} \operatorname{Tr}\left(\overline{\mathbf{b}}_{n+1}^{e, \text { pre}}\right) \mathbf{I},  \quad
	\overline{\mathbf{C}}_{p}^{n+1} = \overline{\mathbf{F}}_{n+1}^{-1} 
	\overline{\mathbf{b}}_{n+1}^{e} \overline{\mathbf{F}}_{n+1}^{-T}\nonumber
	\end{equation}
	\caption{Returning-mapping algorithm for $ J_2 $ flow theory with nonlinear isotropic hardening.}
	\label{al:algorithm1} 
\end{algorithm}

\end{appendices}
\end{document}